%% file: main.tex
\documentclass[pdftex,twocolumn,epjc3]{svjour3}          

\RequirePackage[T1]{fontenc}
\RequirePackage{fix-cm}

\smartqed  

\RequirePackage{graphicx}
\RequirePackage{mathptmx}

\DeclareMathAlphabet{\mathcal}{OMS}{cmsy}{m}{n}

\usepackage{footmisc}
\RequirePackage[numbers,sort&compress]{natbib}
\RequirePackage[colorlinks,citecolor=blue,urlcolor=blue,linkcolor=blue]{hyperref}

\RequirePackage{siunitx}
\RequirePackage{amsmath}
\usepackage{caption}
\usepackage{upgreek}
\usepackage{subfig}
\usepackage{caption}
\usepackage{float}
\usepackage{ragged2e}
\usepackage{pdfpages}

\usepackage{bm}

\input{CommandsAndShortcuts}

\graphicspath{{./Images/}}

\journalname{Eur. Phys. J. C}

\begin{document}
\null 
\includepdf[scale=1]{Images/title-9.pdf}

\raggedbottom

\title{The design of the n2EDM experiment
}
\subtitle{nEDM collaboration}


\author{
    N.~J.~Ayres\thanksref{ETH}  
    \and
        G.~Ban\thanksref{CAEN}    
    \and
        L.~Bienstman\thanksref{Leuven}
    \and
        G.~Bison\thanksref{PSI}    
    \and
        K.~Bodek\thanksref{Cracow}        
    \and
        V.~Bondar\thanksref{ETH, e1}  
    \and
        T.~Bouillaud\thanksref{LPSC} 
    \and
        E.~Chanel\thanksref{Bern}   
    \and
        J.~Chen\thanksref{CAEN}   
    \and
        P.-J. Chiu\thanksref{ETH, PSI} 
    \and
        B.~Cl\'ement\thanksref{LPSC}              
    \and
        C.~Crawford\thanksref{Kentucky}              
    \and
        M.~Daum\thanksref{PSI}      
    \and
        B.~Dechenaux\thanksref{CAEN} 
    \and
        C.~B.~Doorenbos\thanksref{ETH, PSI} 
    \and
        S.~Emmenegger\thanksref{ETH}  
    \and
        L.~Ferraris-Bouchez\thanksref{LPSC} 
    \and
        M.~Fertl\thanksref{Mainz} 
    \and
        A.~Fratangelo\thanksref{Bern}  
    \and
        P.~Flaux\thanksref{CAEN}            
    \and
        D.~Goupillière\thanksref{CAEN}  
    \and
        W.~C.~Griffith\thanksref{Sussex}            
    \and
        Z.~D.~Grujic\thanksref{Serbia}   
    \and
        P.~G.~Harris\thanksref{Sussex}            
    \and
        K.~Kirch\thanksref{ETH,PSI}         
    \and
        P.~A.~Koss\thanksref{Leuven,e4}        
    \and
        J.~Krempel\thanksref{ETH}  
    \and
        B.~Lauss\thanksref{PSI}              
    \and
        T.~Lefort\thanksref{CAEN}            
    \and
        Y.~Lemière\thanksref{CAEN}            
    \and
        A.~Leredde\thanksref{LPSC}  
    \and
        M.~Meier\thanksref{PSI}
    \and
        J.~Menu\thanksref{LPSC}  
    \and
        D.~A.~Mullins\thanksref{Bern}
    \and
        O.~Naviliat-Cuncic\thanksref{CAEN}        
    \and
        D.~Pais\thanksref{ETH,PSI}            
    \and
        F.~M.~Piegsa\thanksref{Bern}             
    \and
        G.~Pignol\thanksref{LPSC, e2}   
    \and
    G.~Quéméner\thanksref{CAEN}
    \and
        M.~Rawlik\thanksref{ETH,e3} 
    \and
        D.~Rebreyend\thanksref{LPSC}  
    \and    
        I.~Rienäcker\thanksref{ETH,PSI}
    \and
        D.~Ries\thanksref{Mainz2}              
    \and
        S.~Roccia\thanksref{LPSC}             
    \and
        K.~U.~Ross\thanksref{Mainz2}          
    \and
        D.~Rozpedzik\thanksref{Cracow} 
    \and
        W.~Saenz\thanksref{CAEN} 
    \and
        P.~Schmidt-Wellenburg\thanksref{PSI}         
    \and
        A.~Schnabel\thanksref{PTB}          
    \and
        N.~Severijns\thanksref{Leuven}             
    \and
        B.~Shen\thanksref{Mainz2} 
    \and
        T.~Stapf\thanksref{PSI} 
    \and
        K.~Svirina\thanksref{LPSC} 
    \and
        R.~Tavakoli Dinani\thanksref{Leuven}  
    \and
        S.~Touati\thanksref{LPSC} 
    \and
        J.~Thorne\thanksref{Bern}
    \and
        R.~Virot\thanksref{LPSC}             
    \and
        J.~Voigt\thanksref{PTB}             
    \and
        N.~Yazdandoost\thanksref{Mainz2} 
    \and
        J.~Zejma\thanksref{Cracow}  
    \and
        G.~Zsigmond\thanksref{PSI}            
    }    

\thankstext{e1}{Corresponding author: bondarv@phys.ethz.ch}
\thankstext{e2}{Corresponding author: guillaume.pignol@lpsc.in2p3.fr}
\thankstext{e3}{Present address: Paul Scherrer Institut, CH-5232 Villigen PSI, Switzerland}
\thankstext{e4}{Present address: Fraunhofer Institute for Physical Measurement Techniques, 79110 Freiburg, Germany}

\institute{
    ETH Zürich, Institute for Particle Physics and Astrophysics, CH-8093 Zürich, Switzerland \label{ETH}
    \and
    Normandie Univ, ENSICAEN, UNICAEN, CNRS/IN2P3, LPC Caen, 14000 Caen, France \label{CAEN}
    \and
    Instituut voor Kern- en Stralingsfysica, University of Leuven, B-3001 Leuven, Belgium\label{Leuven}
    \and
    Paul Scherrer Institut, CH-5232 Villigen PSI, Switzerland \label{PSI}
    \and
    Marian Smoluchowski Institute of Physics, Jagiellonian University, 30-348 Cracow, Poland \label{Cracow}
    \and
    LPSC, Université Grenoble Alpes, CNRS/IN2P3, Grenoble, France\label{LPSC}
    \and
    University of Bern, Albert Einstein Center for Fundamental Physics, CH-3012 Bern, Switzerland\label{Bern}
    \and
    University of Kentucky, Lexington, USA\label{Kentucky}
    \and
    Institut f\"{u}r Physik, Johannes-Gutenberg-Universit\"{a}t, D-55128 Mainz, Germany \label{Mainz}
    \and
    Department of Physics and Astronomy,      University of Sussex, Falmer, Brighton BN1 9QH, UK \label{Sussex}
    \and
    Institute of Physics Belgrade, University of Belgrade, 11080 Belgrade, Serbia \label{Serbia}
    \and
    Department of Chemistry - TRIGA site, Johannes-Gutenberg-Universität, 55128 Mainz, Germany\label{Mainz2}
    \and
    Physikalisch Technische Bundesanstalt, Berlin, Germany\label{PTB}
}

\date{Received: date / Accepted: date}

\maketitle

\begin{abstract}
We present the design of a next-generation experiment, n2EDM, currently under construction at the ultracold neutron source at the Paul Scherrer Institute (PSI) with the aim of carrying out a high-precision search for an electric dipole moment of the neutron. The project builds on experience gained with the previous apparatus operated at PSI until 2017, and is expected to deliver an order of magnitude better sensitivity with provision for further substantial improvements. An overview is given of the experimental method and setup, the sensitivity requirements for the apparatus are derived, and its technical design is described.


\end{abstract}

\section{Introduction and motivation}
\input{Introduction}
\label{sec:intro}

\section{The principle of the n2EDM experiment}
\label{sec:2}
\input{Measurement-intro}

\subsection{The n2EDM concept}
\input{Concept}

\label{sec:2.1}
\subsection{Measurement procedure}
\label{sec:2.2}
\input{Measurement}

\section{Projected statistical sensitivity}
\label{sec:3}
\input{Statistics}

\subsection{UCN counts $N$}
\label{sec:3.1}
\input{UCNs}

\subsection{Electric field strength $E$}
\label{sec:3.2}
\input{Efield}

\subsection{Precession time $T$}
\label{sec:3.3}
\input{Precession}

\subsection{Neutron polarisation $\alpha$}
\label{sec:3.4}
\input{Polarisation}

\subsection{Additional statistical fluctuations and final remarks}
\label{sec:3.5}
\input{AddFluctuations-remarks}

\section{Frequency shifts and systematic effects}
\label{sec:4}
\input{Systematics}

\section{The core systems of the n2EDM apparatus}
\input{Technical-Intro.tex}

\label{sec:5}

\subsection{UCN system}
\label{sec:5.1}
\subsubsection{UCN precession chambers}
\input{Precession-chambers.tex}
\label{sec:5.1.1}
\subsubsection{Electric field generation}
\input{EfieldGeneration.tex}
\label{sec:5.1.2}
\subsubsection{UCN transport}
\input{UCN-apparatus.tex}

\label{sec:5.1.3}
\subsubsection{UCN spin-sensitive detection} 
\input{UCN-detection}

\label{sec:5.1.4}

\subsection{Magnetic field shielding}
\input{Magnetic-shielding-intro.tex}

\label{sec:5.2}
\subsubsection{Passive magnetic shield} 
\input{MSR.tex}

\label{sec:5.2.1}
\subsubsection{Active magnetic shield} 
\label{sec:5.2.2}
\input{AMS.tex}

\subsection{Magnetic field generation}
\input{Magnetic-field-generation}
\label{sec:5.3}

\subsection{Magnetic field measurement}
\label{sec:5.4}

\subsubsection{Magnetometry concept} 
\label{sec:5.4.1}
\input{Magnetometry-concept.tex}
\subsubsection{Hg magnetometry} 
\label{sec:5.4.2}
\label{sec:Hg-magnetometry}
\input{Hg-magnetometry.tex}
\subsubsection{Cs magnetometry} 
\label{sec:5.4.3}
\input{Cs-Magnetometry.tex}
\label{sec:Cs-magnetometry}
\subsubsection{Mapper} 
\input{Magnetometry-mapper}
\label{sec:Mapper}

\section*{Summary and conclusions}
\input{Summary}

\label{sec:6}

\section*{Acknowledgements}
\begin{sloppypar}
We gratefully acknowledge the outstanding support from technicians, engineers and other professional services throughout the collaboration:
B.~Blau, 
K.~Boutellier,
B.~Bougard, 
J.~F.~Cam,
B.~Carniol, 
M.~Chala,
A.~Chatzimichailidis, 
M.~Dill, 
P.~Desrues, 
P.~Erisman, 
A.~Ersin, 
D.~Etasse, 
R.~Faure,
C.~Fourel,
J.~Fulachier,
C.~ Fontbonne, 
C.~ Geraci,
A.~ Gn\"adinger, 
U.~ Greuter, 
J.~ Hadobas, 
L.~Holitzner,
J.~Hommet,  
M.~Horisberger, 
S.~Hauri,
B.~Jehle, 
R.~K\"ach, 
G.~K\"aslin, 
C.~Kramer, 
K.~Lojek,
S.~Major,
M.~M\"ahr, 
J.~Marpaud,
C.~Martin,
M.~Marton,
Y.~Merrer,
O.~Morath, 
R.~Nicolini,
L.~Noorda,
F.~Nourry,
J.~Odier,
J.~Oertli,
C.~Pain, 
J.~Perronnel,  
M.~Philippin,
W.~Pfister,
S.~Roni,
S.~Roudier,
D.~Reggiani,
R.~ Schwarz,
 J.P.~ Scordilis,
M.~St\"ockli,
C.~Str\"assle, 
V.~Talanov, 
V.~Teufel, 
C.~Thomassé,
C.~VanDamme,
C.~Vescovi,
A.~Van Loon, 
R.~Wagner,
X.~Wang,
J.~Welte.
We are grateful for financial support from the
the Swiss National Science Foundation through projects 
200020-188700 (PSI), 
200020-137664 (PSI), 
200021-117696 (PSI), 
200020-144473 (PSI), 
200021-126562 (PSI), 
200021-181996 (Bern), 
200020-172639 (ETH), 
R`EQUIP 139140, 177008
and FLARE 20FL21-186179. 
The support by Emil-Berthele-Fonds is acknowledged.
The LPC Caen and the LPSC Grenoble acknowledge the support of the French Agence Nationale de la Recherche (ANR) under 
reference ANR-14-CE33-0007 and the ERC project 716651-NEDM.
The Polish collaborators acknowledge support from the National Science Center, Poland, Grants No.~2015/18/M/ST2/00056, No.~2018/30/M/ST2/00319, No.~2016/23/D/ST2/00715 and No.~2020/37/B/ST2/02349.
Support by the Cluster of Excellence
``Precision Physics, Fundamental Interactions, and Structure of Matter'' (PRISMA \& EXC 2118/1) funded by the German Research Foundation (DFG) within the German Excellence Strategy 
(Project ID 39083149) is acknowledged. 
This work was partly supported by the Fund for Scientific Research Flanders (FWO) and Project GOA/2010/10 of the KU Leuven.
\end{sloppypar}

\bibliographystyle{spphys}    
\def\urlprefix{}
\bibliography{Ref-short}

\end{document}

%% file: CommandsAndShortcuts.tex
\newcommand{\diff}[1]{\operatorname{d}\ifthenelse{\equal{#1}{}}{\,}{\!#1}}

\newcommand\comagnetometer{{co-mag\-ne\-tome\-ter}}
\newcommand{\ecm}{\ensuremath{\si{\elementarycharge}\!\cdot\!\cm}}
\newcommand{\dn}{\ensuremath{d_\text{n}}}

\newcommand{\TOP}{{\rm TOP}}
\newcommand{\BOT}{{\rm BOT}}

\DeclareSIUnit{\sqrtHz}{\ensuremath{\sqrt{\text{\hertz}}}}

\newcommand{\muT}{\upmu {\rm T}}

\newcommand{\pT}{\mbox{pT}}
\newcommand{\fT}{\mbox{fT}}

\newcommand{\cm}{\ensuremath{\mathrm{cm}}}

\newcommand{\s}{\ensuremath{\mathrm{s}}}
\newcommand{\ms}{\ensuremath{\mathrm{ms}}}
\newcommand{\R}{\mathcal{R}}
\newcommand{\z}{\langle z \rangle}



%% file: Introduction.tex
Searches for permanent electric dipole moments (EDM) of fundamental particles and systems are among the most sensitive probes for CP violation beyond the Standard Model (SM) of particle physics; see e.g.~\cite{Pospelov2005,Engel2013}. 


\begin{sloppypar}
Although the CP-violating complex phase of the CKM matrix is close to maximal, the resulting SM values for EDMs are tiny, while theories and models beyond the SM (BSM) often predict sizeable CP violating effects that lie within the range of experimental sensitivities. Some of these models use specific CP violating mechanisms together with other features to explain the observed baryon asymmetry of the universe (BAU)~\cite{Morrissey_2012}, which is inexplicable by known sources of CP violation in the SM.
\end{sloppypar}
The scale of CP violation in the QCD sector of the SM is experimentally constrained to be vanishingly small, through the non-observation to date of any non-zero hadronic EDM. 
This lack of EDM signals in searches with the neutron ~\cite{nEDM-PhysRevLett} and $^{199}$Hg~atom ~\cite{Graner2016} in particular results in what is known as the "strong CP problem''~\cite{SIKIVIE2012}. 
Theory offers possible explanations for the suppression of the observable CP violation in the strong sector, most elegantly by introducing axions~\cite{Peccei1977,Marsh2016h}. 
Axions are also viable Dark Matter candidates~\cite{Graham2015}, but
aside from the unexpectedly small EDMs there has so far been
no other observations made in support of their existence.

Obviously, nobody today can safely predict where BSM CP violation will first manifest itself in any experiment. 
If it were to show up in an EDM measurement, it is not clear in which system this would be; thus there is a broad search strategy presently being pursued in many laboratories around the world~\cite{Jungmann2013,Chupp2015PR,Chupp2019,pignol2019global}. 
In particular, ongoing efforts target intrinsic particle EDMs, e.g.\ of leptons and quarks, as well as those occurring due to or being enhanced by interactions in nuclear, atomic and molecular systems.
In the current situation any sign of an EDM would be a major scientific discovery. 
In case of a discovery in any one system, however, corresponding EDM measurements in other systems will be needed to clarify the underlying mechanism of CP violation.
The neutron is experimentally the simplest of the accessible strongly interacting systems, and as such remains a prime search candidate. Searches for EDMs of the proton and light nuclei will also become increasingly important.

 The most sensitive neutron EDM search delivered a result of  $d_n = (0 \pm 1.1_{stat} \pm 0.2_{sys}) \times 10^{-26}~e~\mathrm{cm}$, which sets an upper limit of $\lvert d_n \lvert <1.8 \times 10^{-26}~e~\mathrm{cm}$~(90\%~C.L.)~\cite{nEDM-PhysRevLett}. This measurement was performed with the apparatus originally built by the RAL/Sussex/ILL collaboration  ~\cite{Baker2014}, which went through continuous upgrades of almost all subsystems and was also moved to the source of ultracold neutrons (UCNs) ~\cite{Lauss2012,Lauss2014,Bison2020} at the Paul Scherrer Institute (PSI). Arguably the most important of the  upgrades were the addition of an array of atomic cesium magnetometers~\cite{CsM_PRA2020} and of a dual spin detection system~\cite{Afach2015EPJA}.

 With this last measurement, the RAL/Sussex/ILL nEDM apparatus at the PSI UCN source reached its limits with respect both to systematic effects and to statistical sensitivity. Any further increase in sensitivity requires a new apparatus optimally adapted to the UCN source as well as the replacement of numerous subsystems with more modern and higher specification equipment.

There are a number of collaborations around the world \cite{SNS,Ito2018,Ruediger2017,Piegsa2019,Wurm_2019,Serebrov2017} attempting to improve the current neutron EDM limit by at least one order of magnitude. The most ambitious competing project is based on a totally new concept of a cryogenic experiment in superfluid helium, where both the UCN statistics and the electric-field strength could be enhanced~\cite{SNS}. More traditionally, we propose to push and extend the powerful and proven concept of a room-temperature UCN experiment with two separate and complementary magnetometry systems. The n2EDM spectrometer, the subject of this paper, is a next-generation UCN apparatus based on the unification of two concepts: the double-chamber setup pioneered by the Gatchina nEDM spectrometer~\cite{Altarev1980NuPhA}, and the use of Hg co-magnetometry \cite{Green1998}. 

\begin{sloppypar}
The n2EDM apparatus is designed to measure the neutron EDM with sensitivity of $1\times 10^{-27}~e~\mathrm{cm}$, with further possibility to go well into the $10^{-28}~e~\mathrm{cm}$ range. The improvement of statistical sensitivity will arise from the large double-chamber volume as well as an optimized UCN transport arrangement between the source and the spectrometer. The control of systematic effects 
need to shadow these improvements, implying better stability, uniformity and measurement of the main magnetic field. These will be achieved by a dedicated coil system and better magnetic shielding, as well as by substantially improved magnetometry. 
\end{sloppypar}

%% file: Measurement-intro.tex
In this section we present the overall concept of the n2EDM apparatus. The heart of the experiment is a large-volume double storage chamber placed in a new, large, magnetically shielded room. Stable and uniform magnetic-field conditions are of paramount importance for a successful measurement. The magnetic field will be generated by  a main magnetic-field coil in  conjunction with about 70 trim coils, each powered by highly stable power supplies. Monitoring of the magnetic field is provided by atomic mercury co-magnetometry as well as by a large array of optically-pumped Cs~magnetometers.\\

%% file: Concept.tex
The measurement relies on a precise estimation of $f_n$ - the precession frequency of  polarized ultracold neutrons  stored in a weak magnetic field $B$ and a strong electric field $E$. 
The neutron EDM is obtained by comparing the precession frequencies in the anti-parallel ($\uparrow \downarrow$) and parallel ($\uparrow \uparrow$)
configurations of the magnetic and electric fields: 
\begin{equation}
d_n = \frac{\pi \hbar}{2 |E|} (f_{n, \uparrow \downarrow} -f_{n, \uparrow \uparrow}). 
\label{Eq:dn_basic_formula}
\end{equation}

The statistical sensitivity in the former nEDM experiment ~\cite{nEDM-PhysRevLett} was limited by ultracold neutron counting statistics, which depend on the intensity of the UCN source, the efficiency of the UCN transport system, and the size and quality of the storage chambers. 
Independently of possible improvements of the yield of the PSI UCN source, the guideline for the conceptual design of the new apparatus was to maximize the neutron counting
statistics while keeping the systematic effects under control. This will be achieved with a large UCN storage volume and an optimized UCN transport system, placed in a well-controlled magnetic-field environment.

\begin{figure*}
  \centering
    \includegraphics[width=0.8\textwidth]{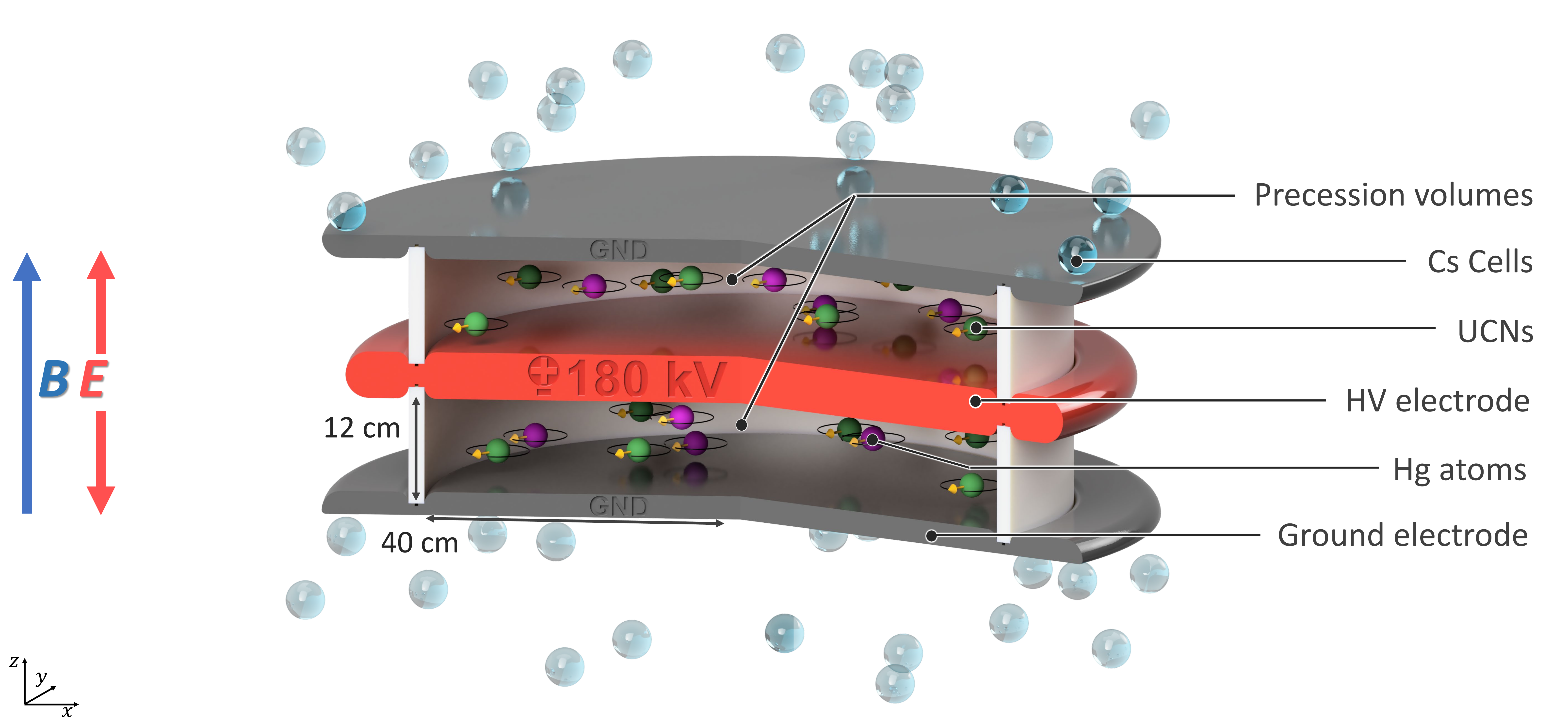}
  \caption[Cut through the central part of the apparatus]
{Cut through the central part of the n2EDM apparatus.
Two vertically stacked storage (Ramsey spin-precession) chambers, filled with polarized UCNs and Hg atoms, are
embedded in the same vertical magnetic field $\vec{B}$, but with opposite electric-field directions~$\vec{E}$. 
}
  \label{fig:Experiment-Scheme}
\end{figure*}

Figure~\ref{fig:Experiment-Scheme} shows the basic concept of the n2EDM experiment. The design of the apparatus is based on two key features: (i) two cylindrical storage (Ramsey spin-precession) chambers, one stacked above the other; (ii) a combination of mercury and cesium magnetometry.~The storage volumes are separated by the shared high-voltage electrode, and are each confined at the opposite end by a ground electrode and radially by an insulating ring.  In addition to doubling the storage volume,  the twin-chamber arrangement also permits the simultaneous measurement of the neutron precession frequencies for both electric-field directions. Below we give short overviews of the core systems of the n2EDM apparatus, which will be presented in more detail in Sec. \ref{sec:5}.

\bigskip

\paragraph{Precession chambers}

\begin{itemize}
\item 
Each of the two precession chambers has internal diameter $D=80$\,cm and height $H=12$\,cm. The choice of the dimensions is explained in Secs. \ref{sec:3.1} and \ref{sec:5.1.1}. 

\item As noted above the upper and lower chambers are separated by the common high-voltage (HV) electrode, which has a thickness of 6\,cm.
As in the previous experiment~\cite{nEDM-PhysRevLett}, the electrodes will be made of aluminum coated with diamond-like carbon and the insulator rings will be of polystyrene coated with deuterated polystyrene.

\item The precession-chamber stack will be installed inside an aluminum vacuum vessel with an internal volume of approximately \SI{1.6 x 1.6 x 1.2}{\meter}. This design allows for the optional installation of a double chamber with inner diameter of up to 100\,cm, for a possible future upgrade of the experiment.  
\end{itemize}

\bigskip

\paragraph{UCN polarization, transport and detection} 

\begin{itemize}
\item Neutrons arriving from the PSI UCN source are fully polarized using a 5\,T superconducting magnet.
\item Neutron guides made of coated glass tubes with ultralow surface roughness connect the precession chambers first to the UCN source and then to the detectors. This is achieved by different operational modes of the so-called UCN switch (see Sec. \ref{sec:5.1.3}).

\item  Neutrons are counted by a spin-sensitive detection system based on fast gaseous detectors (see Sec. \ref{sec:5.1.4})
\end{itemize}

\paragraph{Magnetic shielding} \mbox{} 
\medskip

 To shield the experiment from external variations in the magnetic field, the sensitive part of the apparatus is installed inside a magnetic shield that has both  passive and active components (see Sec. \ref{sec:5.2}):
\begin{itemize}

\item The passive magnetic shield is provided by a large multilayer cubic magnetically shielded room (MSR) with inner dimensions of $2.93\,{\rm m} \times 2.93\,{\rm m} \times 2.93\,{\rm m}$. 

\item The active magnetic shield consists of eight actively-controlled coils placed around the MSR on  a dedicated grid spanning a volume of about \SI{1000}{\cubic\metre}.
\end{itemize}

\paragraph{Magnetic-field generation}

\begin{itemize}

\item A large coil will be installed inside the MSR (but outside the vacuum vessel) in order to produce a highly uniform vertical magnetic field $B_0$ throughout a large volume. The coil is designed to operate in the range $\SI{1}{\muT} < B_0 < \SI{15}{\muT}$. In the short to medium term it is intended to work with $B_0 = \SI{1}{\muT}$, as was the case in the previous single-chamber experiment, but other options are being considered for the future.

\item  In addition to the main coil, an  array  of  56  independent trim coils is used to achieve the required level of magnetic-field uniformity. 

\item A further seven ``gradient coils'' will produce specific field gradients that play an important role in the measurement procedure.

\item RF coils will be installed inside the vacuum tank to generate the oscillating-field pulses applied in the Ramsey measurement cycles.

\end{itemize}
\paragraph{Magnetometry}

\begin{itemize}
\begin{sloppypar}
\item Within each of the storage chambers the volume-averaged magnetic field will be measured using polarized $^{199}$Hg atoms injected into the volume at the beginning of the cycle. The free precession signal is observed using a UV light beam that traverses the chamber. Throughout the period prior to each measurement cycle the mercury gas is continuously polarized by optical pumping within a smaller adjacent volume separated from the main chamber by a shutter. 
\end{sloppypar}

\item An array of 114 Cs magnetometers will measure the field at a 
number of positions surrounding the chambers. 
This will provide instantaneous measurements of the magnetic-field uniformity.

\item An automated magnetic-field mapper will be used offline for B-field cartography of all of the coils as well as for the correction and control of high-order gradients.

\end{itemize}

%% file: Measurement.tex
\begin{figure}[h]
   \subfloat[Polarized UCNs fill the precession chambers]{\label{fig:Filling}\includegraphics[width=1\columnwidth]{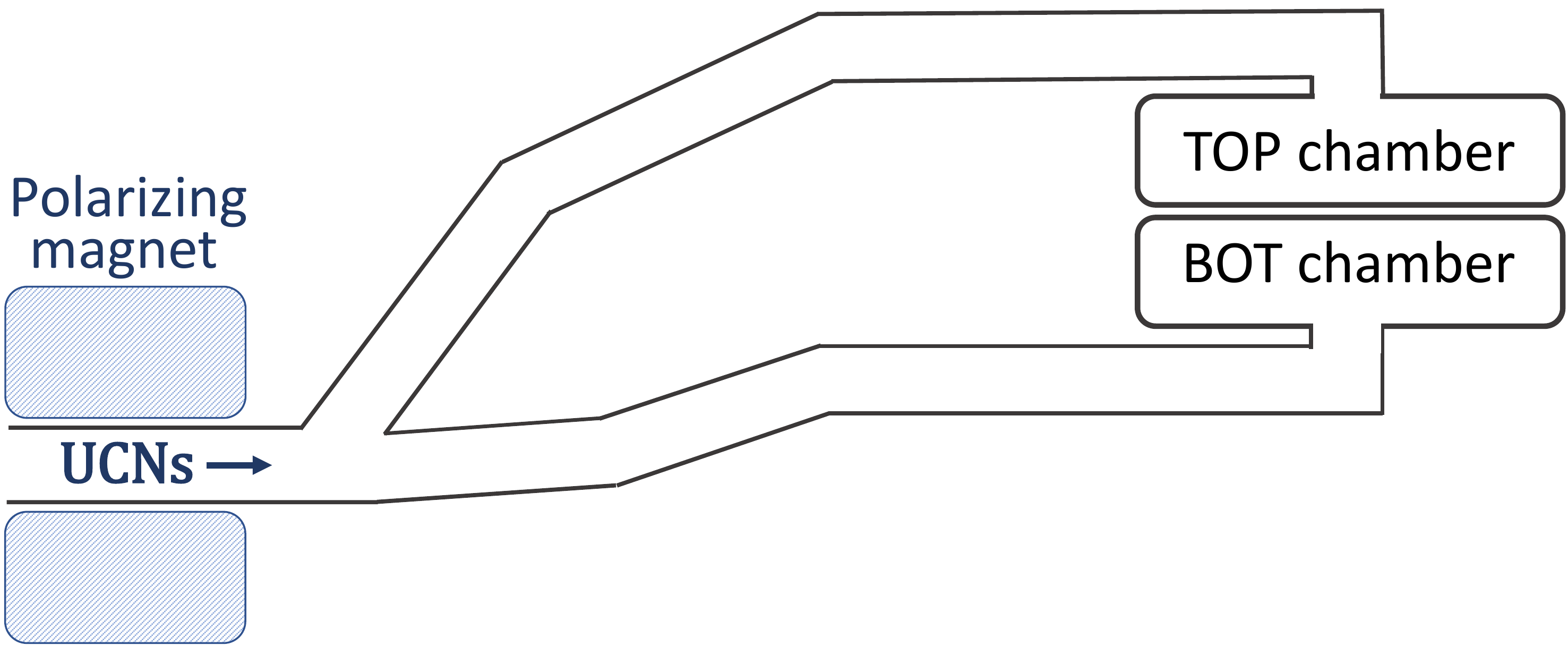}} \\
   \subfloat[UCNs are guided towards the spin-sensitive detectors.]{\label{fig:Detection}\includegraphics[width=1\columnwidth]{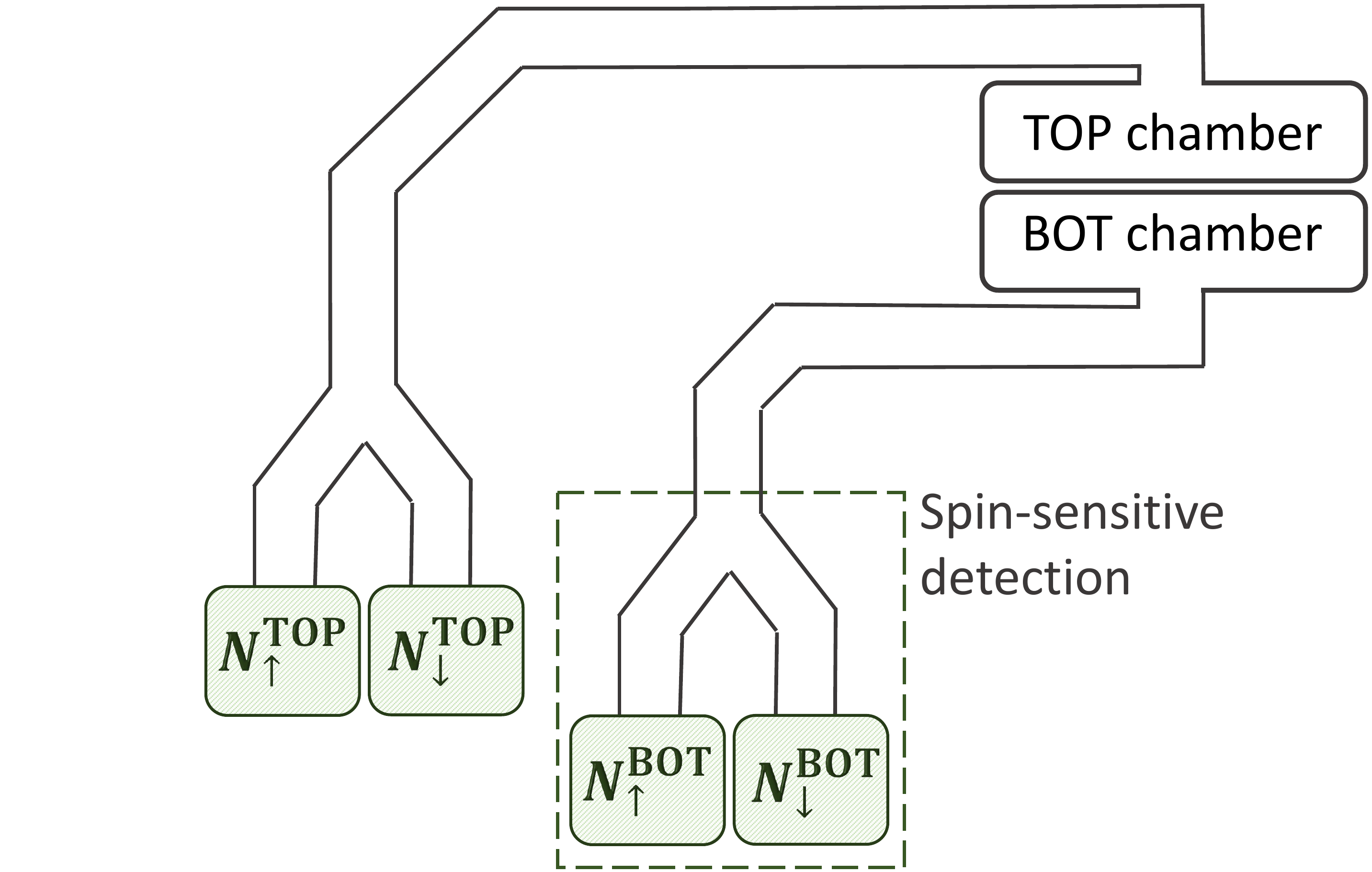}}
   \caption{Schematic view (not to scale) of the UCNs' path (a) while filling the precession chambers, and (b) following the Ramsey cycle, when transporting UCNs towards the spin-sensitive detectors for counting.}
   \label{fig:Filling-detection}
\end{figure}

In the data taking mode, the full PSI proton beam will be kicked to the UCN source for 8\,s  every five minutes, producing a burst of ultracold neutrons.
These UCNs are guided to the apparatus through the UCN transport system (see Fig~\ref{fig:Filling}). 
Along the way they are polarized (to almost 100\% polarization level) by  passage through the 5\,T superconducting magnet. 

Once the precession chambers have been filled with polarized UCNs, the UCN shutters close the chambers and the neutrons are thereby stored. 
Ramsey's method of separated rotating fields is then performed:

\begin{enumerate}

\item A first horizontal rotating field is applied for $t_{\rm RF}=2$\,s 
at a frequency $f_{\rm RF} \approx |\gamma_n|/2\pi B_0 \approx 30$\,Hz 
(for $B_0\approx \SI{1}{\micro T}$). 
The amplitude of the field is chosen such that the neutron spins are tipped by $\pi/2$ into the (horizontal) plane perpendicular to the main magnetic field $B_0$.  

\item  The neutron spins precess freely in the horizontal plane for a duration of $T=180$\,s (referred to as the ``precession time''; see Sec. \ref{sec:3.3})  at a frequency $f_n$ which in principle will be slightly different in the two chambers. 

\item  A second rotating field, in phase with the first, is then applied for another 2\,s. 
The vertical projection of the neutron spins (in units of $\hbar/2$) after the 
whole procedure is
\begin{equation}
A(f_{\rm RF}) = - \alpha \cos \left[ \pi \frac{f_{\rm RF} - f_n}{\Delta \nu} \right], 
\label{Eq:RamseyFringe}
\end{equation}
where $\alpha$ (also referred to as 
the \emph{visibility} of the resonance) is the final polarization of the ultracold neutrons, 
$f_n$ is the neutron Larmor precession frequency to be measured, 
and 
\begin{equation}
\Delta \nu = \frac{1}{2T + 8 t_{\rm RF}/\pi}
\end{equation}
is the half-width of the resonance. 
The quantity $A$ is called the \emph{asymmetry}. 
Since $f_n$ is likely to have a different value in each of the two chambers, 
the asymmetry in the top chamber $A^\TOP$ will not be identical to the asymmetry 
in the bottom chamber $A^\BOT$.
Notice however that the applied frequency $f_{\rm RF}$ is common to the two chambers. 

\item The ultracold neutrons are released from the precession chambers by opening the UCN shutters, and are then guided to the spin analyzers (see Fig.~\ref{fig:Detection}). 
Each chamber is connected to a dedicated spin-sensitive detector. 
These devices simultaneously and independently count
the number of neutrons in each of the two spin states. 
These spin analyzers therefore provide, for each cycle, a measurement of the asymmetries for the top and bottom chambers: 
\begin{equation}
A^\TOP = \frac{N_\uparrow^\TOP - N_\downarrow^\TOP}{N_\uparrow^\TOP + N_\downarrow^\TOP} \ \; {\rm and } \; A^\BOT = \frac{N_\uparrow^\BOT - N_\downarrow^\BOT}{N_\uparrow^\BOT + N_\downarrow^\BOT}, 
\label{Eq:asymmetry2}
\end{equation}

where $N_{\uparrow / \downarrow}^{\rm TOP/BOT}$ are the numbers of neutrons from the top or bottom  chamber, with spin parallel ($\uparrow$) or 
antiparallel ($\downarrow$) to the magnetic field. 
\end{enumerate}

\begin{figure}[ht]
  \centering
  \includegraphics[width=1.\columnwidth]{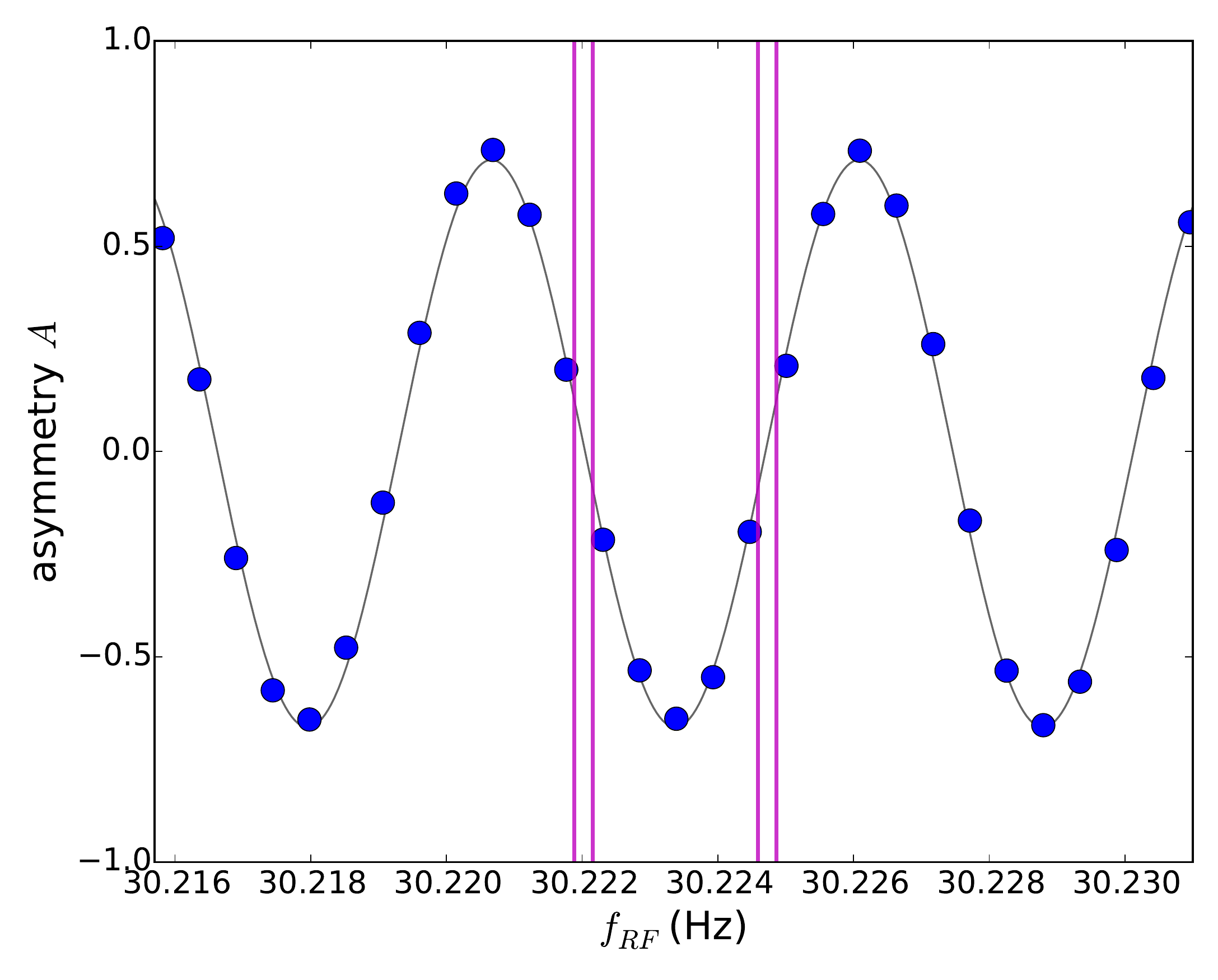}
  \caption[Ramsey fit]
  {
	The asymmetry 
	$A = (N_\uparrow - N_\downarrow)/(N_\uparrow + N_\downarrow)$ as a function of 
	the applied frequency $f_{\rm RF}$. 
	The points represent experimental data where each point is a measurement cycle with a precession time of $T = 180\,\s $ 
	performed with the nEDM apparatus in September 2017 with the standard value 
	of the magnetic field $B_0 = 1036.3$\,nT corresponding to a 	Larmor precession frequency of $30.2235$\,Hz. The error bars are smaller than the size of the points. The line is the fit to the data using the function from Eq.~\ref{Eq:RamseyFringe}. The vertical bars show the position of the four ``working points'' used in the 
	nEDM data-taking in order to maximize the sensitivity. 
}
  \label{fig:RamseyFit}
\end{figure}

\begin{sloppypar}
Figure~\ref{fig:RamseyFit} shows an example of a measurement performed 
with the (single-chamber) nEDM apparatus scanning the Ramsey resonance. 
If the parameter $\alpha$ is known, each cycle provides a measurement 
of $f_n$ by inverting Eq. \eqref{Eq:RamseyFringe}. 
\end{sloppypar}

The statistical error arising from Poisson counting statistics
for one measurement cycle is
\begin{equation}
\sigma(f_n) = \frac{\Delta \nu}{\pi \alpha  \sqrt{N_\uparrow + N_\downarrow}} \left(1- \frac{A^2}{\alpha^2}\right)^{-\frac{1}{2}}. 
\label{Eq:stat_error_fn}
\end{equation}
The maximal sensitivity is obtained for cycles measured 
at $A=0$ where the slope of the resonance curve is highest. 
In fact, in nEDM data production mode the applied frequency $f_{\rm RF}$ 
is set sequentially to four ``working points'' indicated by the 
vertical bars in Fig.~\ref{fig:RamseyFit}. 
Specifically, we set $f_{\rm RF} = f_{n,0} \pm(1\pm0.1) \times \Delta \nu/2$, 
where we calculate $f_{n,0}$ from a measurement of the magnetic field performed in the previous cycle. Hence the four working points follow any possible magnetic-field drifts. 

Magnetic-field drifts induce shifts of the resonance curve that are in 
practice much smaller than the width of the resonance $\Delta \nu$, 
but which nonetheless might be larger than the precision $\sigma(f_n)$ of the measurement; this will be discussed below. 
The slight departure of the working points from the two optimal points $f_{n,0} \pm \Delta \nu/2$ enable the extraction of the visibility $\alpha$ 
(as well as a small vertical offset of the resonance due to the imperfections of 
the spin analyzer system) by combining the data of many cycles of a run. 
This comes at the price of a sensitivity reduction of 2\% in comparison to the optimal points. 

With n2EDM, since the applied frequency is common to the two chambers 
it is important to ensure that the value of $f_{\rm RF}$ is set 
sufficiently close to the optimal points for the two chambers simultaneously. 
This is referred to as the \emph{top-bottom resonance matching condition}.  
It imposes a requirement on the maximum permitted vertical gradient of the magnetic field. 
For n2EDM we require the shift between the resonance curves of the top and bottom chambers to
be at most $0.2 \times \Delta \nu/2$  
in order to limit the resulting sensitivity loss to values lower than 2\%. 

For a precession time of $T = 180\,\s$ this corresponds to a maximum difference 
of $10\,\pT$ between the average field in the top and bottom chambers. 
With a separation between the centers of the two chambers of $H' = 18\,\cm$, the requirement on the vertical magnetic-field gradient is

\begin{equation}
\left| \frac{\partial B_z}{\partial z} \right| < 0.6 \, \pT/\cm. 
\label{Eq:top_bottom_resonance_matching}
\end{equation}

The measurement procedure described for one cycle is repeated continuously to form a sequence of data with many cycles.

\begin{figure}[ht]
  \centering
  \includegraphics[width=1\columnwidth]{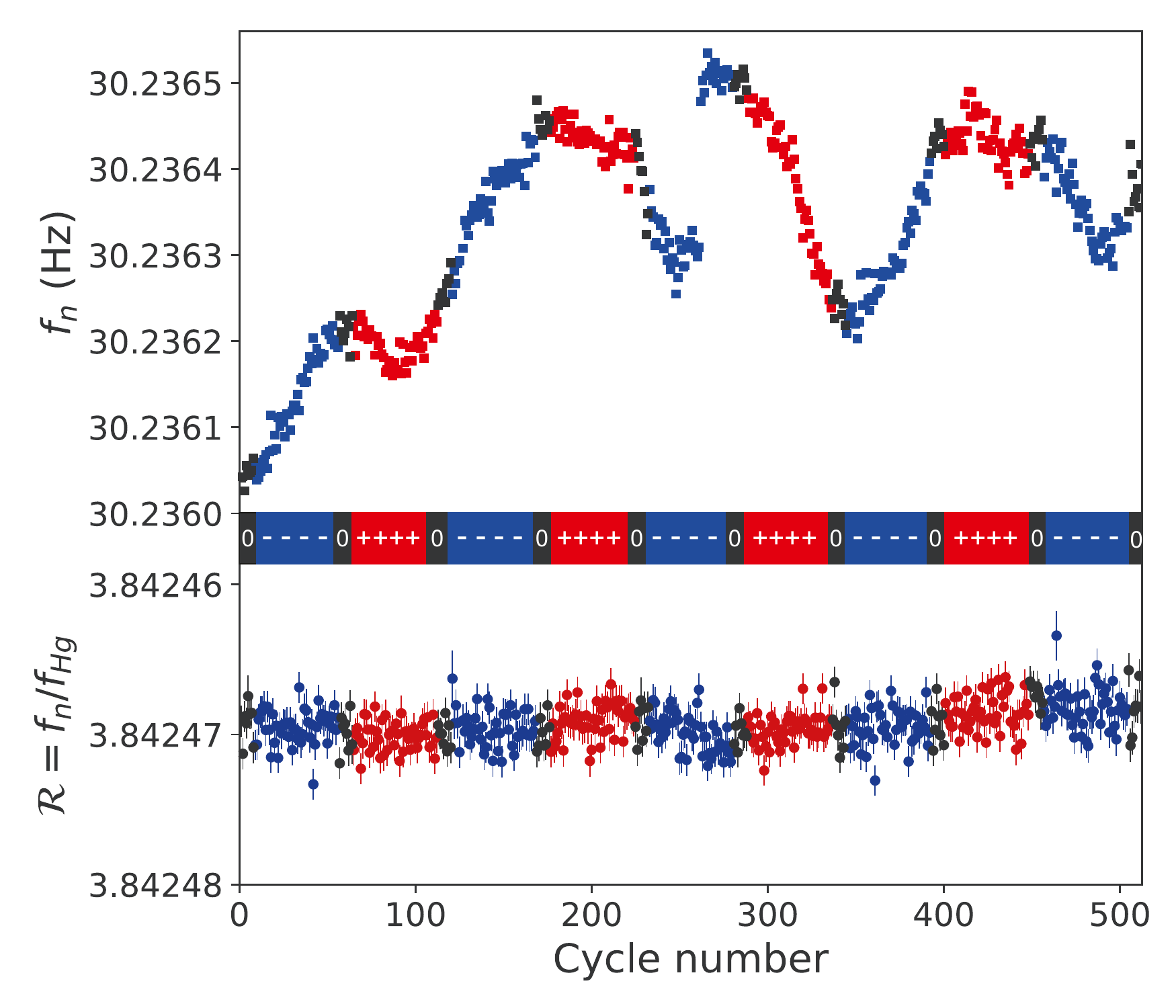}
  \caption[nEDM-sequence-2016]
  {A sequence of nEDM data produced in 2016; each point corresponds to a single measurement cycle. 
  The upper plot shows the neutron frequency as a function of cycle number, whereas the lower plot shows the frequency ratio $\R = f_n/f_{\rm Hg}$. The colors correspond to the high-voltage polarity (blue, negative; red, positive; black, zero). }

  \label{fig:nEDM-sequence-2016}
\end{figure}

Figure~\ref{fig:nEDM-sequence-2016} shows an example of a sequence produced with the nEDM apparatus in 2016. 
For each cycle the neutron frequency was extracted $f_n$ as explained above. 
The electric field polarity was alternated with a period of 112 cycles. 
As is evident from the figure, the neutron frequency is affected by the inevitable small 
drifts of the magnetic field. 
These drifts were compensated by the mercury \comagnetometer. 
At the beginning of a cycle, some mercury from the polarization cell is admitted to the 
precession chamber just after the UCN shutter is closed.  
A $\pi/2$ flip is then applied to the mercury spins, and they start to precess at a 
frequency of $f_{\rm Hg} = \gamma_{\rm Hg}/2\pi B_0 \approx 8$\,Hz  
(for $B_0\approx \SI{1}{\micro T}$). 
The precession is recorded optically, by measuring the modulated transmission of a 
polarized horizontal UV beam. 
From the data one may then extract the mean precession frequency $f_{\rm Hg}$ of the mercury atoms sampling, to a very good approximation, \emph{the same period of time} and \emph{the same volume} as the precessing ultracold neutrons. 

For each cycle we thus obtain simultaneous measurements both of the neutron frequency $f_n$ 
and of the mercury frequency $f_{\rm Hg}$. 
The neutron frequency includes contributions from both the magnetic and the electric dipole moments: 
\begin{equation}
f_n = \left|\frac{\gamma_n}{2\pi} B_0 \right| \mp \frac{\dn}{\pi \hbar} |E|,  
\label{Eq:fn_ideal}
\end{equation}
where the ``-'' sign refers to the parallel $\uparrow \uparrow$ configurations of the 
magnetic and electric fields and the ``+'' sign refers to the anti-parallel 
configuration $\uparrow \downarrow$. 
The electric contribution, which is ultimately the goal of the search, is tiny: 
for $\dn = 10^{-26}~e~\mathrm{cm}$, $E = 15$\,kV/cm and $B_0=\SI{1}{\micro T}$, 
the ratio between the electric and magnetic term is as small as $2\times 10^{-9}$. 
The mercury precession frequency effectively has only a magnetic contribution: since the mercury EDM $d_{Hg} < 10^{-29}~e~\mathrm{cm}$~\cite{Graner2016}, the electrical term can be neglected.  The frequency is then
\begin{equation}
f_{\rm Hg} = \left|\frac{\gamma_{\rm Hg}}{2\pi} B_0 \right|. 
\label{Eq:fHg_ideal}
\end{equation}
The ratio $\R$ of neutron to mercury frequencies
\begin{equation}
\R \equiv \frac{f_n}{f_{\rm Hg}} = \left|\frac{\gamma_n}{\gamma_{\rm Hg}} \right| \mp \frac{|E|}{\pi \hbar f_{\rm Hg}} \dn
\label{Eq:R_ratio}
\end{equation}
is then free from the fluctuations of the magnetic field $B_0$. 
This is illustrated in Fig.~\ref{fig:nEDM-sequence-2016} where the ratio $\R$ is plotted in the lower part. 
Notice that the co-magnetometer 
corrects for random drifts of the magnetic field 
that would spoil the statistical sensitivity \emph{and also} for the B-field variations correlated 
with the electric field (due to leakage currents along the insulator for example) 
that would produce otherwise a direct systematic effect. 
In fact, Eq.~\eqref{Eq:R_ratio} is an idealization.
It is modified by several effects affecting either the neutron or mercury frequencies. 
These will  be described in section \ref{sec:4}, where the associated systematic effects will be discussed. 

Finally, with the single-chamber apparatus, the neutron EDM is calculated as follows: 
\begin{equation}
\dn = \frac{\pi \hbar f_{\rm Hg}}{2|E|} ( \R_{\uparrow \downarrow} - \R_{\uparrow \uparrow}). 
\end{equation}
In n2EDM, both chambers host a mercury \comagnetometer, 
and each cycle will therefore provide the neutron and mercury Larmor precession frequencies 
in the two chambers $f_n^\TOP$, $f_n^\BOT$, $f_{\rm Hg}^\TOP$ and $f_{\rm Hg}^\BOT$. 
Therefore we can form the two ratios
\begin{equation}
\R^\TOP = f_n^\TOP/f_{\rm Hg}^\TOP, \quad \R^\BOT = f_n^\BOT/f_{\rm Hg}^\BOT. 
\label{ratios}
\end{equation}
In principle, since the parallel and anti-parallel configurations of the fields 
are measured simultaneously in n2EDM, a measurement of $d_n$ could be obtained without 
reversing the polarity of the electric field 
by calculating 
\begin{equation}
\dn = \frac{\pi \hbar f_{\rm Hg}}{2|E|} \left( \R_{\uparrow \downarrow}^\TOP - \R_{\uparrow \uparrow}^\BOT \right).
\label{dn_ratios}
\end{equation}
However, in order to compensate for shifts in  $\R$ arising from  various effects described later in section \ref{sec:4}, the electric polarity of the central electrode will be 
reversed periodically as was done in the previous single-chamber nEDM experiment. 
The neutron EDM can then be calculated as follows: 

\begin{equation}
\dn = \frac{\pi \hbar f_{\rm Hg}}{4|E|} \left( \R_{\uparrow \downarrow}^\TOP - \R_{\uparrow \uparrow}^\TOP +  \R_{\uparrow \downarrow}^\BOT - \R_{\uparrow \uparrow}^\BOT \right). 
\label{Eq:dn_n2EDM}
\end{equation}

In the following section we will address the statistical 
and systematic errors of this measurement. 

%% file: Statistics.tex
The statistical sensitivity of the measurement will be limited by the UCN counting statistics. 
By combining the expression for the statistical sensitivity $\sigma(f_n)$ of the neutron 
frequency at the optimal point $A=0$ as given by 
Eq.~\eqref{Eq:stat_error_fn} with the expression Eq.~\eqref{Eq:dn_basic_formula}, 
the following statistical sensitivity on the neutron EDM per cycle may be derived:
\begin{equation}
\sigma (d_n) = \frac{\hbar}{2 \alpha E T \sqrt{N}} ,
\label{Eq:Sensitivity}
\end{equation}
where $\alpha$ is the measured neutron polarization at the end of the Ramsey cycle, 
$T$ is the neutron precession time, $E$ is the magnitude of the electric field and 
$N$ is the total number of neutrons counted from the two chambers.

\begin{table}[h]
\center
\begin{tabular}{c|c|c}
& nEDM 2016 & n2EDM \\
\hline
\hline 
chamber      & DLC \& dPS & DLC \& dPS \\
diameter $D$ & 47 \cm & 80 \cm\\
\hline
$N$ (per cycle) & 15'000 & 121'000\\
$T$   & 180 s & 180 s\\
$E$ & 11 kV/cm & 15 kV/cm\\
$\alpha$ & 0.75 & 0.8\\
\hline
\\[-1em]
$\sigma(f_n)$ per cycle & \SI{9.6}{\micro Hz} & \SI{3.2}{\micro Hz}\\
\hline
\\[-1em]
$\sigma(d_n)$ per day & 11 $\times$ 10$^{-26}~e~\mathrm{cm}$ & 2.6 $\times$ 10$^{-26}~e~\mathrm{cm}$
\\ \hline
\\[-1em]
$\sigma(d_n)$ (final) & $9.5 \times 10^{-27}~e~\mathrm{cm}$ & $1.1 \times 10^{-27}~e~\mathrm{cm}$\\
\hline
\hline
\end{tabular}
\caption[Experimental sensitivity]{
Comparison between (i) the achieved performance of the nEDM apparatus during the datataking at PSI in 2016, (ii) the nominal parameters for the of n2EDM design (see Eq.~\eqref{Eq:Sensitivity} and text). In both cases coatings are made of diamond-like carbon (DLC) for the electrodes and deuterated polystyrene (dPS) for the insulator ring. The number of neutrons $N$ is the total number of UCN counted (in the two chambers in case of n2EDM) after a storage time of  $T=180\,\s$. 
The error on the neutron frequency $\sigma(f_n)$ is given for one cycle and one chamber. 
Also shown are the $d_n$ sensitivities in one day and the final accumulated sensitivities. The final sensitivity listed in the first column is that actually achieved in 2016/2017; that of the second column represents the achievable sensitivity in n2EDM after an assumed 500 days of data taking, which could be achieved in four calendar years.
}
\label{TableStatSens}
\end{table}

Table~\ref{TableStatSens} summarizes the expected values for each of those contributions. 
It is based on the demonstrated sensitivity of the nEDM apparatus, the average UCN source performance in 2016 and on our Monte-Carlo simulation of the n2EDM UCN system. We will next discuss each of the parameters in Eq.~\eqref{Eq:Sensitivity}.

%% file: UCNs.tex
The prediction of the number of detected neutrons in the n2EDM apparatus is based on comprehensive Monte Carlo simulations of the PSI UCN source, guides, and the experiment, treated as one system, performed with the MCUCN code~\cite{Bison2020,Zsigmond2018}. 

As far as the UCN source and guides  leading to the beamports were concerned, the relevant surface parameters of the neutron optics and the UCN flux were calibrated using dedicated test measurements of the achievable  density at the West-1 beamport in 2016~\cite{Ries2016,Bison2017}.  
The simulation parameters and values are listed in Table 4 of~\cite{Bison2020}. These are: optical (Fermi) potential, loss per bounce parameter, fraction of diffuse (Lambertian) reflections, and the attenuation constant of the windows. Separate values were considered for the following  parts: the aluminum lid above the sD2 converter, the vertical NiMo coated guide above the sD2 vessel, the DLC coated storage vessel of the source, the NiMo coated neutron guides to the beamports, and the aluminum vacuum separation windows in the SC magnet and detectors.

For the n2EDM apparatus, the following parameters were used: For the NiMo coated guides, an optical potential 220 neV as measured with cold-neutron reflectometry; a loss per bounce parameter as measured in~\cite{Bondar2017}, with a 1$\sigma$ error of $3\times 10^{-4}$; and an upper limit of 2\% for fraction of Lambertian reflections (as benchmarked for NiMo on glass). The NiMo-coated aluminum guide inserts have a small surface fraction, and were assumed to be highly polished and thus not to increase the overall fraction of diffuse reflections above 2\%. For the loss-per-bounce parameter of the precession chambers we use a value extracted from storage measurements with the single chamber~\cite{Mohanmurthy2020}, adding a 1$\sigma$ error of $2.8\times 10^{-4}$.  This was very close to values reported in~\cite{Atchison2005c} for DLC (on aluminum foil at 300 K),  and in ~\cite{Bodek2008} for dPS. We used optical potentials of 230 neV for DLC~\cite{Bison2020} and 165 neV for dPS, the latter being the average of measured and theoretical values~\cite{Bodek2008} (because of a large measurement error). The diffuse reflection fraction for the electrodes was 2\%, and a maximal roughness was assumed for the insulator ring (Lambertian reflections, corresponding to a diffuse reflection fraction of 100\%).

\begin{figure}
\centering
\includegraphics[width=1.\columnwidth]{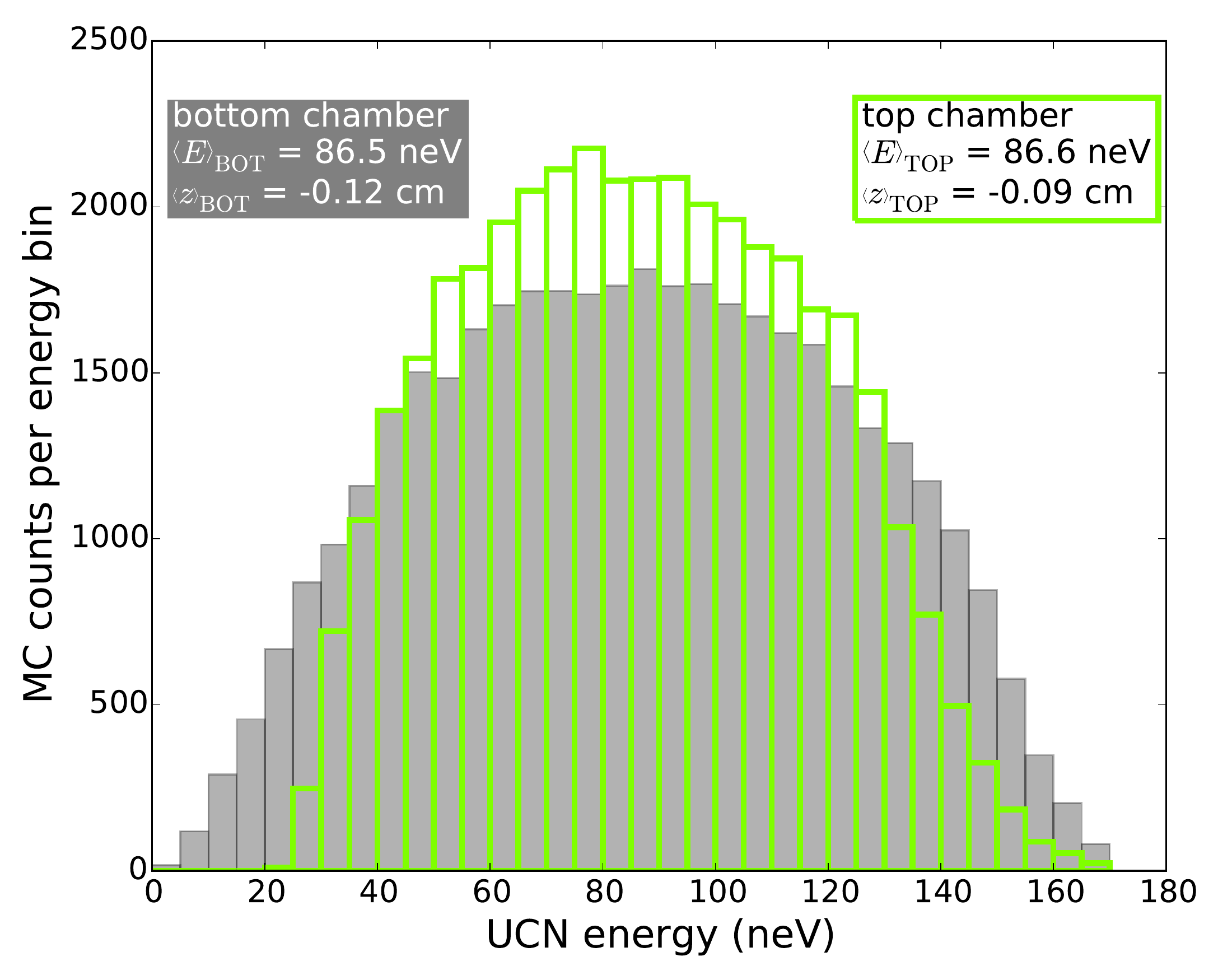}
\caption[Simulated energy spectra]{Simulated energy spectra of the  detected UCNs. The energy is the sum of the kinetic and the gravitational potential energy calculated at the floor levels of the respective chambers.}
\label{fig:MC_spectrum}
\end{figure}

The geometry of the parts of the n2EDM experiment dependent upon UCN optics, and in particular the height of the chambers above the beamline, was optimized in terms of UCN statistics. 
The optimal height is significantly lower than the height of the previous nEDM experiment. 
The simulated energy spectra of detected UCNs calculated at the bottom level of the chambers are shown in Fig.~\ref{fig:MC_spectrum}. 

Due to the lower elevation of the chambers with respect to the beamline, the spectra of the stored UCNs are expected to be harder in comparison with the single chamber nEDM experiment. 
The absence of UCN at lower energies in case of the upper chamber is caused by filling from the top. The maximum attainable energy for the two spectra is determined mainly by the 165 neV optical potential of the insulator ring, and to a lesser extent by the difference in elevation.

The chosen design, with an 80 cm diameter double chamber of 12 cm individual heights, permits an increase of the total number of detected neutrons after 188 s storage time (i.e.\ 180 s precession time) from 15000 in nEDM to 121000 in n2EDM. The uncertainty of the calibration from MC counts to real UCN counts is $\pm 15\%$~\cite{Bison2021}. This considerable gain in UCNs is the result of
(i) a double chamber as compared to a single chamber, 
(ii) an increase of the volume of each individual chamber by a factor of  three,
(iii) an increase in the inner diameter of the UCN guides (6.6~cm to 13~cm), 
(iv) optimization of the vertical position of the precession chambers; the optimum was found to be 80 cm above the beamport. 
None of these estimates include any of the improvements in the performance of the PSI UCN source that have taken place since 2017.

%% file: Efield.tex
\begin{sloppypar}
In the single-chamber nEDM apparatus the electric field was generated by charging the top electrode  using a bipolar high-voltage supply of $\pm 200$~kV.
The top electrode was ramped regularly to $\pm 132$~kV, while the bottom electrode was kept at ground potential.
The maximum voltage was limited by the presence of many optical fibers in contact with both the charged electrode and the grounded vacuum tank. These fibers were used to operate six Cs magnetometers situated on the top electrode. 
\end{sloppypar}
The same system, without the Cs magnetometers and the fibers, could sustain higher 
electric fields; tests were carried out up to 16.6 kV/cm.

In the n2EDM apparatus all Cs magnetometers will be mounted at ground potential, above and below the electrode stack.
The electric field will not be limited by the presence of optical fibers close to the charged central electrode, and we expect to be able to operate the system  at voltages of 200\,kV or higher. 
However, a safe standard operation is anticipated at voltages of $\pm 180$\,kV, 
corresponding to an electric field of $\pm 15$\,kV/cm.  

%% file: Precession.tex
\begin{figure}[H]
  \centering
  \includegraphics[width=0.9\columnwidth]{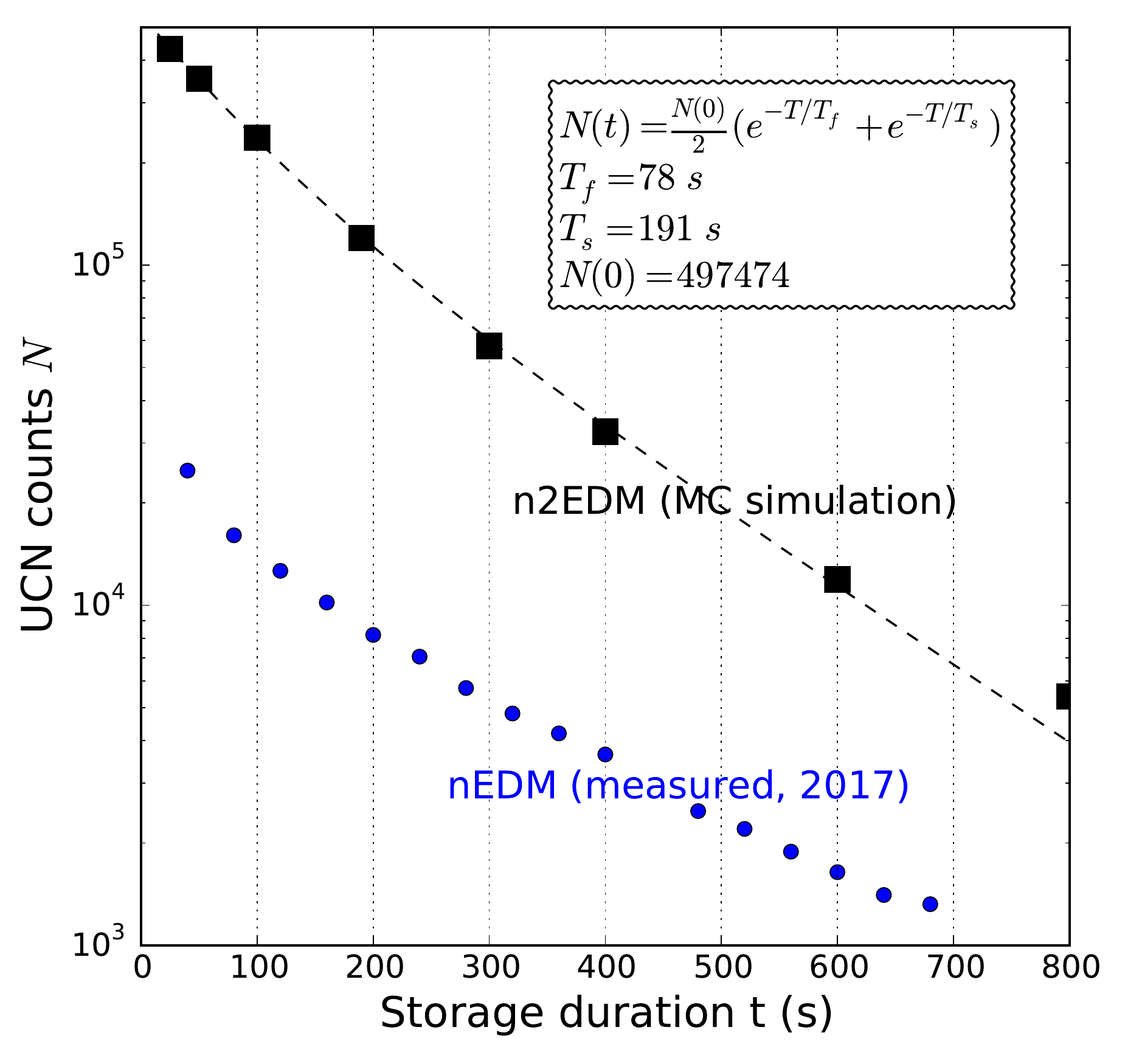}
    \caption[Storage curve]
      {\color{black}Storage curve measured with  nEDM in 2017 (blue points) and simulated storage curve in n2EDM (black squares). The n2EDM storage curve as a function of the precession time $T = t - 8 \, {\rm s}$  is fitted with a double-exponential model; see text for details. 
      }
        \label{fig:StorageCurve}
\end{figure}

The choice of the precession time $T$ results from balancing two dominant effects:  
increasing $T$ reduces the width of the Ramsey resonance and tends to improve the sensitivity, 
but at the same time the number of surviving neutrons $N(T)$ decreases, and this 
decreases the sensitivity. 
Additionally, one has to take into account the fact that increasing $T$ results in fewer measurement cycles per day. 
In detail, the daily sensitivity $\sigma_{\rm day}$ follows from the cycle sensitivity given 
by Eq.~\eqref{Eq:Sensitivity} and has the form 
\begin{equation}
\sigma_{\rm day} = \frac{\sigma(\dn)}{\sqrt{n_{\rm cyc}}} = 
\frac{\hbar}{2 \alpha E} \frac{1}{T \sqrt{N(T)}} \sqrt{\frac{T + T_{\rm dead}}{24 \, {\rm h}}}, 
\label{Eq:DailySensitivity}
\end{equation}
where $n_{\rm cyc}$ is the number of cycles per day, the total length of a cycle being the sum of 
the precession time $T$ and a dead time $T_{\rm dead}$ accounting for filling and emptying 
the chambers as well as ramping the electric field. 
In Eq.~\eqref{Eq:DailySensitivity} we assume that the visibility $\alpha$ is not 
decreasing with time (i.e. we neglect UCN depolarization). 
This important point will be discussed later. 
The loss of neutrons in the chambers is encoded in the function $N(T)$; this is the 
main effect driving the choice of $T$. 
In Fig.~\ref{fig:StorageCurve} we show a simulated storage curve, i.e.\ the number of neutrons counted after a storage duration $t$ as a function of $t$. 

The storage duration $t=T + 8 \, {\rm s}$ within the EDM cycles is a little longer than the precession time $T$ in order to account for the additional time required to fill the mercury atoms and apply the mercury pulse (4s) and to apply the two neutron pulses (4s).

As usual for UCN storage chambers at room temperature, the storage curve departs significantly from a pure exponential decay because the dominating losses originate from wall collisions rather than from beta decay ($\tau_n \approx 880\,\s$). 
Wall collision rates and loss probability per collision are a function of neutron kinetic energy. This results in energy-dependent UCN loss rates and a clear departure from a simple exponential decay.
We fit the storage curve with a double-exponential model assuming only two groups of neutrons 
equi-populated at $T=0$.

\begin{figure}
  \centering
  \includegraphics[width=0.9\columnwidth]{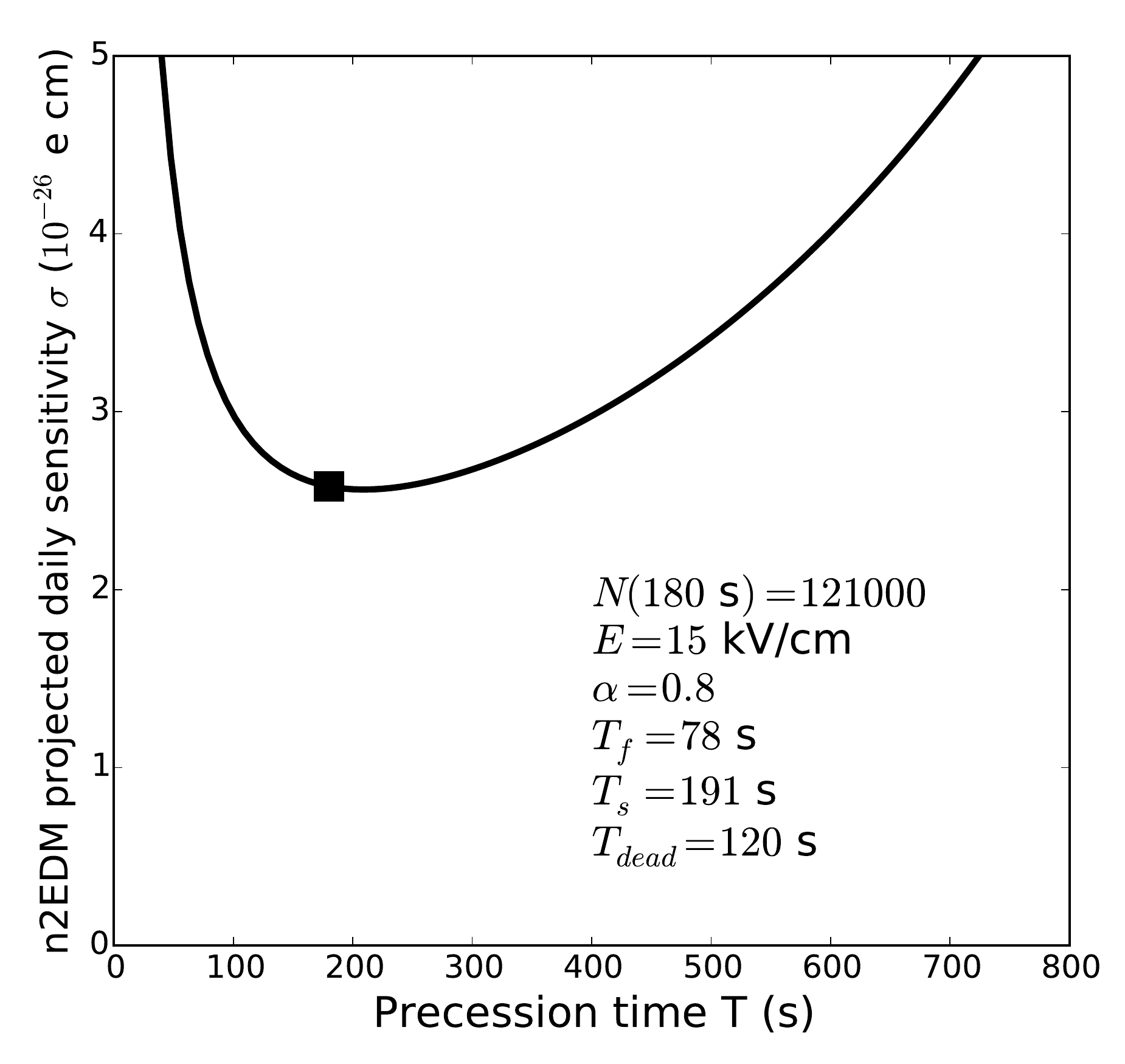}
  \caption[Storage-sensitivity]
  {Projected daily sensitivity with n2EDM as a function of the precession time $T$. The baseline parameter $T=180\,\s$ is indicated by the black square. }
  \label{fig:Storage-sensitivity}
\end{figure}

In Fig.~\ref{fig:Storage-sensitivity} we plot the projection of the daily 
sensitivity, Eq.~\eqref{Eq:DailySensitivity}, for the baseline design of n2EDM. 
For the sensitivity estimation we set $T = 180 \,\s$ (as in ~\cite{nEDM-PhysRevLett}).

%% file: Polarisation.tex
In the perfect case of no depolarization during the precession period, the visibility of the Ramsey resonance would be as high as the initial polarization\footnote {The term ``initial polarization'' is in fact the product of the polarization with the analyzing power of the detection system, and is limited by depolarization in the detection process.}, which was
measured to be $\alpha_0 = 0.86$ in the single-chamber nEDM spectrometer. 
In fact, the final polarization under measurement conditions (T=180\,s) 
was $0.75$ on average.

The decay of polarization during storage $d\alpha/dt$ arises from three different contributions: 
\begin{equation}
\frac{d \alpha}{d t} = - \frac{\alpha}{T_{\rm wall}} + \dot{\alpha}_{\rm grav} - \frac{\alpha}{T_{\rm 2, mag}}. 
\end{equation}
We briefly discuss these effects, and we refer to Ref.~\cite{Uniformity2019} 
for a more complete treatment of this subject. 

\begin{itemize}
\begin{sloppypar}
\item The first contribution $- \alpha/T_{\rm wall}$ is due to depolarization during wall collisions. 
The depolarization rate $1/T_{\rm wall} = \nu \beta $ is given by the product of
 the wall collision rate $\nu$, which is determined by the UCN spectrum, and the depolarization probability at each wall collision $\beta$, which is set by the surface properties of the chamber. 
This depolarization mechanism affects equally the transverse and the 
longitudinal polarization of the neutrons. 
Dedicated measurements performed with the nEDM apparatus in 2016 resulted in a determination of $T_{\rm wall}\approx 4000\,\s$. 
In n2EDM we expect about the same values for $\beta$ and $\nu$. We anticipate that this process will reduce $\alpha$ from 0.86 to 0.83 after 180\,s. 

\item A second contribution $\dot{\alpha}_{\rm grav}$, called \emph{gravitationally enhanced 
depolarization} \cite{Afach2015Grav,SpinEcho2015}, was identified in the nEDM single-chamber experiment. It is due to the vertical striation of UCN under gravity in combination with a 
vertical magnetic-field gradient. 
Neutrons with different kinetic energies have different mean heights $\bar{z}$ due to gravity.
Therefore, in the presence of a vertical field gradient, neutrons with different kinetic energies have different spin precession frequencies. 
This results in a relative dephasing, which in turn is visible as an effective depolarization quantified by 
the following expression: 
\begin{equation}
\dot{\alpha}_{\rm grav} = - \gamma_n^2 \left( \frac{\partial B_z}{\partial z} \right)^2 {\rm Var} [\bar{z}] \, t. 
\end{equation}
\end{sloppypar}
The variance of the distribution of $\bar{z}$ is a quantity that depends on the 
total height of the chamber $H$ and on the energy spectrum of the stored UCNs. It was measured to be ${\rm Var} [\bar{z}] \approx 0.2\,\cm^2$ in the previous nEDM experiment and 
it is expected to be smaller in n2EDM: $0.07\,\cm^2$ in the bottom chamber and $0.002\,\cm^2$ in the top chamber, according to the simulated energy spectra.  
For the data-taking with the nEDM experiment, the vertical gradient was scanned in the 
range $-30\,\pT/\cm < {\partial B_z}/{\partial z} < 30\,\pT/\cm$ as part of the strategy to correct 
for the systematic effects. 
It resulted in a decrease of the $\alpha$ parameter due to the 
gravitationally enhanced depolarization of about 0.05 on average. 
n2EDM will be operated in a much smaller range of vertical gradients, $-0.6\,\pT/\cm < {\partial B_z}/{\partial z} < 0.6\,\pT/\cm$, in 
order to meet the top-bottom resonance matching condition discussed earlier. 
In this case the decrease in $\alpha$ will be negligible. 

\begin{figure}
  \centering
  \includegraphics[width=1.\columnwidth]{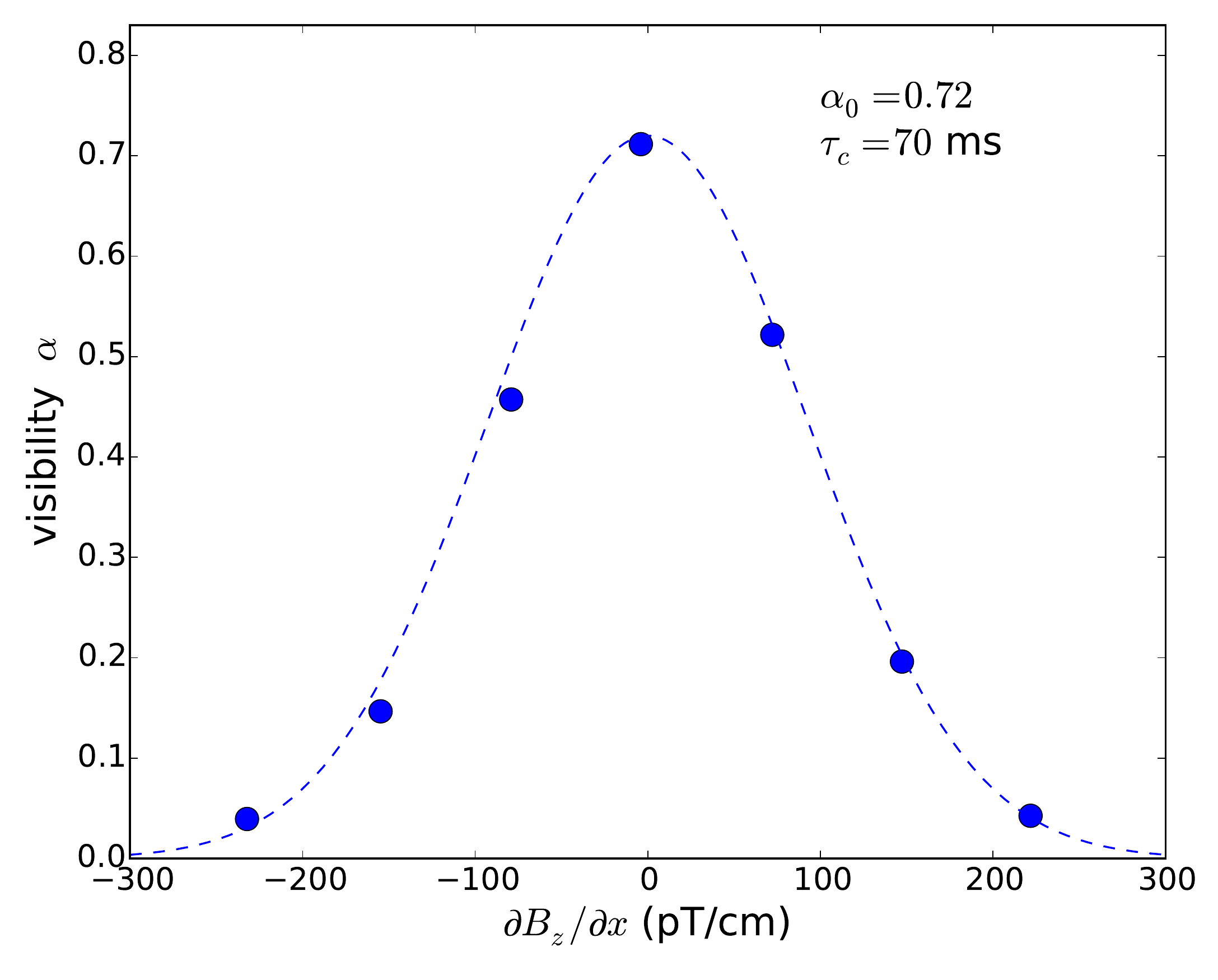}
  \caption[G11 scan]
  {Measurement of the visibility $\alpha$ after a precession time 	of $T=180\,\s$ as a function of an applied horizontal 	gradient $G_{1,1} = {\partial B_z}/{\partial x}$ performed with the nEDM apparatus in 2017. 
  The dashed line corresponds to the exponential decay 
	model $\alpha = \alpha_0 \exp \left(-T/T_{\rm 2, mag}(G_{1,1}) \right)$, 
	where $1/T_{\rm 2, mag}(G_{1,1})$ is given by Eq. \eqref{Eq:IntrinsicDepolarization} 
	with $\tau_c = 70\,\ms$. 
  }
  \label{fig:G11scan}
\end{figure}

\item The last contribution $- \alpha /T_{\rm 2, mag}$ corresponds to the \emph{intrinsic depolarization}, i.e.\ the irreversible process of polarization decay within energy groups due to the random motion in a static but non-uniform field. 
Indeed a neutron sees a longitudinal magnetic disturbance  $b_z(t) = B_z(\vec{r}(t)) - \langle B_z \rangle$ as it moves randomly within the chamber with a trajectory $\vec{r}(t)$. 
Spin-relaxation theory \cite{Redfield1957} allows calculation of the decay rate of the transverse polarization due to this disturbance, to second order in the perturbation, as:

\begin{equation}
\frac{1}{T_{\rm 2, mag}} = \gamma_n^2 \int_0^\infty 
\langle b_z(t) b_z(t+\tau)\rangle d\tau = \gamma_n^2 \, \langle b_z^2 \rangle \, \tau_c, 
\label{Eq:IntrinsicDepolarization}
\end{equation}

where $\tau_c$ is the correlation time, $\langle b_z(t) b_z(t+\tau)\rangle$ is the autocorrelation of the longitudinal disturbance and $\langle b_z^2 \rangle$ is the average of the quantity $b_z^2$ over all particles in the chamber, which in this case is the same as the volume average of $(B_z(r) - \langle B_z \rangle)^2$. 
In fact, Eq.~\eqref{Eq:IntrinsicDepolarization} serves as a definition of the 
correlation time. 
It is important to notice that horizontal gradients $G_{1,1} = {\partial B_z}/{\partial x}$ (fields of the type $B_z = B_0 + G_{1,1} \, x$) 
are much more effective in this depolarization channel compared with vertical gradients $G_{1,0} = {\partial B_z}/{\partial z}$
(fields of the type $B_z = B_0 + G_{1,0} \, z$), due to the aspect ratio of the chambers 
(the height is significantly shorter than the diameter). 
Figure \ref{fig:G11scan} shows a measurement of the  visibility $\alpha$ as a function of an applied (artificially large) horizontal field gradient $G_{1,1} = {\partial B_z}/{\partial x}$ performed with the nEDM apparatus in 2017. 
In this case the mean squared inhomogeneity can be calculated to be 

\begin{equation}
\langle b_z^2 \rangle = G_{1,1}^2 \langle x^2 \rangle = G_{1,1}^2 \frac{R^2}{4}, 
\end{equation}

where $R$ is the radius of the chamber; $R = 23.5\, \cm$ in the case of the previous single-chamber experiment. 
The measurement resulted in a determination of $\tau_c({\rm nEDM}) = 70\,\ms$. 
The correlation time scales as $\tau_c \propto R/v_h$, where $v_h = \langle \dot{x}^2+\dot{y}^2 \rangle \approx 3$~m/s is the horizontal velocity of UCNs. 
However the precise value is complicated to predict; it depends on the velocity spectrum of the stored neutrons, the degree of specularity of the collisions, and also on the shape of the non-uniform field. 
For an estimate of the UCN correlation time in the n2EDM we will simply extrapolate the value measured in the nEDM experiment by taking into account the increase in diameter of the chambers: 

\begin{equation}
\label{Eq:tau_c_n2EDM}
\tau_c({\rm n2EDM}) = \tau_c({\rm nEDM}) \times \frac{80\,\cm}{47\,\cm} = 120\, \ms. 
\end{equation}

For a given field gradient, the depolarization decay rate \eqref{Eq:IntrinsicDepolarization} scales as the third power of the radius of the chamber, because $\tau_c$ is linear in $R$ and $\langle b_z^2 \rangle$ is quadratic in $R$. 
This is a major challenge for the design of n2EDM because of the increased chamber radius; in fact the intrinsic depolarization sets an important requirement for the generation of the magnetic field. 

\end{itemize}

In order to reach a final visibility of $\alpha(180\,\s)>0.80$ in n2EDM, 
we require that the decrease of $\alpha$ due to the intrinsic 
depolarization to be not more than 2\%. 
This corresponds to $T_{\rm 2, mag} > 9000\,\s$. 
Using Eq.~\eqref{Eq:IntrinsicDepolarization} and \eqref{Eq:tau_c_n2EDM} we 
derive the corresponding requirement on the root mean square of the spatial variations 
of the field  in the chamber: 
\begin{equation}
\sigma(B_z) = \sqrt{\langle b_z^2 \rangle} < 170 \, \pT. 
\label{Eq:FieldUniformityReq}
\end{equation}
Notice that this requirement concerns the absolute value of the field, 
and not the relative value $\sigma(B_z)/B_0$.
It applies to the baseline choice $B_0 = \SI{1}{\micro T}$ as well as for
any other $B_0$ field.

%% file: AddFluctuations-remarks.tex
The expression Eq.~\eqref{Eq:Sensitivity} only takes into account the 
statistical error on the neutron frequency. 
In fact, when propagating the error in the $\R$ ratio, the errors on both the 
neutron frequency and the mercury frequency contribute to the neutron EDM 
given by Eq.~\eqref{Eq:dn_n2EDM}. 
Taking into account the mercury error, the total statistical error is increased 
by a factor
\begin{equation}
\sqrt{ 1+ \left( \frac{\sigma(f_{\rm Hg})}{\sigma(f_{\rm n})} \R \right)^2}. 
\end{equation}
In addition there are further sources of statistical fluctuations of the $\R$ ratio, in particular the fluctuations of the magnetic-field gradient (due to the gravitational shift, see Sec.~\ref{sec:4}). 

The goal for the mercury-magnetometer design is to reduce the contribution 
from $\sigma(f_{\rm Hg})$ to less than 2\% of the total statistical error, 
corresponding to 
$\sigma(f_{\rm Hg}) < 0.05 \times \sigma(f_n) = \SI{0.2}{\micro Hz}$ for one cycle of 
measurement in one chamber. 
In terms of magnetic-field sensitivity this corresponds to \SI{25}{fT}. 
In turn, the sensitivity goal of the co-magnetometer sets a goal for the temporal 
stability of the magnetic field during the expected 180~s spin precession time. 
Indeed, the drift of the magnetic field during the precession time has an impact upon the mercury frequency extraction. 
In order to ensure that the accuracy of the co-magnetometer is not reduced by the magnetic-field drifts, it should be of the same order as the magnetometer precision, i.e. $\sigma(B) \sim 25 \,\fT$ over 180~s. 

Assuming the same UCN source performance that was provided in 2016 (See Tab.\,\ref{TableStatSens}) we plan about 500 days of data taking, which can be accomplished within four years of operation. Therefore, after completion of the data taking, the total accumulated statistical sensitivity is expected to be at the level of $\sigma (\dn) = 1 \times 10^{-27}\,\ecm $. Further upgrades and UCN source improvements could allow the measurement to reach sensitivities well into the $10^{-28} \, \ecm$ range.

%% file: Systematics.tex
There are a number of known effects that shift the neutron and mercury frequencies from the ideal case given by Eq. \eqref{Eq:fn_ideal} and \eqref{Eq:fHg_ideal}. 
These are all encapsulated in the following formula, which is valid for individual chambers: 
\begin{equation}
\R = \frac{f_n}{f_{\rm Hg}} = \left| \frac{\gamma_n}{\gamma_{\rm Hg}} \right| \left(1 + \delta_{\rm elec} + \delta_{\rm mag} + \delta_{\rm other} \right),  
\end{equation}
where the three terms $\delta_{\rm elec}$, $\delta_{\rm mag}$ and $\delta_{\rm other}$ are much smaller than one. 
The first contribution 
\begin{equation}
\delta_{\rm elec} = \delta_{\rm EDM}^{\rm true} + \delta_{\rm EDM}^{\rm false} + \delta_{\rm quad} 
\end{equation}
corresponds to the electrical terms (i.e. they are absent in zero electric field). 
The second contribution 
\begin{equation}
\delta_{\rm mag} = \delta_{\rm grav} + \delta_{\rm T}
\end{equation}
corresponds to the nonuniform magnetic terms (i.e. they are absent in a purely uniform magnetic field).  The last contribution $\delta_\mathrm{other}$ corresponds to all other effects. 

The true EDM term $\delta_{\rm EDM}^{\rm true}$ is induced by the linear-in-electric field frequency shifts from the true neutron and mercury EDM: 
\begin{equation}
\delta_{\rm EDM}^{\rm true} = \pm \frac{2}{\hbar |\gamma_n B_0|} |E| \left( d_n + d_{n \leftarrow {\rm Hg}} \right),
\end{equation}
where the $+$ sign corresponds to the anti-parallel ($\uparrow \downarrow$ or $\downarrow \uparrow$) configurations, whereas the $-$ sign corresponds to the parallel ($\uparrow \uparrow$ or $\downarrow \downarrow$) configurations. 
In addition there is the contribution of the mercury EDM:
\begin{equation}
d_{n \leftarrow {\rm Hg}} = \left| \frac{\gamma_n}{\gamma_{\rm Hg}}\right| d_{\rm Hg} = (0.8 \pm 1.2) \times 10^{-29}~e~\mathrm{cm}, 
\end{equation}
where we have used the most recent measurement by the Seattle group $d_{\rm Hg} = (2.2 \pm 3.1) \times 10^{-30}~e~\mathrm{cm}$ ~\cite{Graner2016}.  This term is negligible.

All of the other $\delta$ shifts could generate two types of undesirable consequences. 
First, they could induce random fluctuations of the ratio $\R$ which would increase the statistical error. 
Second, any correlation between the electric-field  polarity and one of these terms will induce a direct systematic effect. 
Although the first type imposes requirements on the stability of environmental variables, in particular the magnetic-field gradients, it is possible to make these additional fluctuations negligible and we will not address them. Here we will concentrate on the latter effects which, following Eq. \ref{Eq:dn_n2EDM}, correspond to a systematic effect of 
\begin{equation}
\delta d_n = \frac{\pi \hbar f_{\rm Hg}}{4 |E|} \left( \delta_{\uparrow \downarrow}^\TOP - \delta_{\uparrow \uparrow}^\TOP +  \delta_{\uparrow \downarrow}^\BOT - \delta_{\uparrow \uparrow}^\BOT \right), 
\label{Eq:ddn_n2EDM}
\end{equation}
when considering the two-chamber extraction of the neutron EDM. 

Before we describe all of the $\delta$ terms in detail, we pause to explain the different conventions used here. 

For the sign conventions, we define an angular frequency $\omega$ as 
an algebraic quantity the sign of which is determined with respect to
the $z$ axis pointing upwards, 
i.e. $\omega > 0$ corresponds to a rotation in the horizontal plane following the right hand rule. 
Note that, since $\gamma_n<0$, the neutron angular frequency $\omega_n = \gamma_n B_0$ is negative when the magnetic field is pointing up. 
It is opposite for the mercury atoms because $\gamma_{\rm Hg} > 0$. 
The quantity $B_0$ is likewise algebraic. 
It is positive when the field is pointing up and negative when the field is pointing down. 
The frequencies are defined as positive quantities, i.e. $f = |\omega|/2\pi$.

\begin{table*}
\centering
\renewcommand{\arraystretch}{1.4}
\begin{tabular}{lll}
\hline \hline
$l$ & $m$ & ${\bm \Pi}_{l,m}$ \\
\hline
0 & -1 & $\vec{e}_y = \sin \phi \ \vec{e}_\rho + \cos \phi \ \vec{e}_\phi$ \\
0 & 0 & $\vec{e}_z$ \\
0 & 1 & $\vec{e}_x = \cos \phi \ \vec{e}_\rho - \sin \phi \ \vec{e}_\phi$ \\
1 & -2 & $\rho(\sin 2 \phi \ \vec{e}_\rho + \cos 2 \phi \ \vec{e}_\phi)$ \\
1 & -1 & $z(\sin \phi \ \vec{e}_\rho + \cos \phi \ \vec{e}_\phi) + \rho \sin \phi\  \vec{e}_z$ \\
1 & 0 & $- \frac{1}{2} \rho \ \vec{e}_\rho+ z \ \vec{e}_z$\\
1 & 1 & $z(\cos \phi \ \vec{e}_\rho - \sin \phi \ \vec{e}_\phi)+ \rho \cos \phi \ \vec{e}_z $ \\
1 & 2 & $\rho(\cos 2 \phi \ \vec{e}_\rho - \sin 2 \phi \ \vec{e}_\phi)$ \\
2 & 0 & $-\rho z \ \vec{e}_\rho+ (z^2 - \frac{1}{2}\rho^2) \ \vec{e}_z$ \\
3 & 0 & $\frac{3}{8}\rho (-4z^2+\rho^2) \ \vec{e}_\rho+ (z^3 - \frac{3}{2}z \rho^2) \ \vec{e}_z$ \\
4 & 0 & $ \frac{1}{2} \rho (- 4 z^3 + 3\rho^2 z) \ \vec{e}_\rho+ (z^4 - 3 z^2 \rho^2 + \frac{3}{8} \rho^4) \ \vec{e}_z$ \\
5 & 0 & $ \frac{5}{16}\rho(-8 z^4 + 12 \rho^2 z^2 - \rho^4) \ \vec{e}_\rho+ (z^5 - 5 z^3 \rho^2 + \frac{15}{8} z \rho^4) \ \vec{e}_z$  \\
\hline \hline
\end{tabular}
\caption{
Expressions for the relevant harmonic modes ${\bm \Pi}_{l,m}$ in cylindrical coordinates. 
}
\label{tab:HarmonicPolynomials}
\end{table*}

To describe the magnetic-field non-uniformities, we use the framework developed in \cite{Uniformity2019} which defines a parametrization of a general field in the form
\begin{equation}
\vec{B}(\vec{r}) = \sum_{l \geq0} \ \sum_{m=-l}^l G_{l, m} {\bm \Pi}_{l,m}(\vec{r}), 
\label{Eq:harmon}
\end{equation}
where $G_{l,m}$ are the generalized gradients and the functions ${\bm \Pi}_{l,m}$, or \emph{modes},  form a basis of harmonic functions constructed from the solid harmonics. 
The modes expressed in Cartesian coordinates are polynomials in $x,y,z$ of degree $l$. 
In cylindrical coordinates $\rho, \phi, z$ the modes take the form
\begin{equation}
{\bm  \Pi}_{l,m}(\vec{r}) = \tilde{ {\bm  \Pi}}_{l,m}(\rho,z) \cdot \vec{y}_m(\phi)
\end{equation}
with $\tilde{ {\bm  \Pi}}_{l,m}(\rho,z)$ a polynomial function of $\rho$ and $z$ of degree $l$, and the azimuthal part is
\begin{equation}
\vec{y}_m(\phi) = 
\begin{cases}
\cos (m \phi) \vec{e}_\rho + \sin (m \phi) \vec{e}_\phi + \cos (m \phi) \vec{e}_z &\text{if} \ m \geq 0, \\ 
 \sin (m \phi) \vec{e}_\rho + \cos (m \phi) \vec{e}_\phi + \sin (m \phi) \vec{e}_z &  \text{if} \ m < 0. \notag
\end{cases}
\end{equation}
Explicit expressions for the relevant modes are specified in Table \ref{tab:HarmonicPolynomials}. 

\subsection{Gravitational shift, uncompensated gradient drift}
The kinetic energy of ultracold neutrons is so low that their spatial distribution is significantly affected by gravity, and their center of mass lies a fraction of a centimeter below the geometric center of the chamber. 
In contrast, the mercury atoms form a gas at room temperature that fills the precession chamber nearly uniformly. 
This results in slightly different average magnetic fields being sampled by the neutrons and the atoms in the presence of a vertical magnetic-field gradient. This effect is called the gravitational shift $\delta_{\rm grav}$. 
In the framework of the harmonic decomposition of the field up to the second order, the volume average of the vertical component is 
\begin{eqnarray}
\delta_{\rm grav}^\TOP & = & (G_{1,0}+H' G_{2,0}) \frac{\z_\TOP}{B_0} ,\\
\delta_{\rm grav}^\BOT & = & (G_{1,0}-H' G_{2,0}) \frac{\z_\BOT}{B_0}, 
\end{eqnarray}
where $\z_\TOP$ and $\z_\BOT$ are the center of mass offset between the neutron and mercury in the top and bottom chamber, and $H' = 18\,\cm$ is the height difference between the centers of the top and bottom chambers.  

The gravitational shift could induce an additional statistical error (due to an instability of the gradients $G_{1,0}$ or $G_{2,0}$) and a systematic effect (due to a direct correlation of the gradients with the electric-field polarity). 
For simplicity we will only discuss the effect of the linear gradient $G_{1,0}$, and will neglect the second order term $G_{2,0}$. 
In the nEDM single-chamber apparatus we measured a value of $\z  = - 0.39\,\cm$. 
For the n2EDM estimates we use the values calculated from the simulated energy spectra in Fig. \ref{fig:MC_spectrum}, $\z_\TOP = -0.09\,\cm$ and $\z_\BOT = -0.12\,\cm$.

A fluctuation of the gradient $G_{1,0}$ with RMS value $\sigma(G)$ induces a contribution to the fluctuation of $\R_\TOP-\R_\BOT$ of 
\begin{equation}
\sigma(\R_\TOP-\R_\BOT) = \left| \frac{\gamma_n}{\gamma_{\rm Hg}}  \frac{\z_\TOP - \z_\BOT}{B_0} \right| \sigma(G). 
\end{equation}
Notice that the effect of the linear gradient drift is reduced when using the double-chamber concept, as compared to the single chamber, because $\z_\TOP \approx \z_\BOT$. 
Still, the residual imperfect compensation of the gradient drifts could generate a direct systematic effect which is called the \emph{uncompensated gradient drift}. 
The application of the electric field might itself generate a magnetic-field change which is correlated to the voltage of the central electrode $V$. 
Such an effect might be due to the leakage current from the high-voltage electrode to the ground electrodes. 
It could also be due to a magnetization of the shield by the charging currents during voltage ramps. 
In principle the mercury co-magnetometer cancels any field fluctuations, including those correlated with the electric field. 
However, the cancellation is not perfect due to the gravitational shift. 
The false EDM due to the correlated part of the gradient $\delta G(V)$ is 
\begin{equation}
\delta \dn = \frac{\hbar \gamma_{\rm n}}{4E} (\z_\BOT-\z_\TOP) \delta G(V). 
\end{equation}

The goal for n2EDM is to have this systematic effect under control at the level of $1\times10^{-28}~e~\mathrm{cm}$, corresponding to a control over the correlated part of the gradient at the level of $\delta G(V)\leq 1.5 $\,fT/cm. 

One possible strategy would be to perform dedicated tests to check for a possible G/V correlation, with frequent reversals of the electric polarity while measuring the magnetic-field gradient with the mercury co-magnetometers and the array of atomic cesium magnetometers. 
For definiteness we consider a series of 1000 polarity reversals, each lasting 5 minutes.
The stability of the field gradient $\sigma (G) [{\rm 5 min}]$ will limit the resolution on the sought effect: $\delta G = \sigma (G) [{\rm 5\,min}] / \sqrt{1000}$. This sets a requirement on short time variations of the gradient to: 
\begin{equation}
\sigma (G) [{\rm 5\,min}] < 50 \, \fT/\cm. 
\label{Eq:req_gradient_variation}
\end{equation}

\subsection{Shift due to transverse fields}

Residual transverse field components $B_x$ and $B_y$ are averaged differently by the neutrons and the mercury atoms.  This produces a shift in $\R$ denominated the transverse shift $\delta_{\rm T}$.  
When a particle (a neutron or a mercury atom) moves in a static but non-uniform field it effectively sees a fluctuating magnetic field $\vec{B}(\vec{r}(t))$, where $\vec{r}(t)$ is the random trajectory. 
In addition to the intrinsic depolarization process already discussed in the previous section, the fluctuation induces a shift of the Larmor precession frequency. 
In fact the shift is induced by the transverse component of the field, which can be described by the complex perturbation
\begin{equation}
b(t) := \vec{B}(\vec{r}(t)) \cdot ( \vec{e}_x + i \vec{e}_y). 
\label{Eq:complex_perturbation}
\end{equation}
We will again make use of the autocorrelation function of the perturbation $\langle b^*(t) b(t+\tau) \rangle$, where the brackets $\langle \cdot \rangle$ denote the ensemble average over all of the particles in the chamber. 
Note that since the motion of the particles is  stationary in the statistical sense, $\langle b^*(t) b(t + \tau) \rangle = \langle b^*(0) b(\tau) \rangle$ is independent of $t$. 
Spin-relaxation theory allows calculation of the angular frequency shift at second order in the perturbation $b$: 
\begin{equation}
\delta \omega = \frac{\gamma^2}{2} \int_0^\infty d\tau \, {\rm Im} \left[ e^{i\omega \tau} \langle b^*(0) b(\tau) \rangle \right]. 
\label{Eq:delta_omega}
\end{equation}
The timescale of the correlation function is set by the correlation time $\tau_c$ that we introduced in the previous section. 
Although \eqref{Eq:IntrinsicDepolarization} defines and \eqref{Eq:tau_c_n2EDM} estimates the correlation time for the longitudinal field $b_z$ and not that of the the transverse field $b$, the quantity of concern here, the two are in general approximately equal. 
The autocorrelation function $\langle b^*(0) b(\tau) \rangle$ decays to zero at times large compared to $\tau_c$. 

Let us consider first the case of the neutrons. 
We have seen in the previous section that the anticipated value for the correlation time of stored UCNs in n2EDM is 
$\tau_c({\rm ucn}) \approx 120\, \ms$ according to the estimation \eqref{Eq:tau_c_n2EDM}. 
In a $B_0 = 1\,\muT$ field the neutron angular frequency is 
$\omega_n = \gamma_n B_0 = -183\,\s^{-1}$. 
Thus we have $|\omega_n \tau_c ({\rm ucn})| = 22 \gg 1$; 
we say that the neutrons are in the \emph{high frequency regime}, sometimes also called the \emph{adiabatic regime}. 
In this regime one can expand Eq. \eqref{Eq:delta_omega} in powers of $1/\omega$, and we find at the lowest order
\begin{equation}
\delta \omega_n = \frac{\gamma_n^2}{2} \frac{\langle b^*(0) b(0) \rangle}{\omega_n}. 
\label{Eq:delta_omega_n_transverse}
\end{equation}
The ensemble average $\langle b^*(0) b(0) \rangle$ is simply the volume average of the quantity $b^* b = B_x^2 + B_y^2$. 
This result \eqref{Eq:delta_omega_n_transverse} can be understood with an intuitive picture of quasi-static neutrons: at any given time each neutron precesses at a frequency 
$|\gamma_n/2\pi| |\vec{B}|$ set by the magnitude of the field at the position of the neutron. 
This picture is correct because the precession frequency is very fast: a neutron stays at the same place throughout several spin rotations. 
At second order in $b$ we have $|\vec{B}| = |B_z| + \frac{B_x^2+B_y^2}{2 |B_z|}$. 
The ensemble of neutron spins precesses on average at a rate 
\begin{equation}
f_n = \frac{|\gamma_n|}{2\pi} \langle |\vec{B}| \rangle = \frac{|\gamma_n|}{2\pi} \left( \langle B_z \rangle + \frac{\langle B_x^2+B_y^2 \rangle}{2 | B_z |} \right). 
\label{Eq:fn_transverse_shift}
\end{equation}
The second term of this expression is consistent with Eq.~\eqref{Eq:delta_omega_n_transverse}.

Now let us consider the case of the mercury atoms. 
They have a mean speed of $180\,{\rm m}/\s$, which is much faster than the neutrons. Therefore, the correlation time is much shorter: $\tau_c({\rm Hg}) \approx 5\, \ms$. 
In a $B_0 = 1\,\muT$ field the mercury angular frequency is 
$\omega_{\rm Hg} = \gamma_{\rm Hg} B_0 = 48\,\s^{-1}$. 
Thus $\omega_{\rm Hg} \tau_c ({\rm Hg}) = 0.24 \ll 1$, 
and therefore the mercury atoms are in the \emph{low frequency regime}, sometimes also called the \emph{non-adiabatic regime}. 
In this regime one can expand Eq. \eqref{Eq:delta_omega} in powers of $\omega$ to get the following order-of-magnitude estimate: 
\begin{equation}
\delta \omega_{\rm Hg} \sim \gamma_{\rm Hg}^2 \omega_{\rm Hg} \tau_c^2 \langle b^* b \rangle. 
\end{equation}
From this estimate one concludes that the relative frequency shift of mercury is much smaller than the relative frequency shift of the neutrons because 
\begin{equation}
\frac{\delta \omega_{\rm Hg}/\omega_{\rm Hg}}{\delta \omega_n/\omega_n} \sim (\omega_{\rm Hg} \tau_c({\rm Hg}))^2 \sim 0.06. 
\end{equation}
The mercury atoms are much less sensitive to the transverse field compared to the neutrons. 
Indeed, during one spin rotation period a mercury atom explores the entire chamber several times and therefore the transverse components of the field effectively average out. 
In the case of the magnetic-field design value of $B_0 = 1 \, \muT$ we will work in the approximation of perfect averaging of the transverse magnetic-field components, and write 
\begin{equation}
f_{\rm Hg} = \frac{|\gamma_{\rm Hg}|}{2 \pi} \langle B_z \rangle 
\label{Eq:fHg_transverse_shift}
\end{equation}
for the mercury frequency, i.e.\ effectively using a volume average of $B_z$, only.
Eqs. \eqref{Eq:fn_transverse_shift} and \eqref{Eq:fHg_transverse_shift} can then be used to compute the shift $\delta_{\rm T}$. 
Note that in these expressions the ensemble average $\langle B_z \rangle$ is in principle different for the neutrons and the mercury atoms, but this difference is already accounted for by the gravitational shift. 
The expression of the transverse shift is therefore 
\begin{equation}
\delta_{\rm T} = \frac{\langle B_{\rm T}^2 \rangle}{2 B_0^2},  
\end{equation}
where $B_{\rm T}^2 = B_x^2 + B_y^2$. 
With transverse fields of the order of $\langle B_{\rm T}^2 \rangle \approx (500 \, {\rm pT})^2$, the transverse shift would give $\delta_{\rm T} \approx 0.1$~ppm. 
Although this is a significant shift relative to the statistical precision of $\R$, it is not a critical concern in the double chamber design. 
This is because a direct systematic effect could arise only through a difference in the shift between the top and bottom chamber; correlated with the electric-field polarity, this is promoted to a direct systematic effect. This in turn is necessarily associated with a non-uniformity of the longitudinal component $B_z$. 

\subsection{Motional field: Introduction}

Let us now come to the important description of the frequency shift induced by the motional field. 
According to special relativity, particles moving with a velocity $\vec{v}$ (in our case $|\vec{v}| \ll c$) in an electric field $\vec{E}$ experience a ``motional'' magnetic field
\begin{equation}
\vec{B}_m = \vec{E} \times \vec{v}/c^2. 
\end{equation}

The effect of the motional field on stored particles was first considered by Lamoreaux  \cite{Lamoreaux1996} who discussed the associated frequency shift quadratic in the electric field. 
Then Pendlebury {\it et al.} \cite{Pendlebury2004} discovered that the motional field also leads to a linear-in-electric-field frequency shift in the presence of magnetic-field gradients. 
Since then this topic has been studied  theoretically  \cite{Lamoreaux2005,Barabanov2006,Clayton2011,Swank2012,Pignol2012,Pignol2015,Golub2015,Swank2016} and experimentally \cite{Afach2015,Uniformity2019}.

This motional field affects both ultracold neutrons and mercury atoms when they are stored in the n2EDM chambers. 
Since the velocities of the particles are changing randomly in time, the motional field is in fact a magnetic noise transverse to $\vec{E}$. 
Let us estimate the magnitude of this noise in a vertical electric field $E = 15$~kV/cm, i.e. the design value for n2EDM. 
For neutrons with RMS horizontal velocity $v_h \approx 3$~m/s, we obtain magnetic fields of about $50$~pT.
For mercury atoms at room temperature, the RMS horizontal velocity is $157$~m/s and the corresponding RMS motional magnetic field is $2.6$~nT. 

\begin{figure*}
  \centering
  \includegraphics[width=\textwidth]{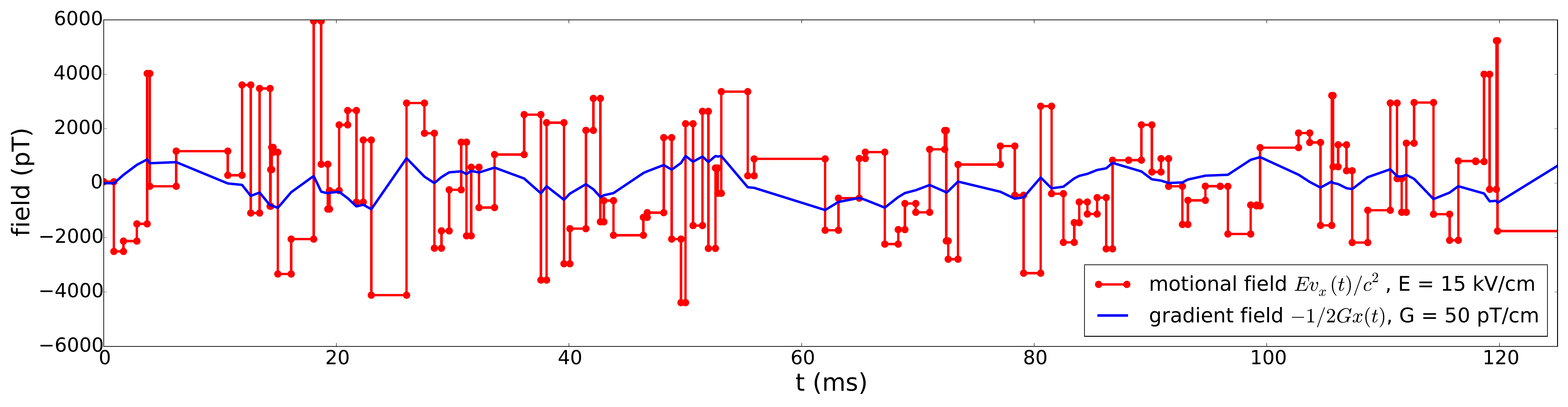}
  \caption[Motional field]
  {Monte-Carlo simulation of the transverse field seen by a mercury atom in thermal ballistic motion inside one n2EDM precession chamber. Red lines with dots: motional field along y induced by the electric field. Blue line: non-uniform field along $x$ in a (very large) gradient of $G_{1,0}=50 \ {\rm pT/cm}$.
	}
  \label{fig:random-field}
  \end{figure*}

The motional field $\vec{B}_m(t)$ adds to the fluctuating field $\vec{B}(\vec{r}(t))$ originating from the random motion of the particle in the non-uniform magnetic field. 
Eq. \eqref{Eq:complex_perturbation} can then be generalized, and the total fluctuating transverse field is described by the complex perturbation 
\begin{eqnarray}
b(t) & = & (\vec{B}_m(t) + \vec{B}(\vec{r}(t))) \cdot ( \vec{e}_x + i \vec{e}_y)  \\
& = & E/c^2 (-\dot{y}(t) + i \dot{x} (t)) + \vec{B}(\vec{r}(t)) \cdot ( \vec{e}_x + i \vec{e}_y). 
\end{eqnarray}
In Figure \ref{fig:random-field} a simulated random realization of the transverse field seen by a mercury atom is shown. 
As discussed before (at which time the motional field was neglected) any transverse magnetic perturbation generates a frequency shift given by Eq. \eqref{Eq:delta_omega}.  The total shift can be decomposed in powers of $E$ as
\begin{equation}
\delta \omega = \delta \omega_{B^2} + \delta \omega_{BE} + \delta \omega_{E^2}. 
\end{equation}
The term linear in $E$ is 
\begin{equation}
\delta \omega_{BE} = \frac{\gamma^2 E}{c^2} \int_0^\infty d\tau \cos (\omega \tau) \langle B_x(0) \dot{x}(\tau) + B_y(0) \dot{y}(\tau) \rangle, 
\label{Eq:delta_omega_BE}
\end{equation}
while the term quadratic in $E$ is 
\begin{equation}
\delta \omega_{E^2} = \left( \frac{\gamma E}{c^2} \right)^2 \int_0^\infty d\tau \sin(\omega \tau) \langle \dot{x}(0) \dot{x} (\tau) \rangle. 
\label{Eq:delta_omega_E2}
\end{equation}
The constant term $\delta \omega_{B^2}$ was discussed previously; it corresponds to the transverse shift $\delta_{\rm T}$. 
Next we will discuss the effects of the other two terms. 

\subsection{Motional field: Quadratic-in-E shift}
Let us first specify the angular frequency shift $\delta \omega_{E^2}$ for the neutrons, by taking the high frequency limit of Eq. \eqref{Eq:delta_omega_E2}. For this purpose we expand the integral in powers of $1/\omega$ by integration by parts and retain the dominant term 
\begin{equation}
\delta \omega_{E^2, n} = \left( \frac{\gamma_n E}{c^2} \right)^2 \frac{\langle \dot{x}^2 \rangle}{\omega_n}. 
\label{Eq:delta_omega_E2_n}
\end{equation}
Second, we specify $\delta \omega_{E^2}$ for the mercury atoms, which are in the low frequency limit if $B_0 = 1 \, \muT$. 
To calculate the low frequency limit of Eq. \eqref{Eq:delta_omega_E2} we first do an integration by parts to obtain  
\begin{equation}
\delta \omega_{E^2} = - \left( \frac{\gamma E}{c^2} \right)^2 \omega \int_0^\infty d\tau \cos(\omega \tau) \langle \dot{x}(0) x (\tau) \rangle. 
\end{equation}
Then, by the stationarity property
\begin{eqnarray}
\nonumber
\langle \dot{x}(0) x(\tau) \rangle = \langle \dot{x}(- \tau) x(0) \rangle
= - \frac{d}{dt} \langle x(-\tau) x(0) \rangle \\
= - \frac{d}{dt} \langle x(0) x(\tau) \rangle = - \langle x (0) \dot{x} (\tau) \rangle, 
\end{eqnarray}
we have 
\begin{equation}
\delta \omega_{E^2} = \left( \frac{\gamma E}{c^2} \right)^2 \omega \int_0^\infty d\tau \cos(\omega \tau) \langle x (0) \dot{x} (\tau) \rangle. 
\end{equation}
Finally,  we obtain the angular frequency shift for the mercury atoms in the low frequency limit by setting $\cos(\omega \tau) = 1$ in the above integral: \begin{equation}
\delta \omega_{E^2, \rm Hg} = - \left( \frac{\gamma_{\rm Hg} E}{c^2} \right)^2 \omega_{\rm Hg} \langle x^2 \rangle = - \left( \frac{\gamma_{\rm Hg} E}{c^2} \right)^2 \omega_{\rm Hg} \frac{R^2}{4}. \label{Eq:delta_omega_E2_Hg}
\end{equation}
The combination of Eqs. \eqref{Eq:delta_omega_E2_n} and \eqref{Eq:delta_omega_E2_Hg} leads to the expression for the quadratic-in-electric-field shift: 
\begin{eqnarray}
\nonumber
\delta_{\rm quad} & = & \frac{\delta \R}{\R} = \frac{\delta \omega_{E^2, n}}{\omega_n} - \frac{\delta \omega_{E^2, \rm Hg}}{\omega_{\rm Hg}} \\
& = & \left( \frac{v_h^2}{2 B_0^2} + \frac{\gamma_{\rm Hg}^2 R^2}{4} \right) \frac{E^2}{c^4}. 
\end{eqnarray}
With $v_h = 3$~m/s, $B_0 = 1 \, \muT$, $R = 40$~cm, $E = 15$~kV/cm we have 
$\delta_{\rm quad} = 2.7 \times 10^{-8}$. 
Notice that the term induced by the mercury atoms is about 20 times larger than the term induced by the neutrons. 

If the strength of the electric field is not exactly the same in the top and bottom chambers, due to a slightly different height of the two chambers, the quadratic frequency shift generates a term $\R^\TOP - \R^\BOT$. 
This generates a systematic effect if we consider the Top/Bottom EDM channel defined as 
$d_{\rm TB} = \frac{\pi \hbar f_{\rm Hg}}{2|E|} \left( \R^\TOP - \R^\BOT \right)$. 
An asymmetry of $\Delta E / E = 10^{-3}$ corresponds to $d_{\rm TB} = 10^{-28}~e~\mathrm{cm}$. 

Similarly, if the strength of the electric field is different in the positive and negative polarity, due to an imperfect polarity reversal of the HV source, the quadratic frequency shift generates a term $\R_+ - \R_-$. 
This generates a systematic effect if we consider the Plus/Minus EDM channel defined as 
$d_{+/-} = \frac{\pi \hbar f_{\rm Hg}}{2|E|} \left( \R_+ - \R_- \right)$. 
An asymmetry of $\Delta E / E = 10^{-3}$ corresponds to $d_n^{+/-} = 10^{-28}~e~\mathrm{cm}$. 

However, in the double-chamber concept, these two types of imperfections are compensated and do not generate a false EDM, as can be deduced from Eq. \eqref{Eq:dn_n2EDM}. 
Nonetheless, we give requirements for the uncompensated channels $d_{\rm TB}$ and $d_{+/-}$. 

First, we set a requirement on the electric-field asymmetry. 
In order to limit the systematic effect due to the quadratic frequency shift in the $d_{\rm TB}$ channel to lower than $1\times10^{-27}~e~\mathrm{cm}$, the electric-field strength must be the same in the top and bottom chamber with a precision better than 1\%  (i.e. $|\Delta E / E| < 10^{-2}$). 
Second, we set a requirement on the voltage reversal. 
In order to limit the systematic effect due to the quadratic frequency shift in the $d_{+/-}$ channel to lower than $1\times10^{-28}~e~\mathrm{cm}$ the absolute value of the voltage applied to the central electrode must be  the same in the positive and negative polarities with a precision better than 0.1\% (i.e. $|\Delta V / V| < 10^{-3}$). 

\subsection{Motional field: false EDM}
Now we will sketch the derivation of the high- and low-frequency limits of the frequency shift linear in electric field given by Eq. \eqref{Eq:delta_omega_BE}. 

The high-frequency limit, which applies for ultracold neutrons, is obtained by using the following approximation: 
\begin{equation}
\int_0^\infty d\tau \, \cos(\omega \tau) \, f(\tau)  = -\frac{1}{\omega^2} \dot{f}(0), 
\end{equation}
which is valid if $f(\tau)$ and $\dot{f}(\tau)$ are smooth functions decaying to $0$ for $\tau \rightarrow \infty$. 
We apply this scheme to the function
\begin{equation}
f(\tau) = \langle B_x(0) \dot{x}(\tau) \rangle = 
\langle B_x(-\tau) \dot{x}(0) \rangle. 
\end{equation}
We have
\begin{equation}
f'(\tau) = \frac{d}{d\tau} \langle B_x(-\tau) \dot{x}(0) \rangle = -\langle \frac{\partial B_x}{\partial x}(-\tau) \dot{x}(-\tau) \dot{x}(0) \rangle. 
\end{equation}
Therefore, at high frequency 
\begin{equation}
\int_0^\infty d\tau \, \cos(\omega \tau) \, \langle B_x(0) \dot{x}(\tau) \rangle = 
\frac{1}{\omega^2} \langle \frac{\partial B_x}{\partial x} \rangle  \langle \dot{x}^2 \rangle.  
\end{equation}
Doing the same with the function $\langle B_y(0) \dot{y}(\tau) \rangle$, and using Maxwell's equation $\frac{\partial B_x}{\partial x} + \frac{\partial B_y}{\partial y} = -\frac{\partial B_z}{\partial z}$, we find
\begin{equation}
\delta \omega_{BE} = - \frac{\gamma^2 E}{2c^2} 
\frac{1}{\omega^2} \langle \frac{\partial B_z}{\partial z} \rangle  v_h^2  \quad {\rm (high~frequency~limit)},
\label{Eq:delta_omega_BE_high_frequency}
\end{equation}
with $v_h^2 = \langle \dot{x}^2 \rangle + \langle \dot{y}^2 \rangle = 2 \langle \dot{x}^2 \rangle$. 

Now, the low-frequency limit, which applies to mercury atoms at low values of $B_0$, is simply obtained by using the approximation $\cos(\omega \tau) = 1$ in the integral Eq. \eqref{Eq:delta_omega_BE}: 
\begin{equation}
\delta \omega_{BE} = - \frac{\gamma^2 E}{c^2} \langle x B_x + y B_y \rangle \quad {\rm (low~frequency~limit)}.
\label{Eq:delta_omega_BE_low_frequency}
\end{equation}

With Eq. \eqref{Eq:delta_omega_BE_high_frequency} and 
Eq. \eqref{Eq:delta_omega_BE_low_frequency}, we can derive the corresponding shift in the $\R$ ratio as
\begin{equation}
\delta_{\rm EDM}^{\rm false} = \pm \frac{2}{\hbar |\gamma_n B_0|} |E| \left(d^{\rm false}_n + d^{\rm false}_{n \leftarrow {\rm Hg}} \right), 
\end{equation}
where the $+$ sign corresponds to the anti-parallel ($\uparrow \downarrow$ or $\downarrow \uparrow$) configurations whereas the $-$ sign corresponds to the parallel ($\uparrow \uparrow$ or $\downarrow \downarrow$) configurations. 
The formula for the false neutron EDM in the high frequency limit is 
\begin{equation}
d^{\rm false}_n = - \frac{\hbar v_h^2}{4 c^2 B_0^2} \langle \frac{\partial B_z}{\partial z} \rangle. \end{equation}
The false EDM transferred from the mercury in the low-frequency limit is
\begin{equation}
d^{\rm false}_{n \leftarrow {\rm Hg}}  =  - \frac{\hbar |\gamma_n \gamma_{\rm Hg}|}{2 c^2} \langle x B_x + y B_y \rangle. 
\label{Eq:false_EDM}
\end{equation}

\subsection{False EDM in a uniform gradient}

At this point it is important to note that the false EDM is really the combined effect of the motional field and the non-uniformities of the static $B_0$ field. 
An assumption of a simple uniform vertical gradient of the form \begin{equation}
\vec{B} = B_0 
\left( \begin{array}{c} 0 \\ 0 \\ 1 \end{array} \right)
 + G_{1,0} 
\left( \begin{array}{c} -x/2 \\ -y/2 \\ z \end{array} \right)
\end{equation}
leads to 
\begin{equation}
\langle \frac{\partial B_z}{\partial z} \rangle = G_{1,0}
\end{equation}
and
\begin{equation}
\langle xB_x + y B_y \rangle = - G_{1,0} \frac{R^2}{4}.
\end{equation}
In this situation, one can estimate the false EDM directly induced on the neutrons $d^{\rm false}_n$ and the one induced via the mercury $d^{\rm false}_{n \leftarrow {\rm Hg}}$: 
\begin{eqnarray}
d^{\rm false}_n & = & - \frac{\hbar v_h^2}{4 c^2 B_0^2} G_{1,0} \\ & = & - \frac{G_{1,0}}{1 \, \pT/\cm} \times 1.65 \times 10^{-28}~e~\mathrm{cm} , \\ 
d^{\rm false}_{n \leftarrow {\rm Hg}} & = &  \frac{\hbar |\gamma_n \gamma_{\rm Hg}| R^2}{8 c^2} G_{1,0} \\ 
& = & \frac{G_{1,0}}{1 \, \pT/\cm} \times 1.28  \times 10^{-26}~e~\mathrm{cm}, 
\end{eqnarray}
where $v_h = 3$~m/s, $B_0 = 1 \, \muT$ and $R = 40$~cm. 
It should be noted that the mercury-induced false neutron EDM is much larger than the directly induced neutron motional false EDM. 

Even if the residual field gradient inside the shield is reduced down to a fraction of a pT/cm,  a systematic effect greater than $10^{-27}~e~\mathrm{cm}$ could still be generated. 
The general strategy to cancel the effect is to split the data production into many runs with different gradient configurations within the allowed range $\pm 0.6$pT/cm. 
In this way we will measure the EDM as function of the gradient, extrapolating to zero gradient in the final step. 
In the nEDM experiment the gradient was inferred from the gravitational shift. 
However, the shift of the $\R$ ratio correlates only imperfectly with the gradient, because of all the other frequency shifts. 
In n2EDM the gradient can be extracted in a more robust way thanks to the double-chamber design. 
We define the \emph{Top/Bottom gradient} as
\begin{equation}
G_{\rm TB} = \frac{\langle B_z \rangle_\TOP - \langle B_z \rangle_\BOT}{H'},
\end{equation}
where $H' = 18$\,cm is the distance between the geometrical centers of the two chambers. 
The $G_{\rm TB}$ will be accurately measured with the mercury co-magnetometers. 

At this point one can identify two possible failures of the extrapolation method that would each produce a residual systematic effect. 

\medskip
\begin{itemize} 
\item 
\begin{sloppypar}
First, a systematic shift of the mercury precession frequency of the upper co-magnetometer relative to the lower co-magnetometer will result in a systematically wrong gradient. 
This is quoted as \emph{co-magnetometer accuracy} in  Table\,\ref{tab:SystematicsTable}. 
The aim is to constrain that error to lower than $1 \times 10^{-28}~e~\mathrm{cm}$. 
This sets a requirement on the accuracy of the magnetometers, which must be $\delta B_{\rm Hg} < 100\,\fT$. 
Note that this requirement is less stringent than the requirement on the precision per cycle of $25\,\fT$ derived in section Sec.~\ref{sec:3.5} . 
All known sources of frequency shifts of the co-magnetometer are listed in the previous section (see also Sec.~\ref{sec:5.4} for magnetometry).
\end{sloppypar}
\item Second, and more importantly, the extrapolation procedure to $G_{\rm TB} = 0$ fails if the field non-uniformities are more complicated than a uniform gradient $G_{1,0}$. 
It is useful to distinguish two types of non-uniformities: 
(i) Large-scale spatial $B$-modes of cubic and higher orders. 
These are generated by the imperfection of the mu-metal shield,  for example due to the openings, and by imperfections of the $B_0$ coil. 
(ii) Magnetic dipole sources localized near the precession chambers,  due to the contamination of the apparatus by small ferromagnetic  impurities. 
\end{itemize}

\subsection{False EDM and phantom modes}
\label{subsec:phatomModes}
To discuss more complicated field non-uniformities, we describe the field by the generalized gradients $G_{l,m}$ as defined in ~\eqref{Eq:harmon}. 
With this formalism we can calculate the Top/Bottom gradient 
\begin{equation}
G_{\rm TB} = G_{1,0} - L_3^2 G_{3,0} + L_5^4 G_{5,0} + \cdots
\end{equation}
where $L_l$ are geometric coefficients. 
Modes with $m\neq0$ do not contribute to the Top/Bottom gradient because the chamber is symmetric by rotation around the magnetic-field axis. 
Modes with even values of $l$ are also absent because the top chamber is the mirror image of the bottom chamber with respect to the plane $z=0$. 
An explicit calculation for the cubic and fifth-order modes gives the geometric coefficients 
\begin{eqnarray}
L_3^2 & = & \frac{3}{4} R^2 - \frac{1}{4} (H^2 + H'^2) = (32.9\,\cm)^2 , \\
\nonumber 
L_5^4 & = & \frac{5}{8} R^4 - \frac{5}{8} R^2 (H^2 + H'^2) + \frac{1}{48} (3 H'^2 + H^2) (H'^2 + 3 H^2) \\ 
 & = & (32.7\,\cm)^4. 
\end{eqnarray}

There are axially symmetric field configurations, i.e. linear combinations of $m=0$ modes, which are invisible in the double-chamber because they satisfy $G_{\rm TB} = 0$. 
We call these field configurations \emph{phantom modes}. 
We define the basis of phantom modes of odd degree as 
\begin{eqnarray}
\acute{\Pi}_3 & = & c_3 \left( \Pi_{1,0} + \frac{1}{L_3^2} \Pi_{3,0} \right), \\ 
\acute{\Pi}_5 & = & c_5 \left( \Pi_{1,0} - \frac{1}{L_5^4} \Pi_{5,0} \right), 
\end{eqnarray}
and similarly for all odd modes
\begin{equation}
\acute{\Pi}_{2k+1}  =  c_{2k+1} \left( \Pi_{1,0} - \frac{(-1)^k}{L_{2k+1}^{2k}} \Pi_{2k+1,0} \right). 
\end{equation}
The normalization of the phantom modes of odd degree are chosen such that 
\begin{equation}
\langle \rho \acute{\Pi}_\rho \rangle_\TOP = \langle \rho \acute{\Pi}_\rho \rangle_\BOT =-R^2/4. 
\end{equation}
In particular, for the phantom modes of degrees $3$ and $5$: 
\begin{eqnarray}
c_3 & = & \frac{4L_3^2}{R^2+2H'^2} , \\ 
c_5 & = & \frac{48 L_5^4}{15R^4+10R^2(3H'^2-H^2)-4H'^2(3H'^2+5H^2)}. 
\end{eqnarray}
The even modes ${\bm \Pi}_{2,0}, {\bm \Pi}_{4,0}$ are also phantom in the sense previously defined, but they do not produce a false EDM and will not be discussed further. 
The odd phantom modes are of particular interest because they generate a false EDM without generating a Top/Bottom gradient. 
Specifically, a field configuration of the type
\begin{equation}
\vec{B} = B_0 \vec{e}_z + G_{\rm TB} {\bm \Pi}_{1,0} +  \acute{G}_3 {\bm \acute{\Pi}}_3 + \acute{G}_5 {\bm \acute{\Pi}}_5 + \cdots
\end{equation}
generates a false EDM through Eq. \eqref{Eq:false_EDM},
\begin{equation}
d^{\rm false}_{n \leftarrow {\rm Hg}} = \frac{\hbar |\gamma_n \gamma_{\rm Hg}| R^2}{8 c^2} \left( G_{\rm TB} + \acute{G}_3 + \acute{G}_5 + \cdots \right). 
\label{Eq:false_EDM2}
\end{equation}
Obviously, the contribution proportional to $G_{\rm TB}$ will be removed by the extrapolation to $G_{\rm TB}=0$, but the contribution proportional to the phantom gradient, $\acute{G} = \acute{G_3} + \acute{G_5} + \cdots$, will remain. 

\begin{figure}
  \centering
  \includegraphics[width=1.\columnwidth]{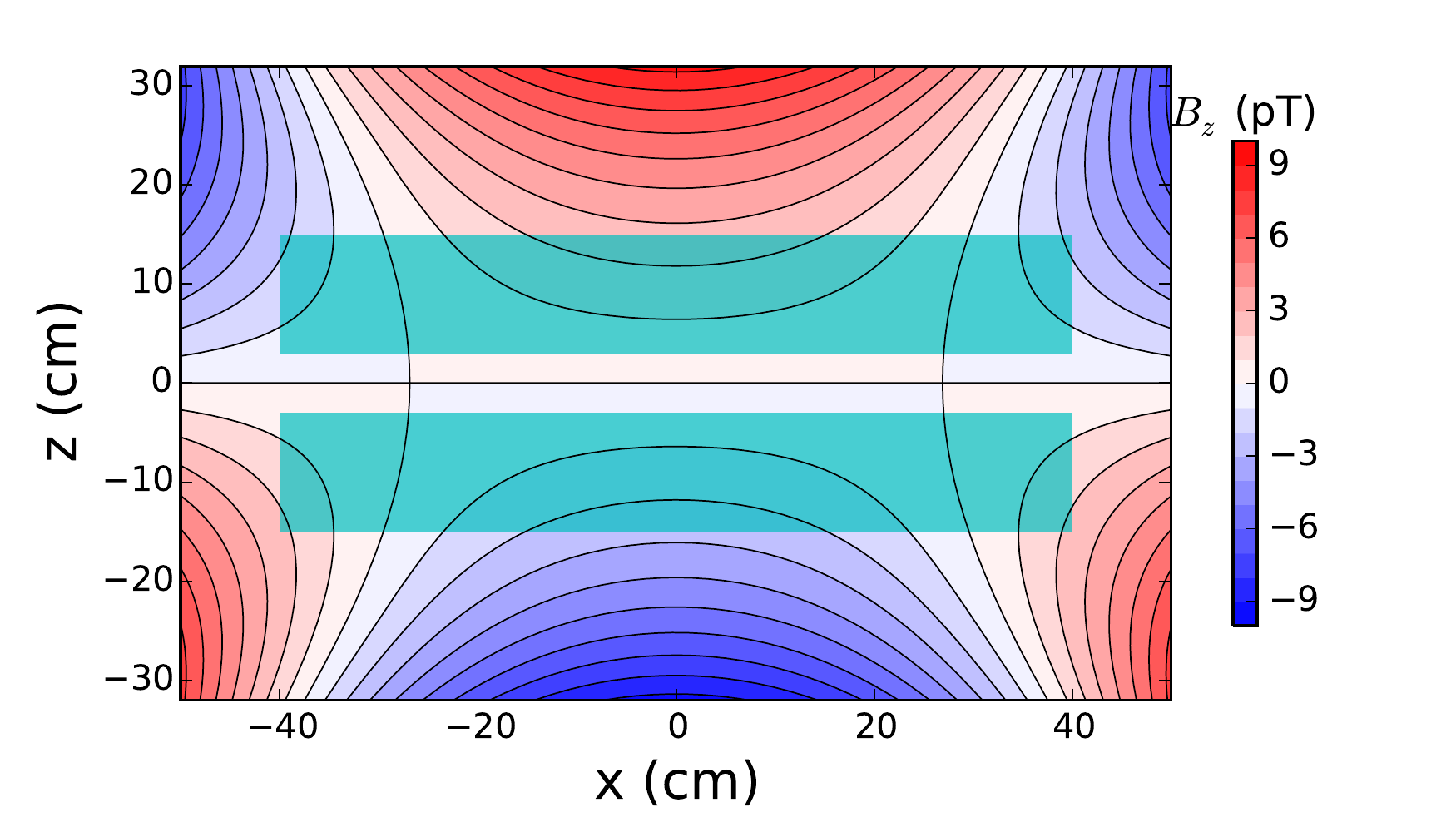}
  \includegraphics[width=1.\columnwidth]{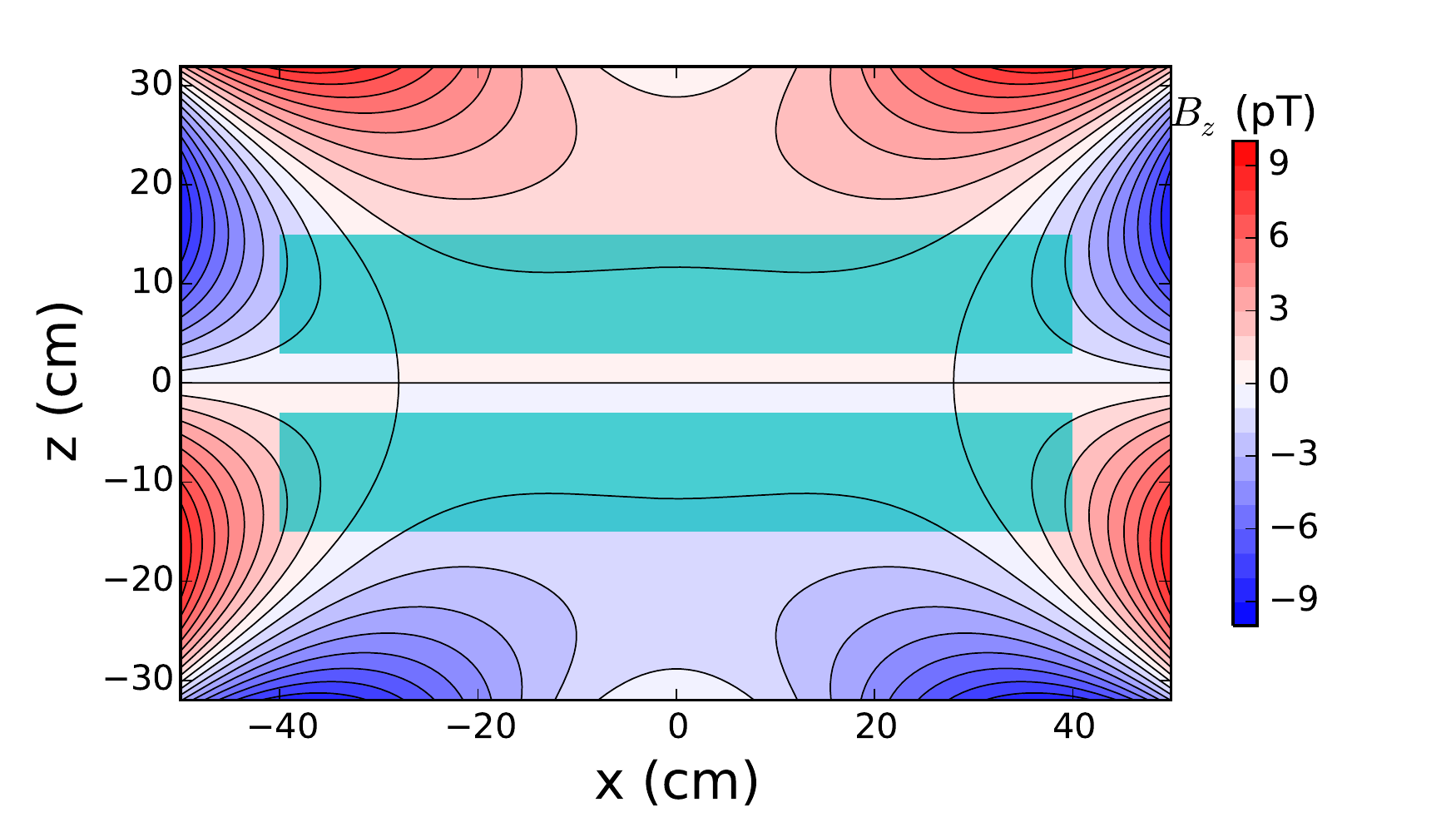}
  \caption
  {Longitudinal component of the phantom modes. 
  Top: phantom mode of order 3 $B_z = \acute{G}_3 {\bm \acute{\Pi}}_{z,3}$ with $\acute{G}_3 = 78\,\fT/\cm$. 
  Bottom: phantom mode of order 5 $B_z = \acute{G}_5 {\bm \acute{\Pi}}_{z,5}$ with $\acute{G}_5 = 78\,\fT/\cm$. 
  Both field configurations generate a false EDM of $d^{\rm false}_{n \leftarrow {\rm Hg}} = 1 \times 10^{-27}~e~\mathrm{cm}$. 
  The light green rectangles represent the inner volume of the precession chambers. 
  }
  \label{fig:Phantom}
\end{figure}

In Figure \ref{fig:Phantom} we show the $B_z$ field configuration corresponding to the phantom modes of order 3 and order 5. 
Our strategy to control the phantom modes is to use a combination of online and offline measurements, the former being more adequate for the low-order modes and the latter more appropriate for the high-order modes. 

The online measurement of the field will be provided by an array of cesium magnetometers, which will be able to extract the gradients $G_{l,m}$ up to order $l=5$. 
In particular the array will provide, online, a measurement of $\acute{G}_3 = \frac{L_3^2}{c_3} G_{3,0}$ that will be used to correct for the corresponding systematic effect. 
As a guide to the design of the magnetometer array, we set  the requirement that the error on the correction for the cubic phantom mode must be lower than $3 \times 10^{-28}~e~\mathrm{cm}$. 
This corresponds to an accuracy of $\delta \acute{G}_3 < 20 \, \fT/\cm$. 

The offline measurement will be performed by a mechanical mapping device.
During the mapping the inner parts of the vacuum vessel, including the precession chambers, will be removed. 
This imposes a requirement on the reproducibility of the field configuration (it needs to be identical during the mapping and during the data-taking), and also a requirement on the accuracy of the magnetic-field mapper. 
As a design guide we set the requirement that the error on the correction for the fifth-order phantom mode $\acute{G}_5 =  \frac{L_5^4}{c_5} G_{5,0}$ must be lower than $3 \times 10^{-28}~e~\mathrm{cm}$. 
This corresponds to an accuracy of $\delta \acute{G}_5 < 20 \, \fT/\cm$. 
The requirements related to the control of the high-order gradients are summarized below in Tables \ref{tab:SystematicsTable} and \ref{tab:RequirementsTable}. 
Note that the requirements on $\delta \acute{G}_3$ and $\delta \acute{G}_5$ concern the magnetic-field measurement and not the magnetic-field generation.

\subsection{False EDM and magnetic-dipole sources}

Contamination of the inner parts of the apparatus by small ferromagnetic impurities generate a second important type of magnetic-field nonuniformity. 
Here we evaluate the induced systematic effect,  and specify the tolerated level of contamination. A small magnetic impurity can be described as a magnetic dipole $\vec{m}$. 
Such a dipole located at distance $\vec{r}_d$ is a source of a dipolar magnetic field of the form $\mathbf{B}_d(\mathbf{r}) = (\mu_0/4\pi)(3(\mathbf{m} \cdot\mathbf{u}) \mathbf{u}-\mathbf{m})/|\mathbf{r}-\mathbf{r}_d|^3$ , 
with $\mathbf{u} = (\mathbf{r}-\mathbf{r}_d)/|\mathbf{r}-\mathbf{r}_d|$ representing the unit vector pointing from the dipole position. 
This will induce a false EDM $d_{n\leftarrow \text{Hg}}^{\text{false}}$ given by Eq.~\eqref{Eq:false_EDM}. 
In addition, the dipole source will generate a Top/Bottom gradient $G_{\rm TB}$ measured by the mercury co-magnetometers, and will also affect the cubic  phantom gradient $\acute{G}_{3, {\rm meas}}$ extracted from the readings of the cesium magnetometers. 
However, the measured correction
\begin{equation}
d^{\rm false}_{\rm meas} = \frac{\hbar |\gamma_n \gamma_{\rm Hg}| R^2}{8 c^2} \left( G_{\rm TB} + \acute{G}_{3, {\rm meas}} \right) 
\end{equation}
will imperfectly estimate the actual false EDM given by Eq.~\eqref{Eq:false_EDM2}, because $\acute{G}_{3, {\rm meas}}$ will be shifted from the true value $\acute{G}_{3}$ and also because the higher-order gradients $\acute{G}_{5}, \acute{G}_{7}, \cdots$ generated by the dipole are not corrected for. 

\begin{figure}
\includegraphics[width=1.\linewidth]{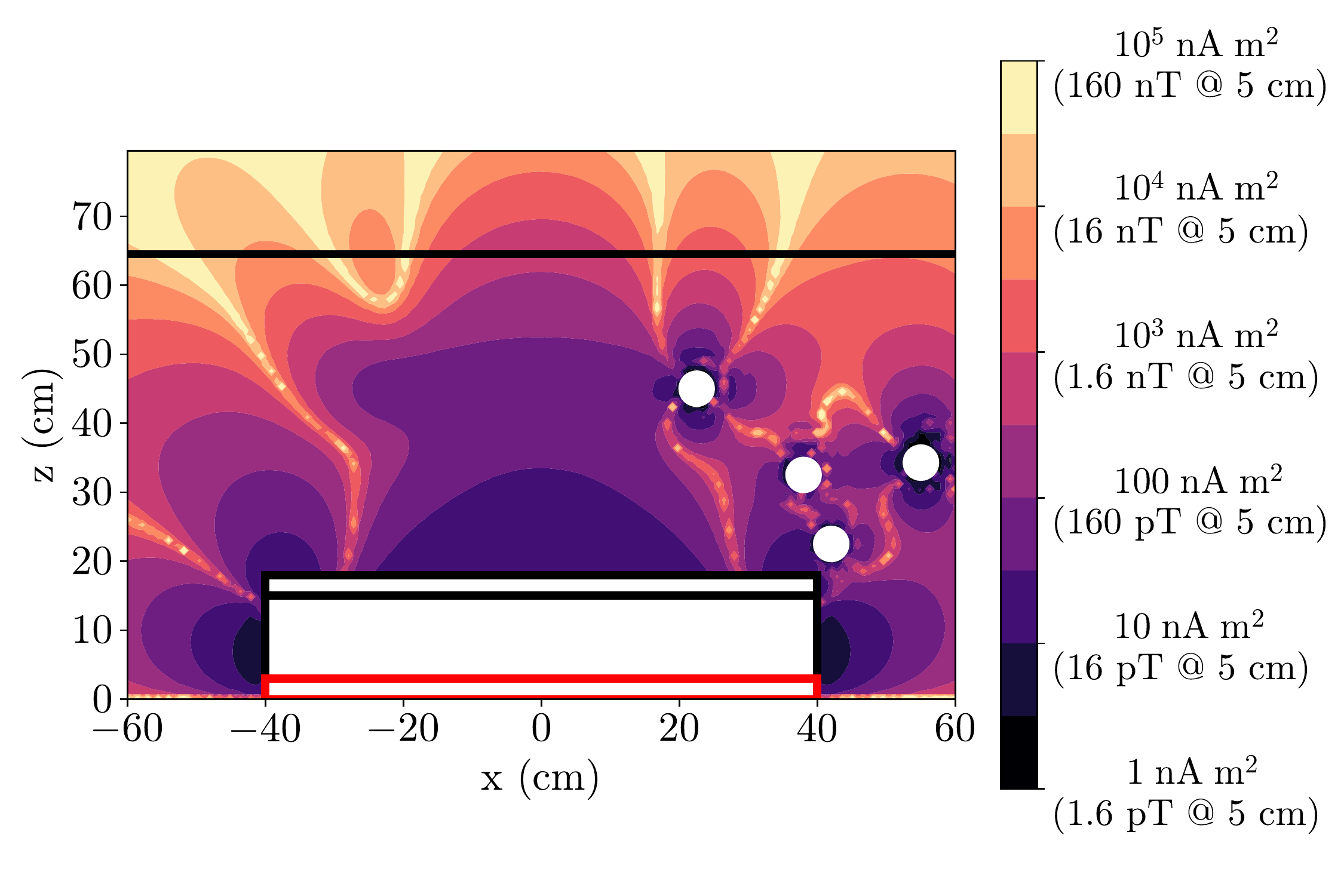}
\caption{Magnetic dipole strength values corresponding  to a residual systematic effect of $d^{\rm false}_{n \leftarrow {\rm Hg}}-d^{\rm false}_{\rm meas}=3\times10^{-29}~e~\mathrm{cm}$ (mean of top and bottom chambers), as a function of the position of the dipole in the $y=0$ plane. 
The direction of the dipole $m$ was chosen to be along the $z$-axis, which is the most sensitive direction. 
This cut ($y=0$) intersects a unit of four magnetometers represented by the white circles. 
The top plate of the vacuum tank is represented by the horizontal black line at $z=65~\mathrm{cm}$. 
The cross section of the electrodes are represented by the black and red-edged rectangles. 
}
\label{fig:dipole_colormap_plot}
\end{figure}

A thorough numerical study of the influence of dipole strength and location was conducted, by considering a given dipole placed at different locations in the experimental volume outside the precession chambers  and calculating the residual effect $d^{\rm false}_{n \leftarrow {\rm Hg}}-d^{\rm false}_{\rm meas}$. 
The value $\acute{G}_{3, {\rm meas}}$ was calculated by considering the field produced by the dipole at the position of each magnetometer (see Sec.\ref{sec:5.4.3} for a description of the designed optimized positions of the magnetometers) and performing the harmonic fit to cubic order (up to $l=3$). 
A sample of the results can be seen in Fig.~\ref{fig:dipole_colormap_plot}, and corresponds to the top half of the $y=0~\mathrm{cm}$ plane of the experiment. 
It shows the dipole strength $|\vec{m}|$ that produces a residual effect of $3\times10^{-29}~e~\mathrm{cm}$, the chosen maximum tolerated contribution for a single dipole. 
We allow for the presence of a maximum of 100 impurities with random and uncorrelated direction, such that the total systematic effect will be $\sqrt{100}$ times the contribution of one individual dipole, i.e. $3\times10^{-28}~e~\mathrm{cm}$ (see Table \ref{tab:SystematicsTable}). 

\begin{sloppypar}
The regions of the apparatus that are most sensitive to the presence of magnetic contamination are the outside of the insulating rings and the immediate proximity of each magnetometer. 
At these locations the critical dipole strength, i.e. the maximum tolerated dipole strength to meet the requirement for the contribution for individual dipole, was found to be $5~\mathrm{nA~m}^2$. 
This dipole strength corresponds to an iron dust particle of diameter $\approx 20 \, \upmu {\rm m}$ magnetized to saturation. It would produce a field of approximately 1~pT at 10 cm distance. 
Other locations are less sensitive but must still be protected against magnetic contamination. 
In fact all components of the apparatus inside the magnetic shield must be magnetically scanned to exclude dipoles larger than specified.
For example, the vacuum tank (represented as a horizontal black line at $z=65~\mathrm{cm}$ in Fig.~\ref{fig:dipole_colormap_plot}) must be carefully quality controlled such that dipoles larger than $500~\mathrm{nA~m}^{2}$ are excluded.
\end{sloppypar}

\subsection{The magic-field option to cancel the false EDM}

We have argued that the significant gain in statistical sensitivity in n2EDM will be obtained by the use of a large double chamber. 
In the described design the diameter of the chambers will be $80\,\cm$, while the vacuum vessel is designed to host a chamber as large as  $100\,\cm$ for a future phase of the experiment. 
This is made possible by the very large magnetically shielded room, with inner dimensions of almost $3\times3\times3$~m$^3$. 
The enlargement of the chambers, as compared to the $47\,\cm$ diameter single-chamber of the previous nEDM apparatus, comes at the price of an increase in the systematic effect due to the mercury motional false EDM. 
This can be clearly seen from Eq. \eqref{Eq:false_EDM}. 
As discussed, controlling the effect induced by the phantom modes brings about a number of challenges:
(i) the cesium magnetometers must reach the required accuracy to measure at least the cubic phantom mode online, 
(ii) the higher-order modes must be reproducible enough to be able to measure these modes offline, 
(iii) magnetic contamination must be kept at a very low level. 

These challenges, and the associated risks for the measurement, prevail if the mercury co-magnetometer operates in the low field regime, as it is the case in the design with $B_0 = 1 \, \muT$. 
There is an alternative possibility that can considerably relax the constraints on the measurement of field nonuniformities. 
It consists of increasing the $B_0$ field to a value that cancels the mercury false EDM ~\cite{Pignol2019}.  
We recall that the false neutron EDM inherited from the mercury is 
\begin{equation}
d^{\rm false}_{n \leftarrow {\rm Hg}} = \frac{\hbar |\gamma_{\rm n} \gamma_{\rm Hg}|}{2c^2} \int_0^\infty d \tau \cos (\omega \tau) \dot{C}(\tau), 
\end{equation}
where $C(\tau)$ is the correlation function 
\begin{equation}
C(\tau) = \langle B_x(0) x(\tau) + B_y(0) y(\tau) \rangle
\end{equation}
and $\omega = \gamma_{\rm Hg} B_0$. 
The correlation function $C(\tau)$ can be calculated with a Monte Carlo  simulation of the thermal motion of mercury atoms in the chamber. 

\begin{figure}[t]
\centering
\includegraphics[width=1.\linewidth]{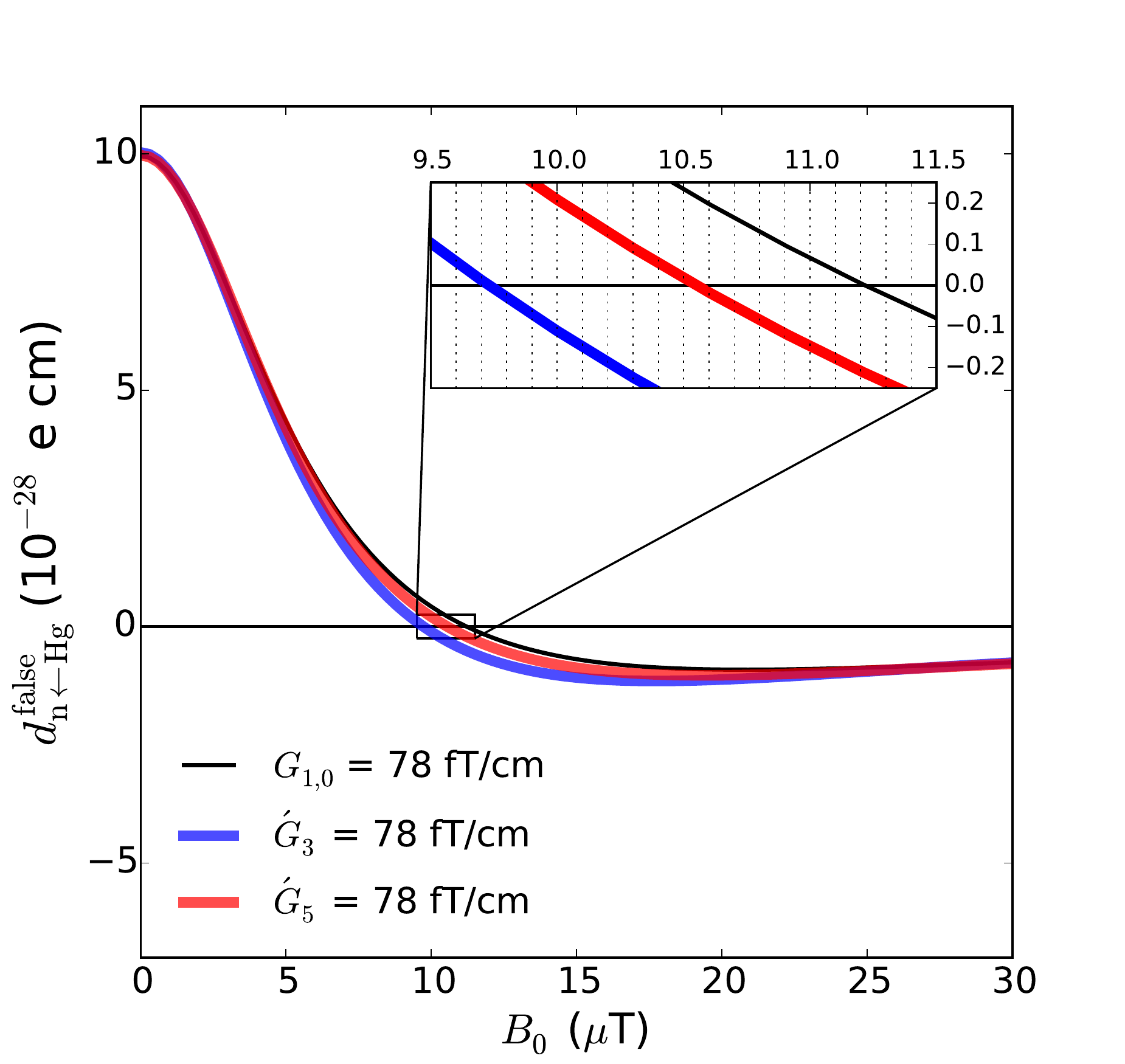}
\caption[False EDM due to a field gradient]
{False EDM due to a uniform field gradient (black), a 3-phantom mode (blue), a 5-phantom mode (red) as a function of the magnitude of the $B_0$ field for $R=40\,\cm$, $H=12\,\cm$, $H'=18\,\cm$. 
}
\label{fig:FalseEDMvsB0}
\end{figure}

In Figure \ref{fig:FalseEDMvsB0} we show the result for the false EDM as a function of the magnitude of the $B_0$ field. 
It is possible to adjust the value of $B_0$ to cancel the systematic effect produced by a given mode. 
We define as  ``magic fields'' the magnetic field values 
\begin{equation}
B_{\rm magic, 3} = 9.7\,\muT , \quad B_{\rm magic, 5} = 10.5\,\muT,
\label{eq:magic}
\end{equation}
which cancel the effect of the respective phantom modes. The magic fields for the different modes are very close. 
This makes the magic option attractive because it allows substantial reduction of the effect of several modes at the same time. 

\emph{The magic-field upgrade option consists of setting the magnetic field to $B_0 = 10.5 \, \muT$}. 
This will suppress the effect of the fifth order phantom mode completely and will also reduce the effect of the cubic phantom mode by a factor of 30. 
The magic field is a factor of ten higher than that of the baseline design, and it therefore increases the difficulty of producing a stable and uniform field by an order of magnitude. 
It should be noted that the requirements of the field uniformity and stability concern the absolute rather than relative values. 
The n2EDM apparatus is designed to allow operation of the apparatus at the magic field and slightly above, after first running in the baseline configuration. 

\subsection{Other frequency shifts}

In addition to the electric and magnetic terms, there are a number of other known shifts of either the neutron or the mercury precession frequencies that correspondingly affect the $\R$ ratio:
\begin{equation}
\delta_{\rm other} = \delta_{\rm AC} + \delta_{\rm Earth} + \delta_{\rm light} + \delta_{\rm pulse} + \delta_{\rm psmag}.
\end{equation}
Below we discuss each individual contribution. 

\subsubsection{Effects of AC fields during the precession:  $\delta_{\rm AC}$ }
Any transverse AC magnetic field during the precession generates a frequency shift for the neutrons and mercury atoms. 
In addition to the AC field seen by the particles moving in a static but non-uniform field (already taken into account by the term $\delta_{\rm T}$) as well as the fluctuating motional $\vec{E} \times \vec{v}/c^2$ field (already taken into account by the terms $\delta_{\rm EDM}^{\rm false}$ and $\delta_{\rm Quad}$), 
the other known possible source of AC fields are  
\\ (i) ripples in the voltage generated by the HV source \cite{Baker2014}; 
(ii) the Johnson-Nyquist noise generated by the metallic parts, in particular by the electrodes \cite{PinJung2021} . 
These were found to be very small effects and will not be discussed in detail here. 

\subsubsection{The effect of Earth's rotation: $\delta_{\rm Earth}$}
Since the precession frequencies of mercury and neutron spins are measured in the Earth's rotating frame, the frequencies are shifted  from the pure Larmor frequency in the magnetic field ~\cite{Golub2007}. 
One can derive the following expression for the associated  shift of $\R$: 
\begin{equation}
\delta_{\rm Earth} = \mp \left( \frac{f_{\rm Earth}}{f_n} + \frac{f_{\rm Earth}}{f_{\rm Hg}}\right) \cos \theta = \mp 1.4 \times 10^{-6}, 
\end{equation}
where $f_{\rm Earth} = 11.6\, \upmu$Hz is the Earth's rotation frequency, $f_n = 29.2\,$Hz and $f_{\rm Hg} = 7.6\,$Hz are the neutron and mercury precession frequencies in a field of $B_0= 1\,\muT$, and $\theta = 42^\circ$ is the angle between the direction of $B_0$ and the rotation axis of the Earth. 
In the previous formula the - sign corresponds to $B_0$ pointing upwards and the + sign corresponds to $B_0$ pointing downwards. 
The shift is large enough to be resolved in principle with a single data cycle (although in fact measurements are needed with both directions of the $B_0$, so two cycles are required), provided the other shifts are constant. 
\begin{sloppypar}
A direct systematic effect could arise in principle if electric-field reversals cause a tilt of the magnetic axis relative to the Earth's rotation axis. 
However, in the double-chamber design this direct systematic effect could arise only in the case of different tilts in the top and bottom chamber (see Eq.~\ref{Eq:ddn_n2EDM}).
Such a magnetic tilt is necessarily associated with a gradient of the longitudinal field, and the requirement set on the control of the gradients in Eq. \ref{Eq:req_gradient_variation} guarantees that the direct systematic effect due to the Earth's rotation will be negligible. 
\end{sloppypar}

\subsubsection{The mercury light shift: $\delta_{\rm light}$ }
This term corresponds to a shift of the mercury precession frequency proportional to the intensity of the UV probe light. 
This small effect should be taken into account in the design of the mercury co-magnetometer (in particular, good monitoring of the light intensity must be foreseen) but does not impose stringent requirements on the magnetic field generation or magnetic field measurement. 

\subsubsection{The effect of the mercury pulse: $\delta_{\rm pulse}$ }
The mercury pulse is generated while the neutrons are already present in the chamber. 
Therefore, the neutron spins are affected by the mercury pulse: they will be slightly tilted  before the first neutron pulse is applied. 
In turn, this could shift the measured Ramsey resonance frequency. 
This effect must be taken into account when designing the generation of the mercury pulse. 
The frequency shift can be reduced by adjusting the duration, phase, and shape of the mercury pulse. 
Care will be taken to avoid indirect cross-talk with the high-voltage polarity. 
However, this effect does not impose stringent requirements on the magnetic field generation or magnetic field measurement. 

\subsubsection{The pseudomagnetic field generated by polarized mercury:  $\delta_{\rm psmag}$ }

Due to the spin-dependent nuclear interaction between the neutron and the mercury-199 nucleus, quantified by the incoherent scattering length $b_{\rm i}(^{199}{\rm Hg}) = \pm 15.5$ fm~\cite{NNews1992}, the UCNs precessing in the polarized mercury medium are exposed to a pseudo-magnetic field  ~\cite{Abragam1975}
\begin{equation}
\vec{B}^\star = -\frac{4 \pi \hbar}{\sqrt{3} m_n  \gamma_n} b_{\rm i} n_{\rm Hg} \vec{P}, 
\end{equation}
where $m_n$ is the neutron mass, $n_{\rm Hg}$ is the number density of atoms in the precession chamber and $\vec{P}$ is the mercury polarization. 
The pseudo-magnetic field is much larger than the genuine magnetic dipolar field generated by the polarized mercury atoms. 
The mercury polarization normally precesses in the transverse plane, but it could have a residual static longitudinal component $P_\parallel$ in the case of an imperfect $\pi/2$ pulse. 
In this case, a shift of the neutron frequency arises that corresponds to a relative shift of the $\R$ ratio of
\begin{equation}
\delta_{\rm psmag} = \pm \frac{2 \hbar}{\sqrt{3} m_n f_n} n_{\rm Hg} b_i P_\parallel. 
\end{equation} 
This small effect will be taken into account in the design of the mercury magnetometer, in particular the control of the mercury pulse, but it does not impose stringent requirements on the magnetic field generation or magnetic field measurement.

\subsection{Summary of the requirements}

In summary, we have described the known sources of systematic effects and discussed how to address them in the n2EDM experiment.
The apparatus is designed to keep the total systematic error below $6 \times 10^{-28}~e~\mathrm{cm}$. The error contributions are expected to be distributed according to the budget shown in Table~\ref{tab:SystematicsTable}. 
The dominant contributions to this budget originate from the mercury false EDM effect, which could be reduced by operating the apparatus at the magic value of the magnetic field in a future upgrade. 

Through consideration of the statistical and systematic errors we have derived the basic requirements on the performance of the n2EDM apparatus. 
For convenience we reproduce in Table \ref{tab:RequirementsTable} the requirements specifically related to magnetic field generation and measurement. 
These requirements constitute the basis for the technical design of the core systems of the n2EDM apparatus, which are described in the next section. 

\begin{table}
\centering
\caption[Goal for the control of systematic effects.]{
Goal for the control of systematic effects in the 2EDM design.
}
\label{tab:SystematicsTable}
\begin{tabular}{lr}
Systematic effect 
&  ($10^{-28}~e~\mathrm{cm}$)  \\
\hline\hline
Uncompensated gradient drift   & $1 $  \\
Quadratic $v \times E$         & $1 $  \\
Co-magnetometer accuracy       & $1 $  \\
Phantom mode of order 3        & $3 $  \\
Phantom mode of order 5        & $3 $  \\
Dipoles contamination          & $3 $  \\
\hline \hline
Total                          & $6 $
\end{tabular}
\end{table}

\begin{table*}
\centering
\begin{tabular}{ll}
\hline \hline
\multicolumn{2}{c}{Related to statistical errors} \\ 
(B-gen) Top-Bottom resonance matching condition      & $-0.6 \, \pT/\cm < G_{1,0}  < 0.6  \, \pT/\cm$ \\
(B-gen) Field uniformity in the chambers    & $\sigma (B_z) < 170 \, \pT$ \\
(B-gen) Field stability on minutes timescale                     & $< 30$ fT   \\
(B-meas) Precision Hg co-magnetometer, per cycle, per chamber       & $< 30$ fT   \\
\hline \hline
\multicolumn{2}{c}{Related to systematical errors} \\
(B-gen) Gradient stability on the timescale of minutes             & $\sigma(G)[5 {\rm min}] < 50$ fT/cm \\
(B-meas) Accuracy mercury co-magnetometer per chamber        & $<100$ fT     \\
(B-meas) Accuracy on cubic mode (Cs magnetometers)       & $\delta \acute{G}_3 < 20 \, \fT/\cm$ \\
(B-gen) Reproducibility of the order 5 mode      & $\sigma (\acute{G}_5) < 20 \, \fT/\cm$ \\
(B-meas) Accuracy of the order 5 mode (field mapper)      & $\delta \acute{G}_5 < 20 \, \fT/\cm$ \\
(B-gen) Dipoles close to the electrode                       & $< 20$~pT at 5~cm  \\
(E-gen) Relative accuracy on E field magnitude & $< 10^{-3}$ \\
\hline \hline
\end{tabular}

\caption{
Summary of the requirements for the magnetic-field measurement (B-meas), magnetic-field generation (B-gen) and electric-field generation (E-gen) for the n2EDM design.
}
\label{tab:RequirementsTable}
\end{table*}

%% file: Technical-Intro.tex
\begin{figure*}
\centering
    \includegraphics[width=1\textwidth]{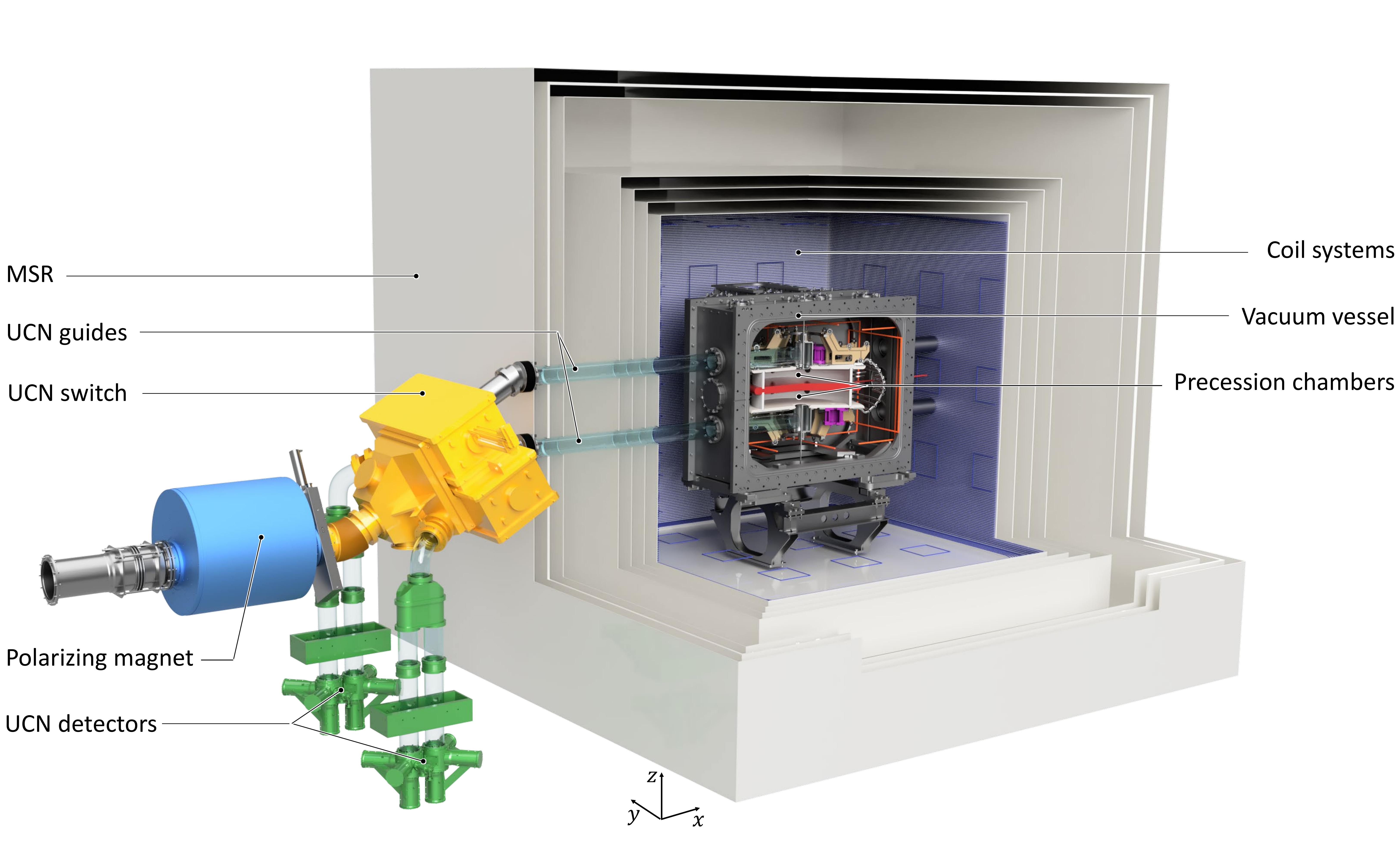}
  \caption{The full model of the n2EDM setup, displaying the core components of the experiment}
\label{fig:n2EDM-setup}
\end{figure*}

In this section we give an overview of the n2EDM baseline design.
Figure~\ref{fig:n2EDM-setup} 
shows the layout of the apparatus positioned in the experimental area south of the UCN source at PSI. We describe
the core n2EDM systems responsible for UCN transport and storage, as well as those for the required magnetic field environment and its control.

%% file: Precession-chambers.tex
\begin{figure}[ht]
\begin{center}
  \includegraphics[width=1\linewidth]{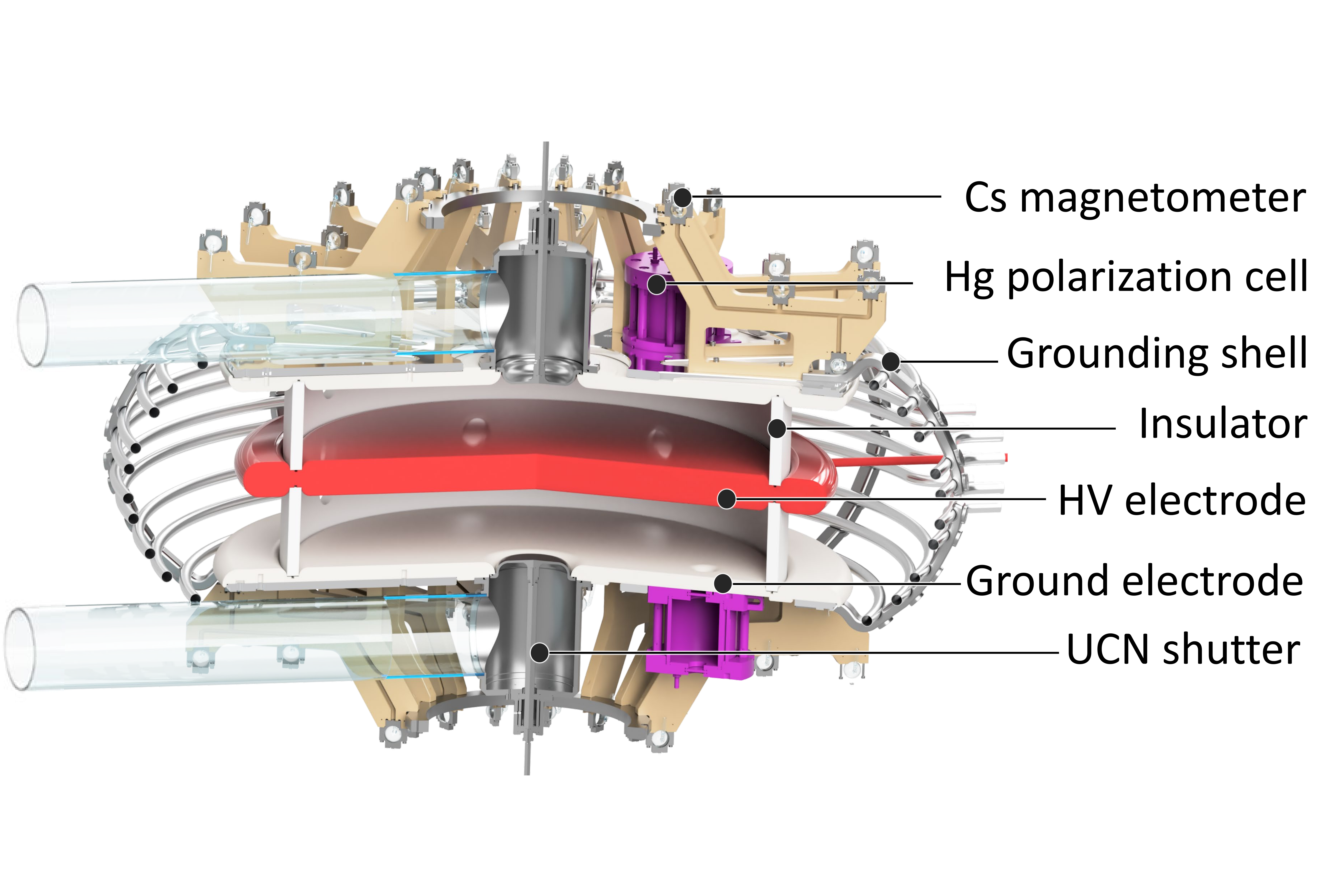} 
  \caption{Central part of the apparatus. Precession volumes are confined by HV and ground electrodes separated by insulator rings. The UCNs enter the chambers via UCN shutters.}
    \label{fig:chambers}
\end{center}
\end{figure}

The two UCN precession chambers lie at the heart of the experiment. They consist of three electrodes separated by two insulator rings, stacked vertically as shown in Fig.\,\ref{fig:chambers}.

\begin{sloppypar}
The precession chambers are cylindrical in shape, of height 12\,cm and inner diameter of 80\,cm, with a design that will allow an upgrade to 100\,cm.  The diameter is increased in comparison to the previous nEDM experiment in order to increase the number of stored neutrons, while the (unchanged) height results from a compromise between the electric field strength and the number of stored neutrons. The dimensions and shape are based on the experience with the previous apparatus and are scaled to the largest possible diameter, currently limited by raw material size and machining capacities.
\end{sloppypar}

The upper and lower chambers are separated by the central HV electrode, which is supplied with $\pm$180\,kV.  The insulator rings  separating the electrodes have a wall thickness of 2\,cm. The design of the electrodes is driven by minimizing UCN losses, optimizing the storage behavior for polarized UCNs and polarized Hg atoms, and withstanding high electric fields (see Sec. \ref{sec:5.1.2}).

The storage of UCNs requires surfaces to have a high neutron optical potential. We use diamond-like carbon (DLC) \cite{Grinten1999,Atchison2006,Atchison2006b,Atchison2007a,Atchison2007b,Atchison2008}, with a measured optical potential $V_{\rm DLC} \approx \SI{230}{neV}$,
as the electrode coating. For the insulator-ring coating it is planned to use deuterated polystyrene (dPS)~\cite{Bodek2008}, with $V_{\rm dPS}\approx \SI{160}{neV}$, or else a coating based on similar deuterated polymers.

The precession chamber stack will be placed inside the vacuum vessel, which is itself manufactured from aluminum with a  usable internal volume of \mbox {1.6 x 1.6 x 1.2 \SI{}{\meter^3}}. The size of the magnetically sensitive area is significantly larger than in any previous or ongoing EDM experiment. This imposes serious challenges in order to ensure a stable and uniform magnetic field environment (see Secs. \ref{sec:5.2}-\ref{sec:5.4}).

%% file: EfieldGeneration.tex
The system of electrodes both confines the UCN precession volumes and provides the electric field in the n2EDM experiment. The central electrode will be connected via a feedthrough in the vacuum tank to the HV power supply, which will provide $\pm{180}$~kV.  The two outer electrodes will be grounded. The optimisation of the electrode design is essential for achieving the highest electric field within the precession chamber, which increases the sensitivity to the neutron EDM. The optimisation process was performed with COMSOL \cite{COMSOL}, a finite element simulation software that allows one to build
complex geometries of different materials and to simulate the resulting fields. The only requirement for the HV system is to provide a stable and uniform 15~kV/cm  electric field, but there are several additional constraints on the design of electrodes.

\begin{itemize}
\item 
The Cs magnetometer arrays are mounted on the outer surfaces of the ground electrodes. The array has components that are sensitive to electric fields, and exposure must be minimized.
\item 
Sharp edges can trigger field emission, limiting the maximum achievable electric field. In particular, the vacuum tank has a structured inside surface so the electric field should be close to zero there. 
\item 
The overall height of the entire precession chamber stack, including the components mounted on the outer surfaces -- namely the UCN shutters, mercury polarization volumes and Cs magnetometer arrays -- must fit in the available space. In total, this means an upper limit of 400 mm from the outer surface of one ground electrode to that of the other.
\end{itemize}

The maximum potential difference attained in the previous nEDM experiment was $\pm{200}$~kV. Using a COMSOL simulation with the nEDM geometry the maximum electric field at any point was found to be 30 kV/cm, which provided a limit for the highest acceptable field in the n2EDM design.

\begin{figure}
\centering
\includegraphics[width=\linewidth]{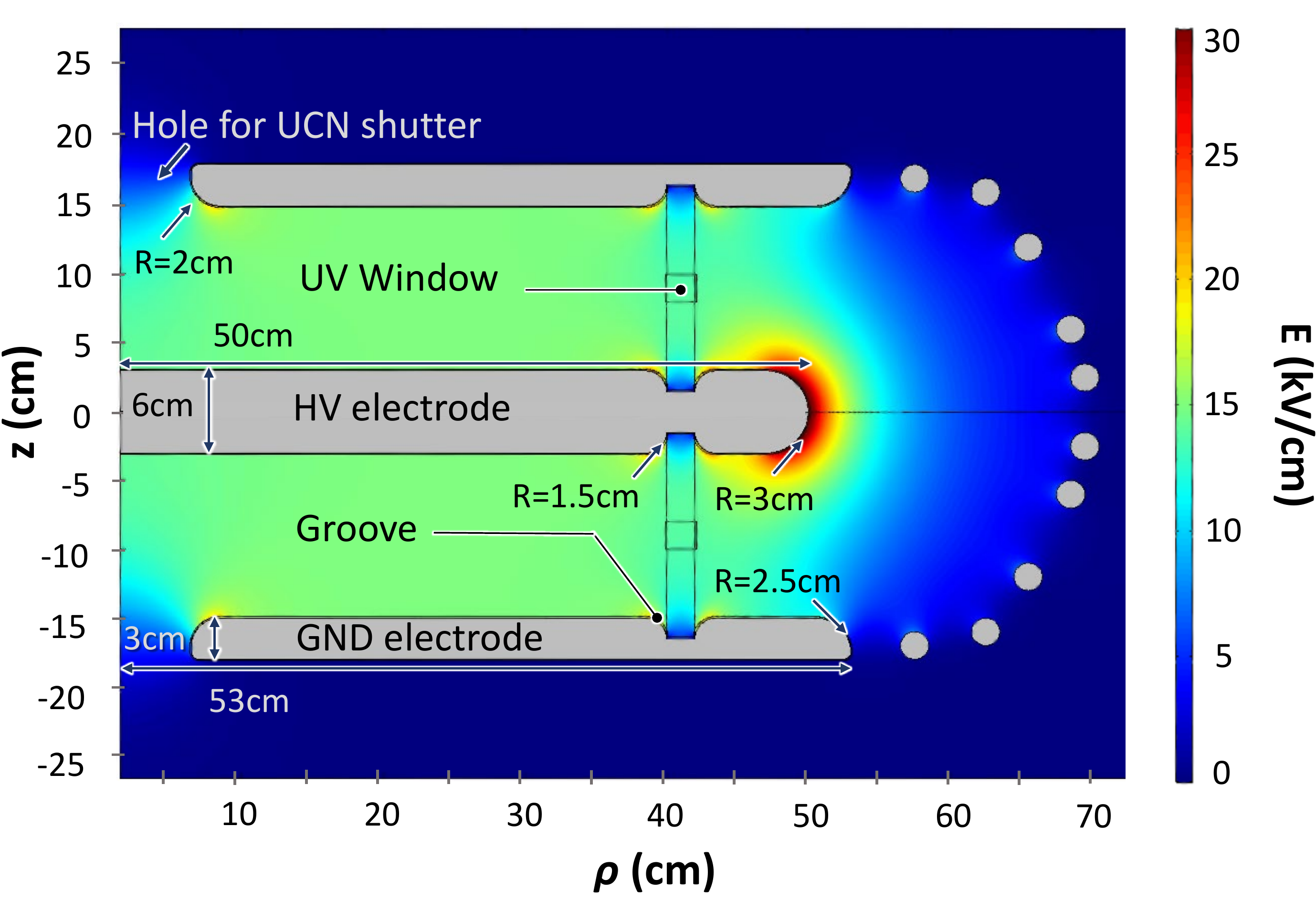}
\caption{COMSOL simulation of the n2EDM optimised geometry. The central (high voltage) electrode is at a potential of 180~kV. The simulation is symmetric on the top and bottom half of the figure.}
\label{n2EDMsim_improved_2}
\end{figure}

In Figure \ref{n2EDMsim_improved_2} the optimized electrode geometry is illustrated. The different parameters of the geometry, listed in the legend, were varied
independently of one another. Particular care was taken during the optimization process to control the electric field strength at the locations indicated by the arrows.

To meet the electric field goal, the thickness
of the HV electrode was set to 6 cm to give a large enough radius on the electrode corona. The diameter of the HV electrode was found to be optimal at 100 cm to separate the influence
of the electric field generated by the corona radius and the presence of the window needed for the UV light beam of the Hg magnetometer. 

The thickness of the ground electrode was determined by the need for moderate radii around the UCN shutter hole (see Fig.~\ref{n2EDMsim_improved_2}), the groove for the insulator ring, and the corona, while still staying within the available space constraints. It was optimized to be 3 cm. The insulator ring groove depth is limited by the available material thickness to 1.5~cm.

A grounded cage of discrete aluminum rods surrounds the central electrodes and insulator in order to minimize the electric fields outside the region of the electrode stack. Several concepts were investigated: an assembly of rings, a fully enclosed shell, or a hybrid of the two designs. Performing COMSOL simulations of the various designs determined that they were all similar in terms of electric field containment. A fully enclosed shell, however, would have caused severe attenuation of the $\pi/2$-flip Ramsey pulses, and therefore a hollow-ring open-cage design was chosen.  This also minimises weight, simplifies the design and installation, and allows better vacuum performance. The simulations optimised the shape and position of the rings while taking into account the need to allow the shell to be split into two halves for mechanical mounting and to have a large enough gap between each ring for effective vacuum pumping and penetration of the Ramsey-pulse fields.

%% file: UCN-apparatus.tex
The n2EDM apparatus is set up at Beamport South of the PSI UCN source. The position of the UCN chambers and the guiding of the UCNs from the UCN source to the chambers was optimized using the MCUCN code \cite{Zsigmond2018}.

\begin{sloppypar}
The UCNs first traverse the open beamport shutter, located just after the superconducting magnet (see Fig.~\ref{fig:SPSA}). Then the UCN guide splits smoothly into two separate tubes, guiding the UCNs towards the two precession chambers through the UCN switch - a major component of the UCN transport system. The switch is located between the superconducting magnet and the MSR, and can operate in filling and counting configurations (see insert in Fig.~\ref{fig:SPSA}).  This is achieved by two movable UCN guides, one for each precession chamber. The UCNs first fill the precession chambers (filling configuration). The chambers are then closed by UCN plugs connected to two shutters: one on the top of the upper chamber, the other below the lower chamber. When emptying the precession chamber, the switch connects the same UCN guides used during filling to the spin-sensitive detection system (counting configuration). The third (test) mode of the switch  permits the guiding of UCNs directly from the source to the detectors in order to monitor the UCN source performance.
\end{sloppypar}
The switch design was based on the common theme of maximizing the transmission efficiency. This resulted in stringent specifications such as the necessity to maintain the same total cross section for UCNs throughout their path in the apparatus, the optimization of the number of bends and the maximization of their radius of curvature, and the minimization of gaps between guides as far as reasonably possible with a target of 0.1 mm.

 The UCN guides are made of glass tubes with an inside diameter of 130~mm and a NiMo coating with ultralow surface roughness \cite{Bison2020}. The manufacturing of the UCN guides follows the process developed and successfully employed during the construction of the PSI UCN source \cite{BLAU2016}. The same process was used to produce the guides of the nEDM apparatus, where UCN transmissions above 97$\%$ per meter were achieved.

%% file: UCN-detection.tex
At the end of the precession time, UCNs stored in the upper and lower precession chambers are released and directed towards two identical spin-sensitive counters. There, UCNs are counted as a function of their spin state, behind distinct simultaneous spin analysers \cite{Hel15}.

\begin{figure}[ht]
\begin{center}
  \includegraphics[width=1\linewidth]{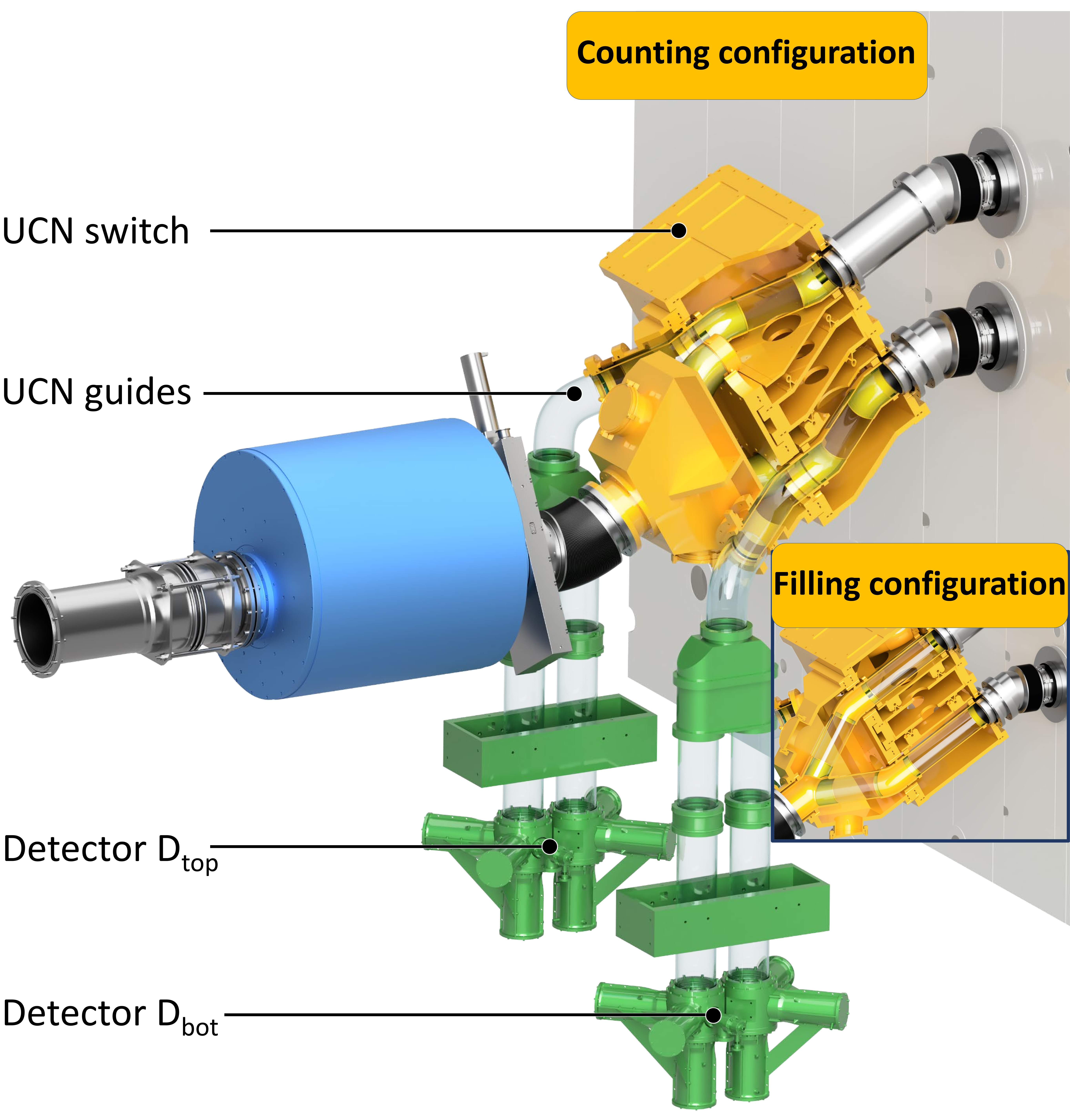} 
  \caption{Scheme of the beamline outside the MSR. The switch with movable UCN guides can operate in three modes: when UCNs are filling the precession chambers (filling configuration); when UCNs from the precession chambers are directed towards the two simultaneous spin-sensitive detectors (counting configuration); and a third (test) mode of the switch (not shown here) that permits guiding of UCNs directly from the source to the detectors in order to monitor the UCN source performance.}
  \label{fig:SPSA}
\end{center}
\end{figure}

\begin{figure}[ht]
\begin{center}
  \includegraphics[width=0.8\linewidth]{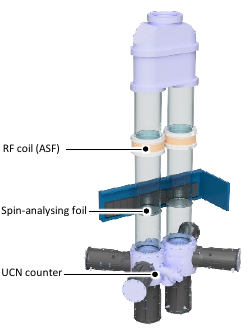} 
  \caption{Simultaneous spin analyzer. Each arm is equipped with an adiabatic spin-flipper (RF coil), a spin-analyzing foil and a UCN counter. }
  \label{fig:USSA-alone}
\end{center}
\end{figure}

A UCN may be detected either in a spin-up state (spin parallel to the main magnetic field) or in a spin-down state (spin antiparallel to the main magnetic field).  In order to detect simultaneously UCN of both spin states, a custom device consisting of two vertical arms, each arm being dedicated to the analysis of one spin state, has been designed and built. The first element of the device splits the UCN guide into two arms (see the Fig.~\ref{fig:SPSA} for the detailed geometry). Each arm consists of an adiabatic spin-flipper (ASF), an analyzing foil and a UCN counter (see Fig.~\ref{fig:USSA-alone}).  The adiabatic spin-flipper consists of a shielded RF coil installed upstream of the foil. Its operating principle is described in~\cite{Grig_97}. Spin-up UCNs are counted in the arm where the ASF is on, and spin-down UCNs in the arm where the ASF is off. The role of each arm is regularly reversed in order to minimize  systematic effects. Ultracold neutrons entering the wrong arm with respect to their polarization (around $50 \%$ of the incoming UCNs) are reflected on the foil, and obtain a second chance to be detected in the correct arm. The internal shape of the unit is specifically designed to guide the reflected UCNs from one arm to the other and hence to improve the efficiency of the spin analysis.

The spin analysis itself is performed by transmission through (or reflection from) an iron foil magnetized to saturation~($B_{sat}~\approx~2~T$) and located below the ASF. Ultracold neutrons are able to cross the foil if their kinetic energy associated with motion perpendicular to the foil ($E_{\perp}$) is larger than
\begin{equation}
\label{spin_analysis_principle}
U = V_{Fe} + \vec{\mu_{n}} \cdot \vec{B_{sat}} = V_{Fe} \pm \lvert \mu_{n} \rvert B_{sat},
\end{equation}
where $V_{Fe}$ is the Fermi potential of iron, $\mu_{n}$ is the neutron magnetic moment, and $B_{sat}$ the magnetic induction inside the iron layer. The $\pm$ sign reflects the spin dependency of the magnetic interaction and stands for the spin-up and spin-down components, respectively.
\begin{sloppypar}
Numerical estimates performed with the Fermi potential of the iron foil ($V_{Fe} = ~210$~neV) and the magnetic potential energy ($\mu_{n} ~ B_{sat} = 120$~neV) show that UCNs with energy $E_{\perp} < 90$~neV are reflected on the foil whatever their spin state while UCNs with $E_{\perp} > 330$~neV are transmitted through the foil. Between $90$~neV and $330$~neV the spin analysis (discrimination) is operational: spin-down UCNs are able to cross the foil while spin-up UCNs are reflected. Finally, the number of UCNs of a given spin state is counted with the counter installed below the foil.
\end{sloppypar}
It is important to emphasize that the height between the precession chambers and the spin analyzing foils is a critical parameter. The maximum kinetic energy of UCNs exiting the precession chambers is given by the Fermi potential of the insulator ring coating, $V_{DPS} = 165$~neV, as shown in Fig.~\ref{fig:MC_spectrum}. As a result, the height difference between the precession chambers and the spin analyzing foils should not exceed $165$~cm in order to prevent UCN exceeding $330$~neV - the maximum analyzable kinetic energy of the UCNs.

\subsubsection{UCN counter}
 The UCN counter is a fast gaseous detector \cite{Sae19}. The neutron detection is based on scintillation occurring in a gas mixture of $\rm{^{3}He}$ and CF$_4$. Neutrons are captured by $\rm{^{3}He}$ nuclei through the reaction
\begin{equation}
{\rm  n +~ ^{3}He  \longrightarrow p~(0.57~MeV) +~ ^{3}H~(0.19~MeV)}
\label{neutron_capture}
\end{equation}
and the emitted proton and triton cause scintillation of the CF$_4$ molecules. The scintillation decay time is only about \SI{10}{ns} \cite{Leh15}, which provides a high count-rate capability up to a few $10^6$\,counts/s. The scintillation light is detected by three photomultiplier tubes working in coincidence. 
The partial {\rm $^3$He} gas pressure required to fully stop the UCN beam is low, between 10 and \SI{20}{mbar}. The gas mixture is completed with CF$_4$, and the detector is sealed. In order to reduce the probability of gamma interaction on CF$_4$ molecules as well as UCN upscattering, the partial CF$_4$ gas pressure is reduced to $P$(CF$_4$)= 400\,mbar.

%% file: Magnetic-shielding-intro.tex
\begin{figure}[htb]%
\centering
\includegraphics[width=1\columnwidth]{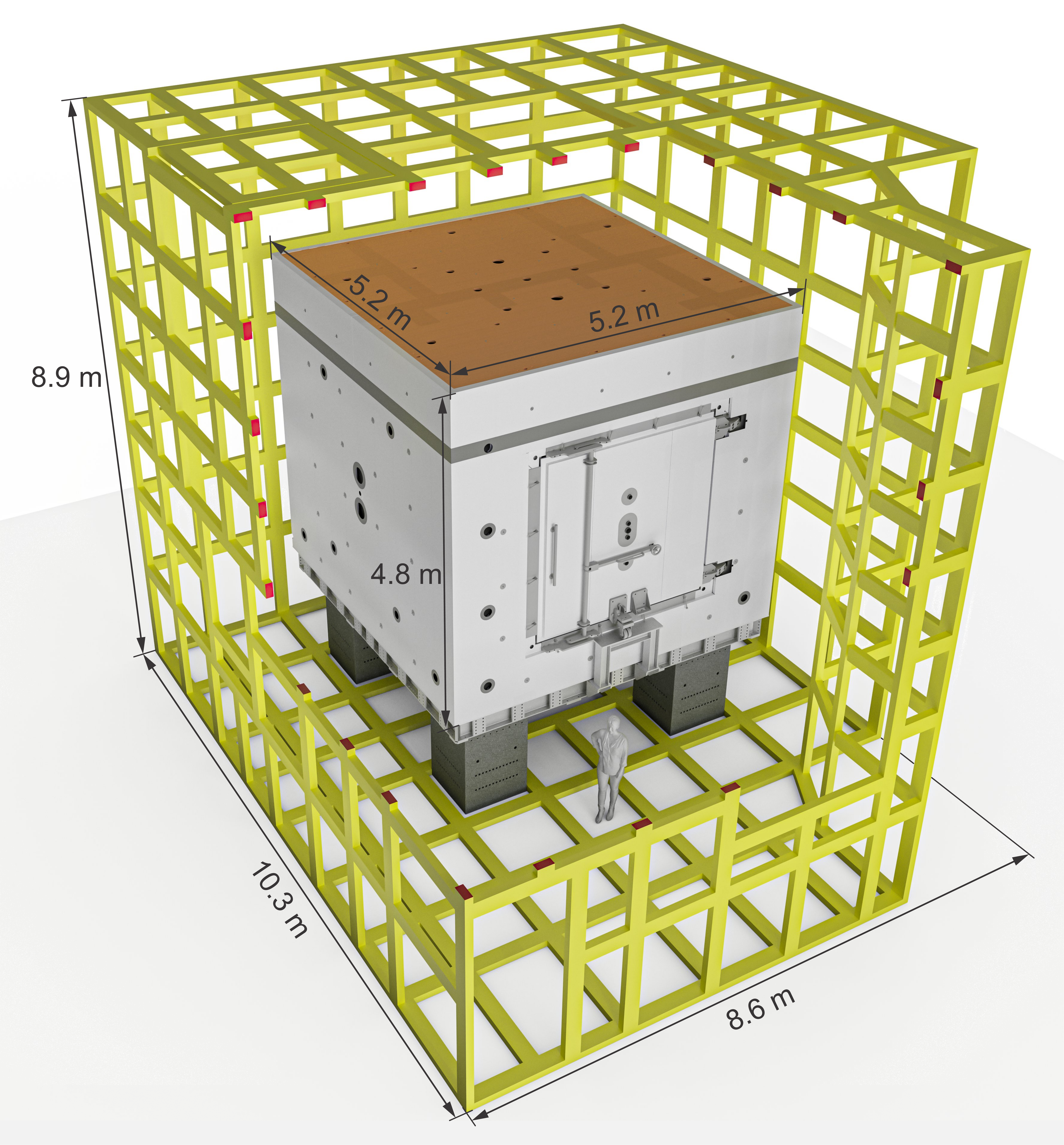}%
\caption{Magnetic shielding. Passive shielding is provided by a large cubic magnetically shielded room (MSR). Active shielding consists of actively-controlled coils mounted on a grid around the MSR.}
\label{fig:Magnetic-shielding}%
\end{figure}

A magnetically stable and uniform field is mandatory in order to exploit fully the statistical reach of the experiment. This is achieved by means of passive and active magnetic shielding, which are illustrated in Fig.~\ref{fig:Magnetic-shielding}.

Passive magnetic shielding is provided by a large cubic magnetically shielded room (MSR). Its performance in the low frequency range ($<5$Hz) is improved by the active magnetic shield (AMS), which consists of a system of actively controlled coils. The AMS is mounted on a grid around the MSR, and it compensates external magnetic field changes at the outermost mu-metal layer of the MSR.

%% file: MSR.tex
The MSR, which was built in partnership with the company VAC\footnote{VAC GmbH, Hanau, Germany (https://www.vacuumschmelze.com){\label{VAC}}}, provides the magnetic environment for the central part of the experiment.  It suppresses external, quasi-static fields by roughly five orders of magnitude: a quasistatic shielding factor of better than 70’000 at 0.01 Hz was specified. After degaussing, the innermost central space was required to have a residual magnetic field smaller than 0.5 nT
and a magnetic field gradient of less than 0.3 nT/m.

The MSR design is based on the magnetic shielding requirements alongside the need to house the n2EDM apparatus in an optimal fashion. It incorporates six cubic layers of mu-metal and one additional layer of aluminum for radio-frequency shielding. 
\begin{figure}[htb]%
\centering
\includegraphics[width=1\columnwidth]{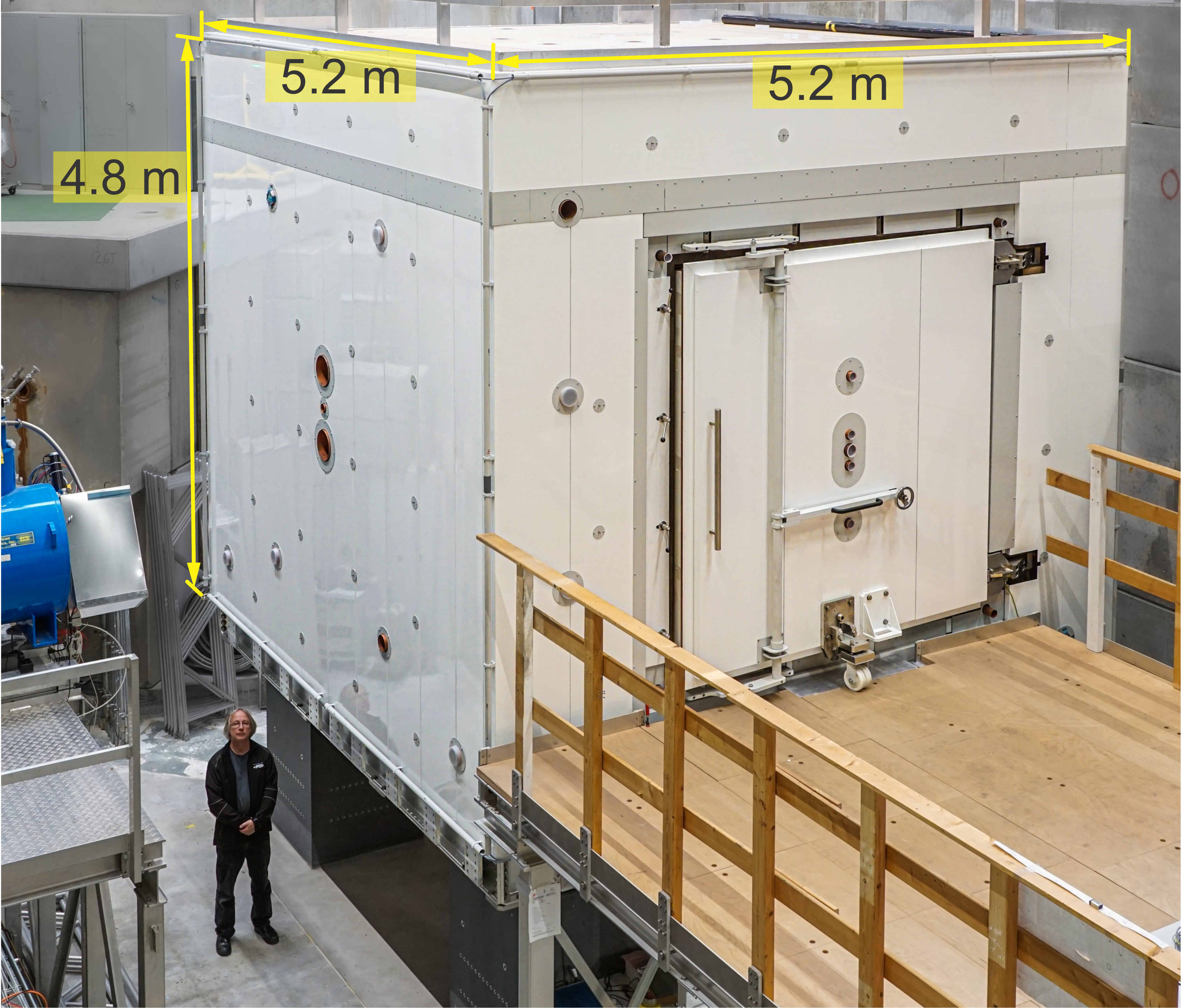}%
\caption[View of the MSR]
{Photo of the magnetically shielded room with indicated outer dimensions
}%
\label{fig:MSR}%
\end{figure}
\begin{sloppypar}
The MSR is composed of two cubic-shaped nested mu-metal rooms, referred to as ``inner'' and ``outer''. The overall outer dimensions are given by a footprint of about 5.2\,m~$\times$~5.2\,m, as indicated in Fig.~\ref{fig:MSR},
and a height of 4.8\,m.  It has a total weight of about 50\,tons.
\end{sloppypar}
The outer room has a two-layer mu-metal wall as well as an
8\,mm thick aluminum layer that serves as an eddy-current shield.
The inner room consists of a four-layer mu-metal shield, 
where the innermost layer is constructed from a  
specially selected high-permeability metal.
It is a cube with inner dimensions of 293\,cm on each axis.
A small but accessible intermediate space between the inner and the outer rooms
creates a useful and moderately magnetically shielded space
close to the central apparatus.
There, sensitive electronics for signal amplification, shaping and measurement
may be located, e.g.\ pre-amplifiers for magnetometers, or precision-current sources.
All parts of the inner cabin were tested for magnetic impurities at PTB's BMSR-2 facility~\cite{Bork2001}.
Each layer of the MSR is equipped with a separate set of degaussing coils.%
The ability to degauss each layer in this manner helps to provide uniform residual magnetic fields.

An air-conditioned thermal enclosure maintains the outer MSR at a temperature stable to 1$^\circ$C, and the innermost magnetically shielded room together with the apparatus has a temperature stability of better than 0.1$^\circ$C. This prevents thermal gradients across the MSR or temperature changes that would otherwise lead to magnetic-field changes through thermal expansion of materials and/or thermal currents.

The position of the MSR was chosen so as to allow for a straight path for UCNs from the source in order to minimize transport distance and losses.

%% file: AMS.tex
The n2EDM experiment is located in the vicinity of other facilities generating variable magnetic fields of similar strength to the Earth's own field. Our experiment therefore experiences a magnetically noisy environment, subject to changes in the ambient magnetic field of up to tens of \SI{}{\muT} on timescales from minutes to hours. In order to realize the required magnetic field conditions in the inner part of n2EDM, shielding from external magnetic field changes is of key importance. 

\begin{sloppypar}
 The stability of the magnetic field within the MSR is directly dependant on the stability of the field around the MSR. There are two complementary mechanisms for this. First, attenuation of external field fluctuations before they affect the MSR will improve the overall shielding factor multiplicatively. Second, avoiding changes in the magnetization of the outer passive shielding layers eliminates long-term drifts of the magnetic field inside the MSR.

In order to provide stable magnetic-field conditions around the MSR, the Active Magnetic Shielding (AMS) system was designed. It consists of a system of actively controlled coils.

Before the n2EDM construction, the magnetic field in the empty experimental area was mapped in 3D several times with different combinations of nearby superconducting magnets from other research installations switched on or off. It was found that the external field can be described with a precision of approximately \SI{1}{\micro T} using a set of eight harmonic polynomials: three homogeneous components with five first-order gradients were able to match the reproducibility of the measurement. The measured field values were up to \SI{50}{\micro T} in each spatial direction, and the gradients were up to  \SI{5}{\micro T \per m} in each of the five linear components, thus specifying the required field strength for each component needed fully to compensate the large external field changes. The \SI{1}{\micro T} field mapping accuracy was chosen as a target for the field-compensation accuracy.
\end{sloppypar}

The space available for the placement of coils is  limited, being approximately 8~m in each dimension. A complex coil geometry is required to produce the desired compensating field. Additionally, coil elements cannot be placed at arbitrary locations due to conflicts with other parts of the apparatus as well as other practical considerations. An algorithm was developed to allow the design of geometry-confined coils that would produce arbitrary field configurations \cite{Rawlik2018}. This allows the placing of current-carrying wires along a predetermined but not completely uniform grid that is mounted on the inside of the thermal shell around the MSR (see Fig.~\ref{fig:Magnetic-shielding}).

%% file: Magnetic-field-generation.tex
\begin{sloppypar}
Ramsey's method of oscillating fields requires polarized UCNs, a static B$_{0}$ field and two RF field pulses. The UCN polarization is achieved with a 5\,T superconducting solenoid. The static field is mainly generated by a single large coil (the ``B$_{0}$ coil'') and its coupling to the innermost layer of the shield.  An array of 56 independent correcting coils is used  to tune the field to the required level of uniformity. Seven coils produce specific gradients that play an important role in the measurement procedure. Finally, the RF pulses of the Ramsey cycle are generated by RF coils installed inside the vacuum tank. 
\end{sloppypar}

\subsubsection{B$_{0}$ field generation}
\begin{figure}[htb]
\centering
\includegraphics[width=\columnwidth]{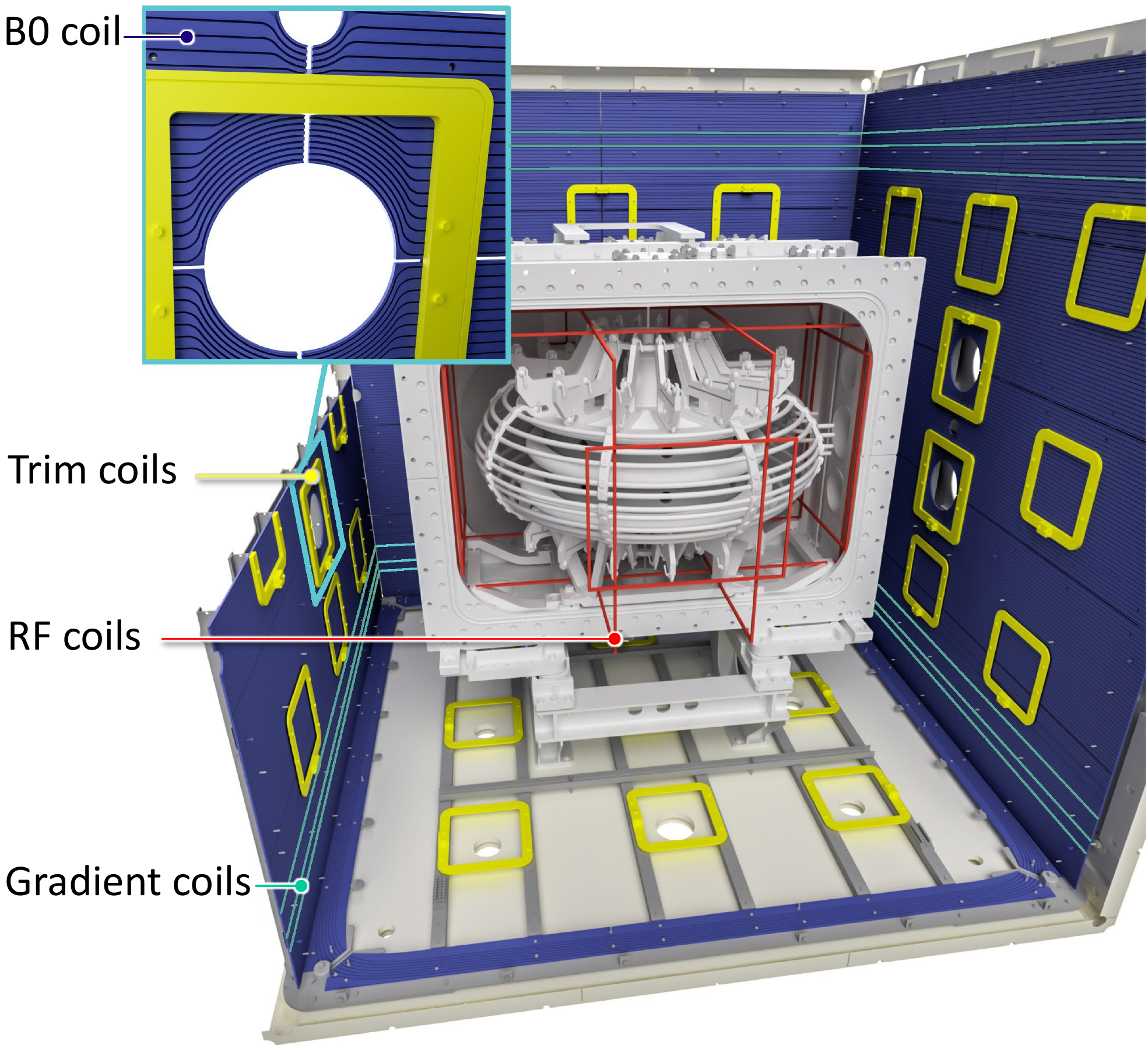}
\caption{The coil system inside the MSR (close to the innermost layer of the shield). For clarity, only the G10 coil is shown as an example of a gradient coil.}
\label{fig:coils_syst}
\end{figure}

A B$_{0}$ field of  \SI{1}{\micro T} is produced by a vertical cubic solenoid complemented with two sets of seven horizontal loops symmetrically located on the top and the bottom. These end-cap loops help to suppress the field nonuniformities induced by the finite size of the magnet. The coil is fixed on a cubic support outside the vacuum tank, located at about 10~cm from the innermost mu-metal layer of the shield (Fig. \ref{fig:coils_syst}).  On one side, a large rotating door of  size of 2~m x 2~m allows access to the central part of the experiment. 
 
\begin{figure}
\centering
\includegraphics[width=1.0\linewidth]{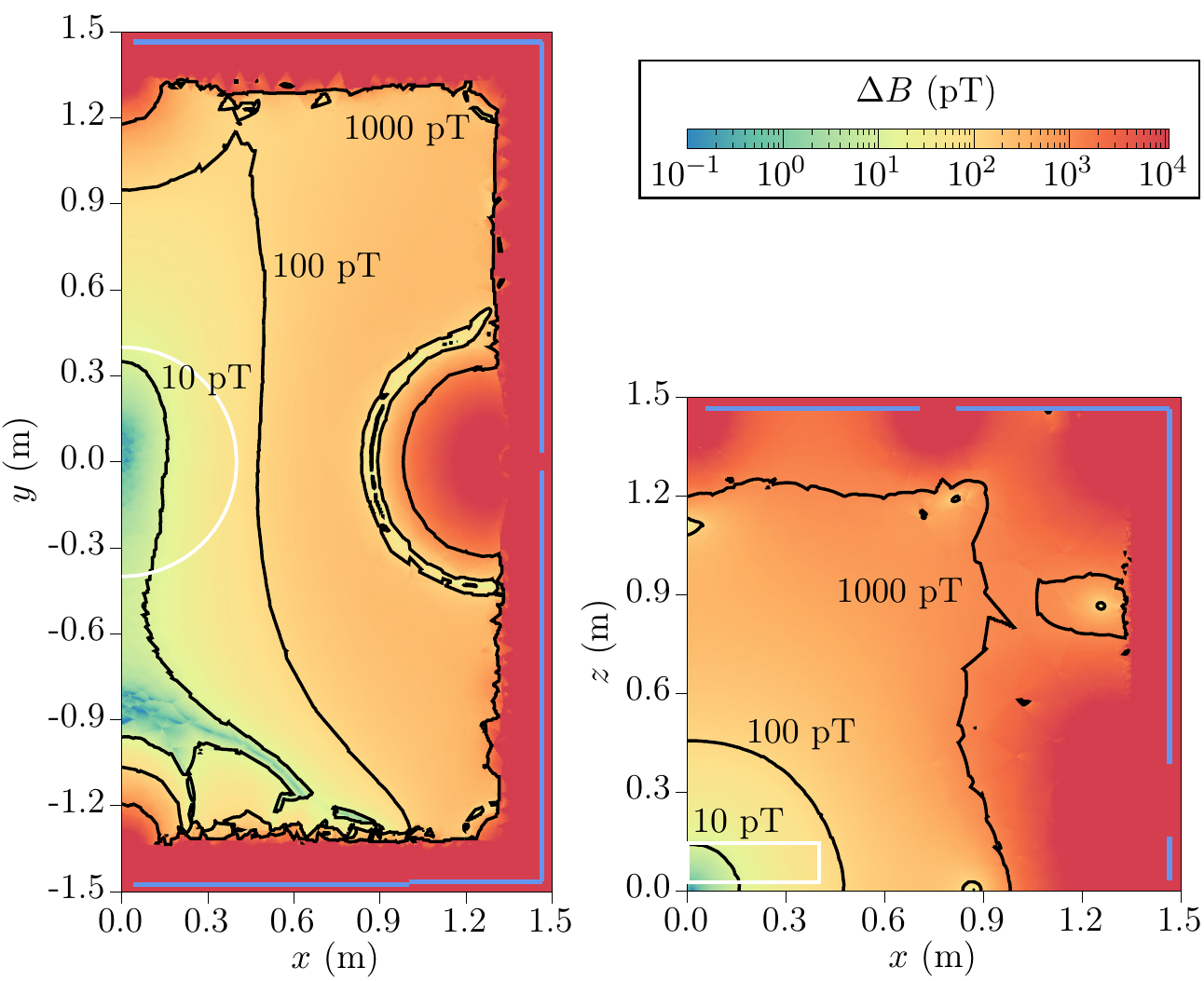}
\caption{COMSOL B-field simulation, showing 
field variations produced by the B$_0$ coil in the horizontal plane
at $z = 0$\,m (left) and the vertical plane $y=0$\,m (right). The rainbow scale corresponds to $\log_{10}(\Delta B)$.
The contours of the precession chambers, which have a diameter 
of 80\,cm, are shown in white, while the innermost layer of the MSR is shown in blue on the edges of the plots. The black bold contour lines highlight $\Delta$B = 10\,pT, 100\,pT, and 1000\,pT. The simulations are performed with the dimensions of the innermost layer of the MSR measured in situ. The $\mu_r$  values for the mu-metal material were communicated by the producer VAC\footref{VAC} and are proprietary information.}
\label{fig:fieldvar}
\end{figure}

The field produced by the B$_{0}$ coil and its coupling to the innermost layer of the shield
was simulated with the COMSOL software package. For a current of \SI{12}{mA} through the B$_{0}$ coil the magnitude of the field at the center is about \SI{1}{\micro T}, with approximately one third arising from the magnetization of the innermost mu-metal layer. The B$_{0}$ field is expected to increase a few percent after equilibration (degaussing within a non-zero surrounding and/or inner field). The field variations around the central value $B(\vec{0})$ have been estimated by computing $\Delta B(\vec{r}) = {|\vec{B}(\vec{r}) - \vec{B}(\vec{0})|}$. The variations, shown in Fig.~\ref{fig:fieldvar}, do not exceed 100\,pT in a large volume that includes the precession chambers. The observed nonuniformities come from the openings that are present in the MSR walls as well as from a recess of the MSR door with respect to its surrounding wall.

From the simulated field maps it is also possible to estimate the field uniformity $\sigma(B_z)$ in the region of the precession chambers. The achieved uniformity, $\sigma(B_z)$ = \SI{16}{\pico T}, is well below the requirement of 170 pT (see Sec.~\ref{sec:3.4}). The requirement is also fulfilled when the magnetic field is increased to the ``magic'' value of ~\SI{10}{\micro T}.


Such a high degree of uniformity is very sensitive to the reproducibility of the equilibration, and also to imperfections in the shielding material due to the construction from single sheets and the coupling of the B$_{0}$ field with the innermost shielding layer. Mechanical alignment is also critical; for instance, a vertical misalignment of one  millimeter between the entire $B_0$ coil and the MSR triples the field non-uniformities as compared to the ideal symmetric case. While the B$_{0}$ coil will be installed with great care, unavoidable imperfections will remain. In order to suppress the induced nonuniformities a set of 56 independent rectangular trim coils is used. They are fixed on the same cubic support as the B$_{0}$ coil, with nine or ten coils per side.  
These can produce all generic field gradients up to the 5th order.

With a B$_{0}$ field of \SI{1}{\micro T}, a stability of a few dozen fT on the timescale of a minute is 
required to achieve an efficient operation of the Hg co-magnetometer (see Sec.~\ref{sec:5.4.1}). The coil is therefore powered by an ultra-stable current source with a relative stability of a few $\times 10^{-8}$. 

\subsubsection{Generation of specific gradients}

To control the magnetic-field gradient during data taking, as well as to study various systematic effects, seven additional field and gradient coils are mounted to the B$_{0}$ coil support. 

A constant offset value for the three field components can be generated by three independent coils, with the underlying uniform B$_{z}$ component being produced by the B$_{0}$ coil. 
The linear gradients of the B$_{z}$ component, ${\partial B_z}/{\partial x}$, ${\partial B_z}/{\partial y}$, and ${\partial B_z}/{\partial z}$, can also be generated. Besides the optimisation of $\alpha$ (see Sec.~\ref{sec:3.4}), these are used to monitor and/or control {\it in situ} the positioning of every Cs magnetometer at the mm level (see Sec.~\ref{sec:5.4.2}). The field measured by each magnetometer probes their position in the three directions. The ${\partial B_z}/{\partial z}$ field is also used to perform the vertical tuning of the B$_{z}$ component in order to fulfill the top/bottom matching condition (see Sec.\ref{sec:3.4}). The power supply of the $\partial_{z}$B$_{z}$ coil allows variation in the vertical linear gradient with a resolution of 0.01 pT/cm.

Finally, it is important to control the gradients responsible for the most significant systematic effect, the motional EDM. Therefore, in addition to the ${\partial B_z}/{\partial z}$ gradient, two other gradients that are of particular interest, $G_{2,0}$ and $G_{3,0}$,  are produced by two additional independent coils. 

\subsubsection{RF field generation}
Rotating fields perpendicular to the B$_{0}$ field are used at the beginning and end of the Ramsey cycles to flip the spins of the UCNs and of the Hg atoms into and out of the horizontal plane.
These fields, of frequencies $\sim$\SI{30}{\hertz} and $\sim$\SI{8}{\hertz} respectively, are generated by the eight RF coils: four along the $x$ axis and four along the $y$ axis. 
The coils are located inside rather than outside the vacuum tank because of the pronounced damping that would be caused by the thick aluminum walls.
Finite-Element Method simulations using ANSYS ~\cite{ANSYS} were performed to study the impact of the electrodes and other conductive components close to the coils, and to optimize the setup.
The simulated spatial homogeneity inside the precession chamber for the UCN pulse is $\sigma_{\mathrm{RF}}<$ ~\SI{120}{\pico\tesla}, well below the \SI{170}{\pico\tesla} upper limit requirement (see Eq. \ref{Eq:FieldUniformityReq}).

\subsubsection{UCN spin transport}
The 5\,T superconducting magnet (SCM) acts as an almost perfect polarizer ($P>99$\%), producing an axial (horizontal) polarization. The transport of the UCN spin from the SCM to the precession chambers has two parts: outside the MSR, the SCM fringe field is sufficiently large to fulfil the adiabatic transport condition. At the MSR, the field is rotated from axial (horizontal) to transverse (vertical) and is adapted to the B$_{0}$ field strength between the shield entrance and the inner cabin of the MSR.

%% file: Magnetometry-concept.tex
Statistical and systematic uncertainties in a neutron EDM experiment depend on the homogeneity and the stability of the main magnetic field $B_0$, in which the neutrons precess.
The overall goal of the magnetometry systems in the n2EDM experiment is to ensure that all magnetic-field-related uncertainties are small compared to the fundamental statistical uncertainty given by the UCN counting.
The two major magnetometry systems are the Hg co-magnetometer and an array of Cs magnetometers.

The magnetic-field information provided by the magnetometers of n2EDM 
is used in three sequential phases: before, during, and after the actual measurement. 
In an initial phase, before the neutron measurements start,
information about the magnetic field has to be acquired 
in order to provide a magnetic environment that allows for long neutron precession times.
Magnetic-field inhomogeneities increase the neutron's depolarization rate and thus lead to a smaller visibility $\alpha$, which in turn decreases the statistical sensitivity 
(see Eq.\,\eqref{Eq:stat_error_fn}).
Since it is impossible to correct for a faster loss of neutron spin polarization after the measurement,
the $B_{0}$ field must be sufficiently tuned for a high visibility $\alpha$.
We plan to use a magnetic-field mapper to study the distribution of the field inside the MSR in dedicated measurements once per year, usually during the accelerator shutdown period.
We will also employ the array of Cs magnetometers (see Sec.~\ref{sec:Cs-magnetometry}) to fine-tune the field homogeneity during UCN data taking after each change of the magnetic-field polarity.
This concept proved to be successful in our previous nEDM experiment and routinely provided neutron spin relaxation times of more than 1200\,s\@ \cite{CsM_PRA2020}.

During the data taking with neutrons, magnetic-field information is essential in order to keep the neutrons in a magnetic resonance condition.
Ramsey's method provides a unique sensitivity 
to the Larmor precession frequency only if the final measurement of the neutron spin is on the steep slope of the interference pattern. 
In order to stay at these positions, it is necessary to correct for  drifts of the magnetic field by adjusting the rf-pulse frequency or the relative phase between first and second pulse.
In the previous experiment the  field value measured in the previous Ramsey cycle using the Hg \comagnetometer\  (see Sec.~\ref{sec:Hg-magnetometry}) was used to compute the frequency of the Ramsey pulses.
In n2EDM we will use the magnetic-field values deduced from the two Hg magnetometers to stabilize the working points. 
To achieve that, two parameters need to be controlled, for example the
$\partial B_z / \partial z$ gradient and the frequency of the Ramsey pulses.
We plan to keep the parameters that influence the working points static during a Ramsey cycle. 
They are chosen before the cycle starts based on the information from the previous Ramsey cycle.
A dynamic compensation that uses information gained during the current  cycle to update those parameters will only be considered if EDM sensitivity is lost in significant amounts due to drifting working points.

Last but not least, the entire time resolved, synchronously recorded information on the magnetic field will be used in the offline analysis to correct for the effect of magnetic-field fluctuations on the nEDM result.
All magnetometer systems are involved in this process.
The Hg co-magnetometer provides the primary magnetic reference measurement for the neutrons that helps us to distinguish changes in the neutron spin precession frequency due to magnetic-field changes or due to a possible EDM.
A second magnetic reference is provided by the Cs magnetometers that surround the two neutron volumes.
All magnetometers will be used to determine magnetic-field gradients that cause systematic errors in the nEDM measurement.

%% file: Hg-magnetometry.tex
The n2EDM Hg magnetometry will follow the 
same operation principle as the original Hg co-magnetometer used in our previous 
experiment and introduced by the RAL/Sussex collaboration~\cite{Green1998}.
Atomic vapor of $^{199}$Hg is polarized by optical pumping in a 
polarization cell placed on each of the  ground electrodes of the precession chamber stack.
The vapor in the polarization cell enters through a small valve into the precession chambers once they are filled with UCN. 
The application of a $\pi/2$ pulse starts the precession of the $^{199}$Hg spins in the 
same volume as the neutrons. 
During the precession time, a photodetector records the power of a beam of resonant light 
traversing the chamber, which is modulated at the Larmor frequency by the interaction of 
the probe beam with precessing Hg atoms.
%

\begin{figure}
\centerline{\includegraphics[width=\columnwidth]{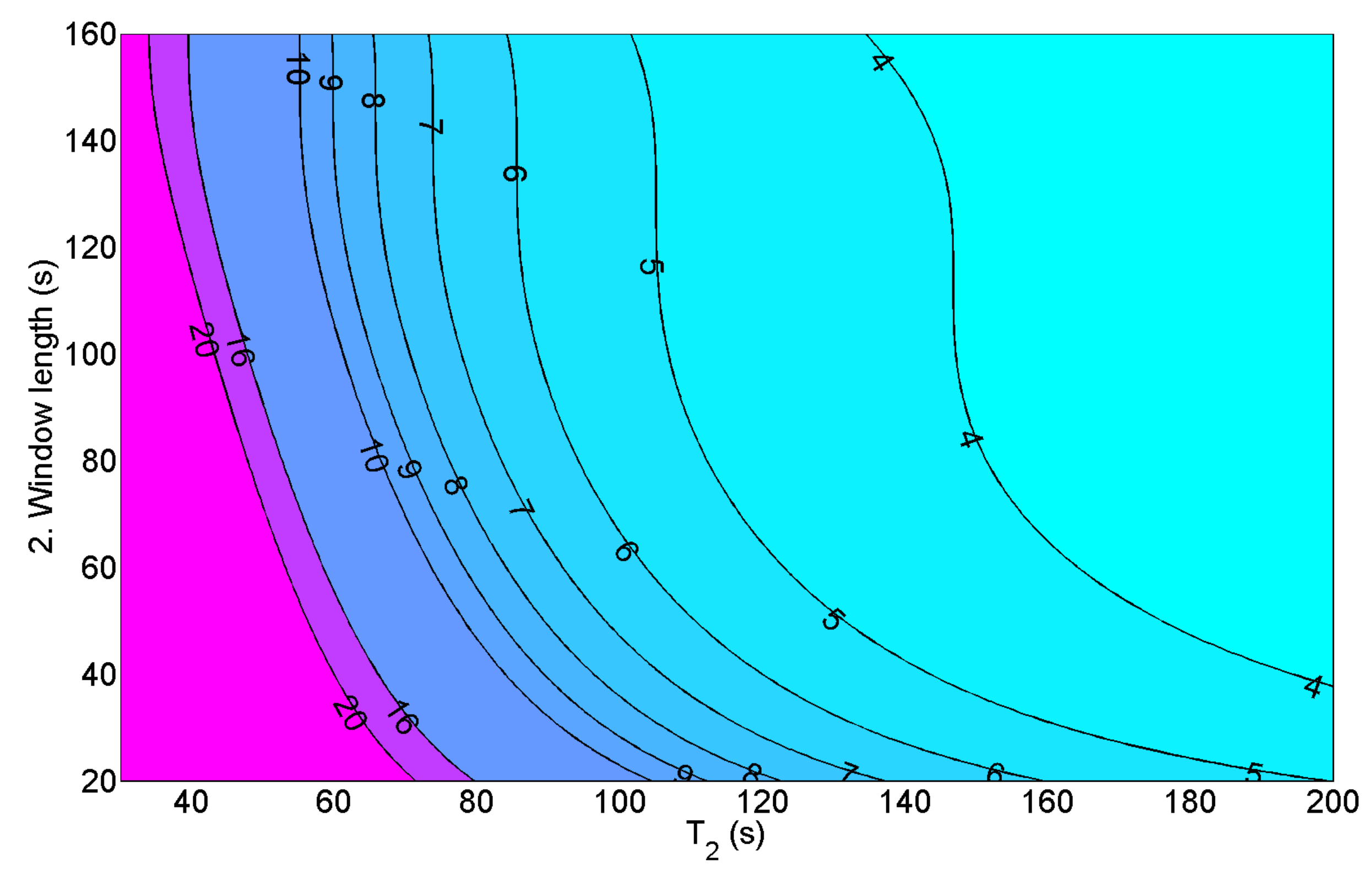}}
\caption {\label{fig:Hgperformance}
Statistical magnetometer uncertainty based on a signal/noise measurement using laser light to polarize and probe the Hg atoms. The values are given in \SI{}{fT} as a function of the Hg spin depolarization time ($T_2$) and the length of the signal windows used for the analysis. 
}
\end{figure}

The sensitivity requirement per cycle of 25~${\fT}$ (0.03 ppm at \SI{1}{\micro T}) 
was already demonstrated for a 180~s precession time  with our previous apparatus as part of our Hg R\&D program~\cite{Ban2018}. 
This was made possible by replacing the Hg discharge-lamps used so far for the probe beam by a 
tunable UV laser. 
The analysis only uses data during two analysis windows at the beginning and the end of the signal. 
While the first window is always \SI{20}{s} long the length of the second window (and correspondingly the amount of data used) can be varied. 
Figure \ref{fig:Hgperformance} shows that statistical uncertainties are smaller than the required \SI{30}{fT} for most combinations of Hg $T_2$ time and window length. 
Even stricter requirements, which might be necessary for a potential upgrade of n2EDM, can be fulfilled if the same performance as in our former experiment with $T_2$ times around \SI{100}{s} can be achieved. 
Our experience shows clearly that long $T_2$ times of the Hg atoms can only be achieved  if the precession chambers are periodically discharge cleaned. 

%% file: Cs-Magnetometry.tex

Cesium magnetometers were introduced to the nEDM experiments
as auxiliary magnetometers in order to monitor the main magnetic field and its gradients.
We plan to mount a set of such magnetometers in close vicinity to the neutron precession
volume (see Fig.~\ref{fig:chambers}).
The sensors of choice are optically pumped magnetometers (OPM) that detect the spin precession of Cs atoms
and so gain an optical signal that is modulated at the Larmor frequency~\cite{Aleksandrov2006}.
The basic sensor principle has been known for more than 50 years~\cite{Bloom1962} and was initially studied using discharge lamps as light sources.
Since affordable lasers for the required near-infrared wavelengths have become available,
this measurement principle has gained a renewed interest that has led to the development of many different OPM variants~\cite{Budker2007}.
The OPMs in previous nEDM experiments \cite{Serebrov2014, CsM_PRA2020} used a mode of operation, called the $M_x$ mode, that is sensitive to the magnitude of the magnetic field.
%

In past experiments \cite{CsM_PRA2020} it was realized that the accuracy of the Cs sensors is the most critical factor limiting the usefulness of the measurements.
Accurate sensor readings are necessary in order to extract information about the field gradients, which have to be known on an absolute scale.
As a consequence our research and development efforts in Cs magnetometry has for  several years focused on sensor stability and accuracy.
We have developed highly stable vector magnetometers~\cite{Afach2015OExpress} and
magnetically silent (all optical) magnetometers~\cite{Grujic2015}.
These designs are based on a pulsed approach that allows us to monitor the free-spin precession, in contrast to the $M_x$ mode which is based on a continuously driven magnetic resonance.
The free precession has significant advantages for the sensor accuracy, since it avoids a whole class of systematic effects.
There is, however, a class of systematic errors related to the complex atomic spin structure that is present in all tested magnetometer schemes.
The shift is significantly smaller if the magnetometer is operated with linearly polarized light, which creates and detects atomic spin alignment, in contrast to circularly polarized light, which interacts with atomic spin orientation.
The offset is further suppressed if the light is propagating parallel to the magnetic field, since there is no first-order dependence on misalignment in this geometry.
A prototype of a scalar magnetometer that combines the features mentioned above has been realized, showing a statistical uncertainty of \SI{1}{pT/Hz^{1/2}}.  This will be sufficient to meet the requirements of n2EDM.

\begin{figure}
  \centering
  \includegraphics[width=8cm]{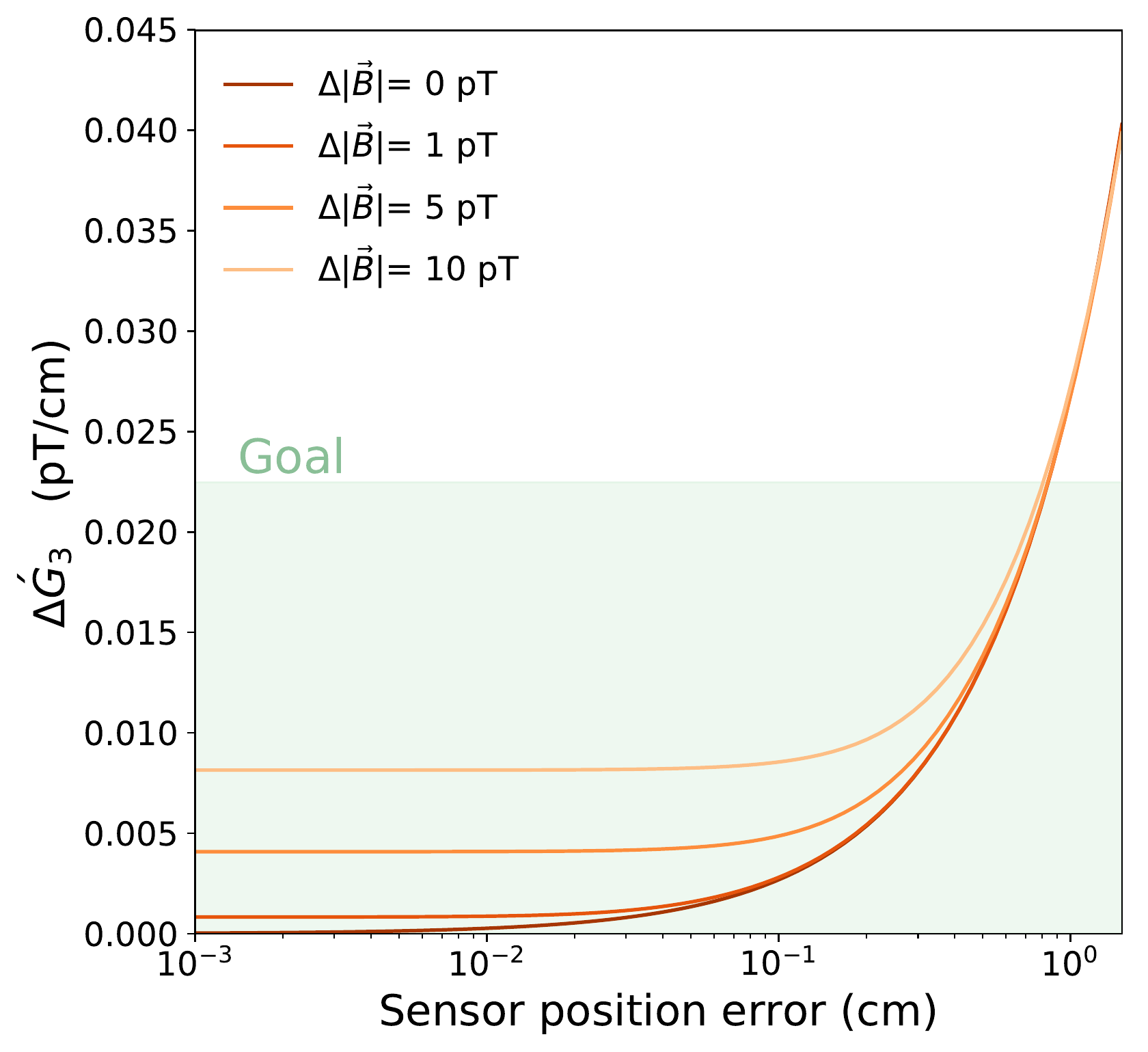}
  \caption[Cs accuracy]{
  Uncertainty of the calculation of the phantom gradient $\acute{G}_{3}$.
	The different curves show simulations for different assumptions of Cs magnetometer accuracy.  }
\label{fig:CsAccuracy}
\end{figure}

In n2EDM we plan to mount an array of 114 Cs sensors above and below the two ground electrodes.
The ability to extract all relevant gradients from the measurements of the Cs sensors depends
largely on the placement of the sensors.
Such placement must minimise the effects of the uncertainty of their position and field readings, and is a non-trivial task.
For this reason, a genetic algorithm was developed to output optimised coordinates of the CsM array.
Its fitness function includes all $G_{l,0}$ gradients up to $l=7$ weighted appropriately.
These positions optimize the extraction of all gradient components responsible for
systematic shifts in the neutron EDM and thus facilitate the correction of EDM results based on spatially resolved magnetic-field measurements.
This optimized set of positions has the advantage that the correction is significantly less dependent on the accuracy of the Cs sensor readings and on the errors in their position.
Figure~\ref{fig:CsAccuracy} shows the remaining error in the most important gradient, $G_{3,0}$,  after the correction with the Cs array as a function of placement accuracy.
The light green area indicates the goal necessary for the projected initial performance of n2EDM. 
The curves show simulation results with an assumed magnetic measurement accuracy ranging from \SI{0}{pT} (perfectly accurate) to \SI{10}{pT}.
Our goal is to achieve a geometrical placement accuracy of \SI{\pm 0.5}{mm}, which leads to virtually no increase in extraction uncertainty of the gradient. 
Our goal for the magnetometric accuracy is \SI{5}{pT}, which would give us a certain headroom for later upgrades of n2EDM.
These goals, necessary for corrections to the neutron EDM measurements, are by far the most stringent requirements for the Cs magnetometer array.
Requirements deduced from other types of measurements for which the CsM array will be used, like the homogenization of the magnetic field in order to avoid gradient-induced depolarization of neutrons, is thus automatically fulfilled if the simulated performance is achieved.

Since accuracy is so important, we plan to evaluate individually the accuracy of each Cs sensor in the array.
For that purpose a calibration setup is currently being installed in the magnetic shield of the previous singe-chamber spectrometer at PSI.
The setup consists of a rotating platform that can accommodate up to seven Cs sensors and a reference magnetometer based on $^3$He \cite{Koch2015}.
The setup permits the comparison of the reading of each  of the Cs sensors and of the $^3$He magnetometer to calibrate every Cs sensor that will be deployed.

%% file: Magnetometry-mapper.tex
An automated magnetic-field mapper will be used for the coil system commissioning and its cartography as well as for
offline control of high-order gradients and searches for magnetic contamination within the apparatus.
These measurements require an empty vacuum vessel in which to install the mapper, and will be performed once per year during the accelerator shutdown period.
Such a mapper apparatus has already been in use in the previous nEDM experiment. Although the design has evolved substantially over the years, the concept remains the same: a remote motion system allows movement of a magnetometer inside a large volume of interest.

\begin{figure}[htb]%
\centering
\includegraphics[width=1\columnwidth]{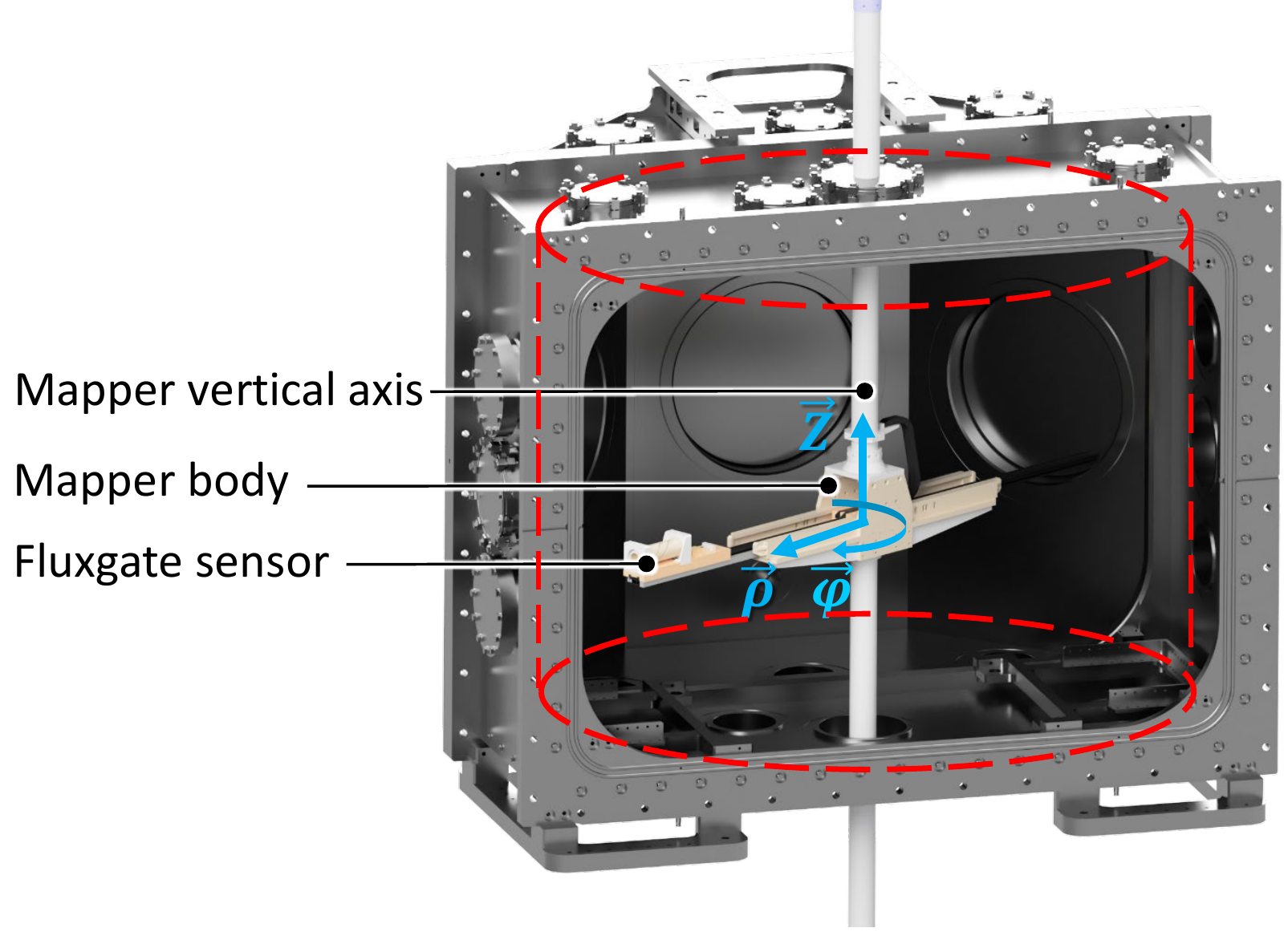}
\caption{Cutaway view of the mapper installed inside the vacuum vessel. The fluxgate can move along the $\rho$, $\varphi$, and $z$ axes and can explore almost the entire volume of the vacuum vessel.}
\label{fig:mapper}
\end{figure}

The sensor, usually a three-axis low-noise fluxgate, will explore a cylinder of \SI{80}{\cm} diameter and \SI{90}{\cm} height, thus covering the majority of the vacuum vessel's inner volume (see Fig. \ref{fig:mapper}).

The fluxgate can also be turned along the $\rho$ axis in $\pi/2$ steps to determine the overall DC offset of each single fluxgate for absolute field measurements with $< 200$~pT accuracy. 
Every part of the robot mounted inside the MSR is made of non-magnetic material (e.g. PEEK, POM and ceramics) and, in order to avoid Johnson noise, there are no metallic parts  close to the magnetometer.
The motorization block will be located outside, below the MSR, and is composed of three motors that are coupled to encoders for relative positioning of the magnetometer along the  $\vec{\rho}$, $\vec{\varphi}$ and $\vec{z}$ axes with a respective resolution of around \SI{100}{\micro m}, \SI{2}{\milli \radian} and \SI{5}{\micro m}.
The absolute position of the mapper will be determined after each installation and before each dismounting using photogrammetry, with an accuracy on the order of \SI{100}{\micro m}.
The combination of the relative and absolute sensor position knowledge is well within the requirements needed to extract the fifth-order phantom mode $\acute{G}_5$ detailed in Sec.~\ref{sec:4}.

A typical map acquisition lasts a few hours for a few thousand measurement points, and therefore several magnetic field configurations can be measured in a single day.
The complete analysis routine described in~\cite{Mapping2020} will be used to extract the magnetic field gradients within a few seconds once the measurement sequence is complete.

%% file: Summary.tex
We presented details of the new n2EDM apparatus being developed and built by the international nEDM collaboration based at the ultracold neutron source at PSI, Switzerland, with a view to significantly improving the sensitivity of the ongoing search for an electric dipole moment of the neutron.
The concept employed is based upon a room-temperature measurement of the spin-precession frequency of stored ultracold neutrons, using Ramsey’s method of separated oscillatory fields in combination with an atomic mercury co-magnetometer. This principle lies behind the most successful measurement that has been made to date.

 The concepts and requirements for the development of the new components, which are based on our experience with the previous apparatus, have been presented in detail. The expected increase in statistical sensitivity of a single Ramsey cycle for the chosen new design is stated. Advances in our understanding of systematic effects have been elaborated, and from these are derived the planned strategies to keep such effects under control.

The technical design of the core components is complete, and construction is ongoing. The various components developed by the collaborating institutions are gradually arriving at PSI for integration into the new apparatus. 
It has been demonstrated that a sensitivity of 1$\times$ 10$^{-27}$e·cm can be reached after 500 days of data taking. Possible future modifications are expected to lead to a sensitivity well within the 10$^{-28}$ e·cm range.

%% file: main.bbl
\begin{thebibliography}{10}
\providecommand{\url}[1]{{#1}}
\providecommand{\urlprefix}{URL }
\expandafter\ifx\csname urlstyle\endcsname\relax
  \providecommand{\doi}[1]{DOI \discretionary{}{}{}#1}\else
  \providecommand{\doi}{DOI \discretionary{}{}{}\begingroup
  \urlstyle{rm}\Url}\fi

\bibitem{Pospelov2005}
M.~Pospelov, A.~Ritz, Ann. Phys. \textbf{318}(1), 119  (2005).
\newblock
  \urlprefix\url{http://www.sciencedirect.com/science/article/pii/S0003491605000539}

\bibitem{Engel2013}
J.~Engel, M.J. Ramsey-Musolf, U.~{van Kolck}, Prog. Part. Nucl. Phys.
  \textbf{71}, 21  (2013).
\newblock
  \urlprefix\url{http://www.sciencedirect.com/science/article/pii/S0146641013000227}

\bibitem{Morrissey_2012}
D.E. Morrissey, M.J. Ramsey-Musolf, New Journal of Physics \textbf{14}(12),
  125003 (2012).
\newblock
  \urlprefix\url{https://doi.org/10.1088%2F1367-2630%2F14%2F12%2F125003}

\bibitem{nEDM-PhysRevLett}
C.~Abel et~al., Phys. Rev. Lett. \textbf{124}(8), 081803 (2020).
\newblock
  \urlprefix\url{https://link.aps.org/doi/10.1103/PhysRevLett.124.081803}

\bibitem{Graner2016}
B.~Graner et~al., Phys. Rev. Lett. \textbf{116}(16), 161601 (2016).
\newblock
  \urlprefix\url{https://link.aps.org/doi/10.1103/PhysRevLett.116.161601}

\bibitem{SIKIVIE2012}
P.~Sikivie, Comptes Rendus Physique \textbf{13}(2), 176  (2012).
\newblock
  \urlprefix\url{http://www.sciencedirect.com/science/article/pii/S1631070511002039}.
\newblock Flavour physics and CP violation / Physique de la saveur et violation
  de CP

\bibitem{Peccei1977}
R.D. Peccei, H.R. Quinn, Phys. Rev. Lett. \textbf{38}(25), 1440 (1977).
\newblock \urlprefix\url{https://link.aps.org/doi/10.1103/PhysRevLett.38.1440}

\bibitem{Marsh2016h}
D.J. Marsh, Physics Reports \textbf{643}, 1  (2016).
\newblock
  \urlprefix\url{http://www.sciencedirect.com/science/article/pii/S0370157316301557}

\bibitem{Graham2015}
P.W. Graham et~al., Annual Review of Nuclear and Particle Science
  \textbf{65}(1), 485 (2015).
\newblock \urlprefix\url{https://doi.org/10.1146/annurev-nucl-102014-022120}

\bibitem{Jungmann2013}
K.~Jungmann, Annalen der Physik \textbf{525}(8-9), 550 (2013).
\newblock
  \urlprefix\url{https://onlinelibrary.wiley.com/doi/abs/10.1002/andp.201300071}

\bibitem{Chupp2015PR}
T.~Chupp, M.~Ramsey-Musolf, Phys. Rev. C \textbf{91}(3), 035502 (2015).
\newblock \urlprefix\url{https://link.aps.org/doi/10.1103/PhysRevC.91.035502}

\bibitem{Chupp2019}
T.E. Chupp et~al., Rev. Mod. Phys. \textbf{91}(1), 015001 (2019).
\newblock \urlprefix\url{https://link.aps.org/doi/10.1103/RevModPhys.91.015001}

\bibitem{pignol2019global}
G.~Pignol, arXiv 1912.07876  (2019).
\newblock \urlprefix\url{https://arxiv.org/abs/1912.07876v1}

\bibitem{Baker2014}
C.~Baker et~al., Nucl. Instrum. Methods A \textbf{736}, 184  (2014).
\newblock
  \urlprefix\url{http://www.sciencedirect.com/science/article/pii/S0168900213013193}

\bibitem{Lauss2012}
B.~Lauss et~al., AIP Conference Proceedings \textbf{1441}(1), 576 (2012).
\newblock \urlprefix\url{https://aip.scitation.org/doi/abs/10.1063/1.3700622}

\bibitem{Lauss2014}
B.~Lauss, Physics Procedia \textbf{51}, 98  (2014).
\newblock
  \urlprefix\url{http://www.sciencedirect.com/science/article/pii/S1875389213007104}

\bibitem{Bison2020}
G.~Bison et~al., Eur. Phys. J. A \textbf{56}(2), 33 (2020).
\newblock \urlprefix\url{https://doi.org/10.1140/epja/s10050-020-00027-w}

\bibitem{CsM_PRA2020}
C.~Abel et~al., Phys. Rev. A \textbf{101}(5), 053419 (2020).
\newblock \urlprefix\url{https://link.aps.org/doi/10.1103/PhysRevA.101.053419}

\bibitem{Afach2015EPJA}
S.~Afach et~al., Eur. Phys. J. A \textbf{51}, 143 (2015).
\newblock \urlprefix\url{https://doi.org/10.1140/epja/i2015-15143-7}

\bibitem{SNS}
M.~Ahmed et~al., Journal of Instrumentation \textbf{14}(11), 11017 (2019).
\newblock
  \urlprefix\url{https://doi.org/10.1088%2F1748-0221%2F14%2F11%2Fp11017}

\bibitem{Ito2018}
T.M. Ito et~al., Phys. Rev. C \textbf{97}(1), 012501 (2018).
\newblock \urlprefix\url{https://link.aps.org/doi/10.1103/PhysRevC.97.012501}

\bibitem{Ruediger2017}
R.~Picker, JPS Conf. Proc. \textbf{13}, 010005 (2017).
\newblock \urlprefix\url{https://journals.jps.jp/doi/10.7566/JPSCP.13.010005}

\bibitem{Piegsa2019}
E.~Chanel et~al., Proc. Int. Workshop on Particle Physics at Neutron Sources
  (PPNS) \textbf{219}, 02004 (2019).
\newblock \urlprefix\url{https://doi.org/10.1051/epjconf/201921902004}

\bibitem{Wurm_2019}
D.~Wurm et~al., Proc. Int. Workshop on Particle Physics at Neutron Sources
  (PPNS) \textbf{219}, 02006 (2019).
\newblock \urlprefix\url{http://dx.doi.org/10.1051/epjconf/201921902006}

\bibitem{Serebrov2017}
A.~Serebrov, POS (INPS2016) \textbf{281}, 179 (2017).
\newblock \urlprefix\url{https://doi.org/10.22323/1.281.0179}

\bibitem{Altarev1980NuPhA}
I.S. {Altarev} et~al., Nucl. Phys. A \textbf{341}(2), 269 (1980).
\newblock \urlprefix\url{http://adsabs.harvard.edu/abs/1980NuPhA.341..269A}

\bibitem{Green1998}
K.~{Green} et~al., Nucl. Instrum. Methods A \textbf{404}(2-3), 381 (1998).
\newblock \urlprefix\url{http://adsabs.harvard.edu/abs/1998NIMPA.404..381G}

\bibitem{Zsigmond2018}
G.~Zsigmond, Nucl. Instrum. Methods A \textbf{881}, 16  (2018).
\newblock
  \urlprefix\url{http://www.sciencedirect.com/science/article/pii/S0168900217311476}

\bibitem{Ries2016}
D.A. Ries, The source for ultracold neutrons at the paul scherrer institute -
  characterisation, optimisation, and international comparison.
\newblock Ph.D. thesis, ETH Zurich, Zürich (2016).
\newblock \urlprefix\url{10.3929/ethz-a-010795050}

\bibitem{Bison2017}
G.~Bison et~al., Phys. Rev. C \textbf{95}(4), 045503 (2017).
\newblock \urlprefix\url{https://link.aps.org/doi/10.1103/PhysRevC.95.045503}

\bibitem{Bondar2017}
V.~Bondar et~al., Phys. Rev. C \textbf{96}(3), 035205 (2017).
\newblock \urlprefix\url{https://link.aps.org/doi/10.1103/PhysRevC.96.035205}

\bibitem{Mohanmurthy2020}
P.~Mohanmurthy, {A Search for Neutron to Mirror-Neutron Oscillations}.
\newblock Ph.D. thesis, ETH Z{\"u}rich, No.26525 (2020).
\newblock \urlprefix\url{10.3929/ethz-b-000417951}

\bibitem{Atchison2005c}
F.~Atchison et~al., Phys. Lett. B \textbf{625}, 19 (2005).
\newblock
  \urlprefix\url{https://www.sciencedirect.com/science/article/abs/pii/S0370269305011846?via%3Dihub}

\bibitem{Bodek2008}
K.~Bodek et~al., Nucl. Instrum. Methods \textbf{597}(2), 222  (2008).
\newblock
  \urlprefix\url{http://www.sciencedirect.com/science/article/pii/S0168900208014381}

\bibitem{Bison2021}
G.~Bison et~al., Experimental and simulation study of the energy dependent
  ultracold neutron transport at the PSI UCN source, in preparation for Phys.
  Rev. A.  (2021)

\bibitem{Uniformity2019}
C.~Abel et~al., Phys. Rev. A \textbf{99}(4), 042112 (2019).
\newblock \urlprefix\url{https://link.aps.org/doi/10.1103/PhysRevA.99.042112}

\bibitem{Afach2015Grav}
S.~Afach et~al., Phys. Rev. D \textbf{92}, 052008 (2015).
\newblock \urlprefix\url{https://link.aps.org/doi/10.1103/PhysRevD.92.052008}

\bibitem{SpinEcho2015}
S.~Afach et~al., Phys. Rev. Lett. \textbf{115}, 162502 (2015).
\newblock
  \urlprefix\url{https://link.aps.org/doi/10.1103/PhysRevLett.115.162502}

\bibitem{Redfield1957}
A.G. {Redfield}, IBM Journal of Research and Development \textbf{1}(1), 19
  (1957).
\newblock \urlprefix\url{https://ieeexplore.ieee.org/abstract/document/5392713}

\bibitem{Lamoreaux1996}
S.K. Lamoreaux, Phys. Rev. A \textbf{53}(6), R3705 (1996).
\newblock \urlprefix\url{https://link.aps.org/doi/10.1103/PhysRevA.53.R3705}

\bibitem{Pendlebury2004}
J.M. Pendlebury et~al., Phys. Rev. A \textbf{70}(3), 032102 (2004).
\newblock \urlprefix\url{https://link.aps.org/doi/10.1103/PhysRevA.70.032102}

\bibitem{Lamoreaux2005}
S.K. Lamoreaux, R.~Golub, Phys. Rev. A \textbf{71}(3), 032104 (2005).
\newblock \doi{10.1103/PhysRevA.71.032104}.
\newblock \urlprefix\url{https://link.aps.org/doi/10.1103/PhysRevA.71.032104}

\bibitem{Barabanov2006}
A.L. Barabanov, R.~Golub, S.K. Lamoreaux, Phys. Rev. A \textbf{74}(5), 052115
  (2006).
\newblock \urlprefix\url{https://link.aps.org/doi/10.1103/PhysRevA.74.052115}

\bibitem{Clayton2011}
S.M. Clayton, Journal of Magnetic Resonance \textbf{211}(1), 89  (2011).
\newblock
  \urlprefix\url{http://www.sciencedirect.com/science/article/pii/S1090780711001418}

\bibitem{Swank2012}
C.~Swank, A.~Petukhov, R.~Golub, Phys.~Lett.~A \textbf{376}(34), 2319  (2012).
\newblock
  \urlprefix\url{http://www.sciencedirect.com/science/article/pii/S0375960112006330}

\bibitem{Pignol2012}
G.~Pignol, S.~Roccia, Phys. Rev. A \textbf{85}(4), 042105 (2012).
\newblock \urlprefix\url{https://link.aps.org/doi/10.1103/PhysRevA.85.042105}

\bibitem{Pignol2015}
G.~Pignol et~al., Phys. Rev. A \textbf{92}(5), 053407 (2015).
\newblock \urlprefix\url{https://link.aps.org/doi/10.1103/PhysRevA.92.053407}

\bibitem{Golub2015}
R.~Golub et~al., Phys. Rev. A \textbf{92}(6), 062123 (2015).
\newblock \urlprefix\url{https://link.aps.org/doi/10.1103/PhysRevA.92.062123}

\bibitem{Swank2016}
C.M. Swank, A.K. Petukhov, R.~Golub, Phys. Rev. A \textbf{93}(6), 062703
  (2016).
\newblock \urlprefix\url{https://link.aps.org/doi/10.1103/PhysRevA.93.062703}

\bibitem{Afach2015}
S.~Afach et~al., Eur. Phys. J. D \textbf{69}(10), 225 (2015).
\newblock
  \urlprefix\url{https://link.springer.com/article/10.1140/epjd/e2015-60207-4}

\bibitem{Pignol2019}
G.~Pignol, Physics Letters B \textbf{793}, 440  (2019).
\newblock
  \urlprefix\url{http://www.sciencedirect.com/science/article/pii/S0370269319303235}

\bibitem{PinJung2021}
C.~Abel et~al., Johnson-Nyquist Noise Effects in Neutron Electric-Dipole-Moment
  Experiments, in preparation for Phys.Rev.A  (2021)

\bibitem{Golub2007}
S.K. Lamoreaux, R.~Golub, Phys. Rev. Lett. \textbf{98}(14), 149101 (2007).
\newblock
  \urlprefix\url{https://link.aps.org/doi/10.1103/PhysRevLett.98.149101}

\bibitem{NNews1992}
P.~Schofield, Neutron News \textbf{3}(3), 29  (1992).
\newblock
  \urlprefix\url{https://ncnr.nist.gov/resources/n-lengths/elements/hg.html}

\bibitem{Abragam1975}
A.~Abragam et~al., Journal de Physique Lettres \textbf{36}(11), 263  (1975).
\newblock \urlprefix\url{https://hal.archives-ouvertes.fr/jpa-00231204}

\bibitem{Grinten1999}
M.~{van der Grinten} et~al., Nucl. Instrum. Methods A \textbf{423}(2), 421
  (1999).
\newblock
  \urlprefix\url{http://www.sciencedirect.com/science/article/pii/S0168900298013382}

\bibitem{Atchison2006}
F.~Atchison et~al., Phys. Lett. B \textbf{642}(1), 24  (2006).
\newblock
  \urlprefix\url{http://www.sciencedirect.com/science/article/pii/S037026930601166X}

\bibitem{Atchison2006b}
F.~Atchison et~al., Phys. Rev. C \textbf{74}(5), 055501 (2006).
\newblock \urlprefix\url{https://link.aps.org/doi/10.1103/PhysRevC.74.055501}

\bibitem{Atchison2007a}
F.~Atchison et~al., Phys. Rev. C \textbf{76}(4), 044001 (2007).
\newblock \urlprefix\url{https://link.aps.org/doi/10.1103/PhysRevC.76.044001}

\bibitem{Atchison2007b}
F.~Atchison et~al., Diamond and Related Materials \textbf{16}(2), 334  (2007).
\newblock
  \urlprefix\url{http://www.sciencedirect.com/science/article/pii/S0925963506002160}

\bibitem{Atchison2008}
F.~Atchison et~al., Nucl. Instrum. Methods A \textbf{587}, 82  (2008).
\newblock
  \urlprefix\url{{https://www.dora.lib4ri.ch/psi/islandora/object/psi%3A18824}}

\bibitem{COMSOL}
{COMSOL Multiphysics$\textregistered$,} ver. 5.3.
\newblock \url{www.comsol.com}

\bibitem{BLAU2016}
B.~Blau et~al., Nucl. Inst. Method. A \textbf{807}, 30  (2016).
\newblock
  \urlprefix\url{http://www.sciencedirect.com/science/article/pii/S0168900215013091}

\bibitem{Hel15}
S.~Afach et~al., Eur. Phys. J. A \textbf{51}(11), 143 (2015).
\newblock \urlprefix\url{https://doi.org/10.1140/epja/i2015-15143-7}

\bibitem{Grig_97}
S.~Grigoriev, A.~Okorokov, V.~Runov, Nucl. Inst. Method. A \textbf{384}(2), 451
   (1997).
\newblock
  \urlprefix\url{http://www.sciencedirect.com/science/article/pii/S0168900296009199}

\bibitem{Sae19}
W.~Saenz, A gaseous detector for ultracold neutrons in the n2edm project.
\newblock Master's thesis, Universite de Caen Normandie (2019).
\newblock \urlprefix\url{http://hal.in2p3.fr/in2p3-02957354}

\bibitem{Leh15}
G.~Lehaut et~al., Nucl. Instrum. Methods A \textbf{797}, 57 (2015).
\newblock
  \urlprefix\url{https://www.sciencedirect.com/science/article/abs/pii/S0168900215007093?via%3Dihub}

\bibitem{Bork2001}
J.~Bork et~al., Proc. 12th Int. Conf. on Biomagnetism \textbf{Biomag2000}, 970
  (2001)

\bibitem{Rawlik2018}
M.~Rawlik et~al., American Journal of Physics \textbf{86}(8), 602 (2018).
\newblock \urlprefix\url{https://doi.org/10.1119/1.5042244}

\bibitem{ANSYS}
{ANSYS$\textregistered$}.
\newblock \url{www.ansys.com}

\bibitem{Ban2018}
G.~Ban et~al., Nucl. Instrum. Methods A \textbf{896}, 129 (2018).
\newblock \urlprefix\url{https://doi.org/10.1016/j.nima.2018.04.025}

\bibitem{Aleksandrov2006}
E.~Aleksandrov et~al., Tech. Phys. Lett. \textbf{32}(7), 627 (2006).
\newblock \urlprefix\url{https://doi.org/10.1134/S1063785006070236}

\bibitem{Bloom1962}
A.L. {Bloom}, Applied Optics \textbf{1}, 61 (1962).
\newblock \urlprefix\url{https://doi.org/10.1364/AO.1.000061}

\bibitem{Budker2007}
D.~{Budker}, M.~{Romalis}, Nature Physics \textbf{3}, 227 (2007).
\newblock \urlprefix\url{https://doi.org/10.1038/nphys566}

\bibitem{Serebrov2014}
A.P. Serebrov et~al., Phys. Rev. C \textbf{92}(5), 055501 (2015).
\newblock \urlprefix\url{https://link.aps.org/doi/10.1103/PhysRevC.92.055501}

\bibitem{Afach2015OExpress}
S.~Afach et~al., Opt. Express \textbf{23}(17), 22108 (2015).
\newblock
  \urlprefix\url{http://www.opticsexpress.org/abstract.cfm?URI=oe-23-17-22108}

\bibitem{Grujic2015}
Z.D. Gruji\'{c} et~al., Eur. Phys. J. D \textbf{69}, 135 (2015).
\newblock \urlprefix\url{https://doi.org/10.1140/epjd/e2015-50875-3}

\bibitem{Koch2015}
H.C. Koch et~al., EPJ D \textbf{69}, 202 (2015).
\newblock \urlprefix\url{https://doi.org/10.1140/epjd/e2015-60018-7}

\bibitem{Mapping2020}
C.~Abel et~al., Mapping of the magnetic field to correct systematic effects in
  a neutron electric dipole moment experiment, in preparation for Phys. Rev. A.
   (2021)

\end{thebibliography}
